\documentclass[structabstract]{aa}  
\usepackage{graphicx}
\usepackage{txfonts}
\usepackage{natbib}
\newcommand{\oiii}{[O\,{\footnotesize III}]}
\newcommand{\nii}{[N\,{\footnotesize II}]}
\newcommand{\niii}{[N\,{\footnotesize III}]}
\newcommand{\cii}{[C\,{\footnotesize II}]}
\newcommand{\hii}{H\,{\footnotesize II}}
\newcommand{\neiii}{[Ne\,{\footnotesize III}]}
\newcommand{\suiii}{[S\,{\footnotesize III}]}
\newcommand{\pcm}{\,cm$^{-1}$}
\newcommand{\cc}{\,cm$^{-3}$}
\newcommand{\micron}{\,$\mu$m}
\bibpunct{(}{)}{;}{a}{}{,}
\begin{document}
   \title{Properties of active galactic star-forming regions probed by imaging spectroscopy with the Fourier transform spectrometer (FTS) onboard AKARI}

\author{Yoko Okada\inst{1,2}
\and Mitsunobu Kawada\inst{3}
\and Noriko Murakami\inst{3,4}
\and Takafumi Ootsubo\inst{2}
\and Hidenori Takahashi\inst{5}
\and Akiko Yasuda\inst{2,3}
\and Daisuke Ishihara\inst{6,3}
\and Hidehiro Kaneda\inst{3}
\and Hirokazu Kataza\inst{2}
\and Takao Nakagawa\inst{2}
\and Takashi Onaka\inst{6}
}

\institute{I. Physikalisches Institut, Universit\"{a}t zu K\"{o}ln, Z\"{u}lpicher Str. 77, 50937 K\"{o}ln, Germany\\
\email{okada@ph1.uni-koeln.de}
\and Institute of Space and Astronautical Science, Japan Aerospace Exploration Agency, 3-1-1 Yoshinodai, Kanagawa 229-8520, Japan
\and Division of Particle and Astrophysical Science, Graduate School of Science, Nagoya University, Furo-cho, Chikusa-ku, Nagoya 464-8602, Japan
\and Bisei Astronomical Observatory, 1723-70 Ookura, Bisei-cho, Ibara, Okayama 714-1411, Japan
\and Gunma Astronomical Observatory, 6860-86 Nakayama, Takayama, Agatsuma, Gunma 377-0702, Japan
\and Department of Astronomy, Graduate School of Science, University of Tokyo, 7-3-1 Hongo, Bunkyo-ku, Tokyo 113-0033, Japan}

   \date{Received ; accepted }

% \abstract{}{}{}{}{} 
% 5 {} token are mandatory

  \abstract
   {}
   {We investigate the structure of the interstellar medium (ISM) and identify the location of possible embedded excitation sources from far-infrared (FIR) line and mid-infrared continuum emission maps.}
   {We carried out imaging spectroscopic observations of four giant Galactic star-forming regions with the Fourier Transform Spectrometer (FTS) onboard AKARI.   We obtained \oiii\ 88\micron\ and \cii\ 158\micron\ line intensity maps of all the regions: G3.270-0.101, G333.6-0.2, NGC~3603, and M17.}
   {For G3.270-0.101, we obtained high-spatial-resolution \oiii\ 88\micron\ line-emission maps and a FIR continuum map for the first time, which imply that \oiii\ 88\micron\ emission identifies the excitation sources more clearly than the radio continuum emission.  In G333.6-0.2, we found a local \oiii\ 88\micron\ emission peak, which is indicative of an excitation source.  This is supported by the 18\micron\ continuum emission, which is considered to trace the hot dust distribution.  For all regions, the \cii\ 158\micron\ emission is distributed widely as suggested by previous observations of star-forming regions.}
   {We conclude that \oiii\ 88\micron\ emission traces the excitation sources more accurately than the radio continuum emission, especially where there is a high density and/or column density gradient.  The FIR spectroscopy provides a promising means of understanding the nature of star-forming regions.}

   \keywords{infrared:ISM---ISM: HII regions---ISM: lines and bands}

   \titlerunning{Star-forming regions with FTS onboard AKARI}
   \authorrunning{Y. Okada et al.}

   \maketitle
%
%________________________________________________________________

\section{Introduction}
Far-infrared (FIR) spectroscopy of star-forming regions provides us with a great deal of information about the interstellar medium (ISM) without being affected by the uncertainty of the extinction correction.  FIR forbidden lines are less sensitive to the electron temperature than optical forbidden lines in \hii\ regions, and observations with balloon-borne and airborne telescopes have been used to determine physical properties such as the electron density and elemental abundance \citep{Rubin88,Simpson95,Simpson04}.  The Long-Wavelength Spectrometer \citep[LWS;][]{Clegg96} onboard the Infrared Space Observatory \citep[ISO;][]{Kessler96} provided continuous spectra of the wavelength range 43--197\micron, enabling us to investigate the physical conditions of the gas in the ionized and/or photodissociation regions (PDRs) by measuring several emission lines and the continuum emission (\citealt{MartinHernandez02}, see \citealt{Abergel05} and \citealt{Peeters05} for reviews).  Since the LWS was a single-beam instrument, raster-scan observations were needed to investigate the spatial distribution of emission spectra and physical properties \citep[e.g.][]{Rodriguez01,Mizutani02,Okada03,Okada06}.  In the present paper, we present FIR imaging spectroscopic observations with the Fourier Transform Spectrometer (FTS) onboard AKARI \citep{Murakami07,Kawada08}, the Japanese infrared astronomical satellite launched in 2006.  The FTS is the spectroscopic part of the Far-Infrared Surveyor \citep[FIS;][]{Kawada07}.  Since the spectroscopic capability of the Multiband Imaging Photometer for {\it Spitzer} (MIPS) onboard {\it Spitzer} \citep{Werner04} is restricted to wavelengths of $<97$\micron\ at low spectral resolutions of $R=15$--$25$, the FIS-FTS is a unique FIR spectroscopic instrument in space for studying FIR forbidden-line emission after ISO/LWS and before the Herschel Space Observatory \citep{Pilbratt08}.  With two dimensional arrays, the FIS-FTS has the capability of obtaining spectral maps over a wide area in a few pointing observations.

\section{Observations and data reduction}
\subsection{Observation parameters}

\begin{table*}
\caption{The summary of the observations.}\label{obs_summary}
\centering
\begin{tabular}{ccccccc}
\hline
Target & Number of & Observation ID & Observation Date & AOT & Reset interval & Resolution\\
& pointings & & (UT) && [s] & \\
\hline
G3.270-0.101 & 1 & 5110044 & Mar. 20 2007 & FIS03 & 0.25 & Full \\
G333.6-0.2 & 2 & 5110052/53 & Mar. 3 \& 4 2007 & FIS03 & 0.1 & Full\\
NGC3603 & 3 & 1403056/57/59 & Jul. 21,22,23 2007 & FIS03 & 0.1 & Full\\
M17 & 1 & 5110024 & Sep. 27 2006 & FIS03 & 0.1 & Full\\
\hline
\end{tabular}
\end{table*}

The observations are summarized in Table~\ref{obs_summary}.  All observations were carried out with the Astronomical Observation Template (AOT) of FIS03, which is a spectroscopic mode of the AKARI/FIS.  The FIS-FTS adopts a Martin-Puplett-type interferometer \citep[see][for details of the instrument]{Kawada08}.  To investigate emission lines, we use the full-resolution mode with the spectral resolution of 0.19\pcm\ without apodization and 0.38\pcm\ with `Hanning' apodization.  We obtain 7 sets of forward and backward scans of the moving mirror with a total $\sim 11$ minute integration time for each pointing observation.  The FIS has a short wavelength array, SW, and long wavelength array, LW, each of which has narrow and wide bands.  In the spectroscopic mode, only a wide band of each array is used.  They are $3\times 20$ and $3\times 15$ pixel arrays that cover 85--130\pcm\ and 60--88\pcm, on pixel scales of 26.8\arcsec and 44.2\arcsec, respectively \citep[][Murakami et al. in prep.]{Kawada07,Kawada08}.  Some of the present observations were carried out as Director's Time observations, and the others as the Mission Program of ISMGN (Interstellar dust and gas in various environments of out Galaxy and nearby galaxies: P.I. H. Kaneda).  The field-of-view (FOV) of SW and LW is $1.5^\prime \times 9.8^\prime$ and $2.4^\prime \times 12.2^\prime$, respectively.  Because of the misalignment of the arrays, the FOVs of SW and LW are not completely matched with each other and cover slightly different areas of the sky \citep{Kawada07}. The point spread function (PSF) of the FIS-FTS can be described well by a combination of two Gaussian profiles, which are located along the major axis of the detectors with the separation of $14^{\prime\prime}$ in each detector array \citep{Kawada08}.  The FWHM of each Gaussian is $39^{\prime\prime}$ and $44^{\prime\prime}$ along the minor and major axis, respectively, for SW, and $53^{\prime\prime}$ and $57^{\prime\prime}$ along the minor and major axis, respectively, for LW.  The position accuracy of the FTS mode is less than a half of the pixel size (Murakami et al. in prep.).

\subsection{Targets}

To take advantage of the imaging spectroscopy of the FIS-FTS, we selected four active star-forming regions that are bright and extended enough to detect the interferogram with a high signal-to-noise ratio (S/N) over an area of $\sim 10$\arcmin. %Figures~\ref{fig:obspos_G3}--\ref{fig:obspos_M17} show the observed positions in each target.

G3.270-0.101 is a giant \hii\ region located 14.3~kpc from the Sun.  It was observed in the radio and in the mid-infrared (MIR) with the Mid-course Space Experiment (MSX) \citep{Conti04}.  The number of the Lyman continuum photons $N(\mathrm{Lyc})$ is estimated to be $3\times 10^{50}$\,s$^{-1}$.  \citet{Okada08} detected several MIR ionic fine-structure lines with the Infrared Spectrometer (IRS) onboard {\it Spitzer} in this region.  We completed one pointing observation, covering the positions observed by the IRS.

G333.6-0.2 is a bright \hii\ region at radio and infrared wavelengths \citep{Goss70,Becklin73,Hyland80,Fujiyoshi98,Fujiyoshi01,Fujiyoshi05,Fujiyoshi06} that is totally obscured in the visible.  It is located at 3.1~kpc \citep{Conti04} and likely to be excited by a cluster of O and B stars \citep{Fujiyoshi98}, behind which dust grains with a large variety of temperatures emit strong infrared emission \citep{Hyland80}. \citet{Fujiyoshi06} discovered the complex structure of the central 10 arcsec region using high spatial resolution radio observations.  \citet{Colgan93} and \citet{Simpson04} detected several MIR and FIR forbidden lines and examined the electron density and the abundance ratio at the center and several northern positions.  \citet{Okada08} performed MIR spectroscopy with the {\it Spitzer}/IRS as for G3.270-0.101, and detected 14 emission lines from metal ions and molecular hydrogen, indicating the coexistence of gas in various excitation states.

NGC~3603 is one of the most massive, optically visible \hii\ regions in our Galaxy.  It has an integrated H$\alpha$ flux of $L$(H$\alpha$) $\sim 1.5\times 10^{39}$\ erg\ s$^{-1}$ , equivalent to 20 O5 V stars \citep{Kennicutt84}.  It is located in the Carina spiral arm at a distance of about 7~kpc \citep[][and references within]{Nurnberger02}.  The central cluster, HD~97950, contains 3 Wolf-Rayet (WR) stars \citep{Moffat94} and it is understood it experienced a single burst of star formation 1 Myr ago \citep{Stolte04}.  It shows a remarkable similarity to the dense core R136 in 30 Dor in the LMC \citep{Moffat94}.  \citet{Nurnberger02} determined the properties and structure of the molecular gas using CS observations.  The structure of the ISM and infrared sources were examined with MIR observations \citep{Nurnberger03,Nurnberger_Stanke03}.  MIR spectroscopic observations were performed using the {\it Spitzer}/IRS and spatial variations in both polycyclic aromatic hydrocarbons (PAHs) and elemental abundances were examined \citep{Lebouteiller07,Lebouteiller08}.

M17 is also one of the brightest giant \hii\ regions in our Galaxy and has been extensively studied at various wavelengths. It contains a dozen of O- or early B-type stars \citep{Hanson97} and the age of the central cluster is $<3$~Myr \citep{Hanson97,Jiang02}.  \citet{Felli84} presented radio maps of M17 and discussed the structures and nature of the ISM.  The northern bar is clearly seen in optical emission, which is indicative of a small amount of foreground extinction, whereas the southern bar has no clear optical counterpart and is highly obscured.  Emission at \oiii\ 88\micron\ was detected by \citet{Ward75}, and the electron density and nature of the ionizing sources were investigated based on observations of the FIR \oiii\ and \niii\ emission lines \citep{Emery83}.  The \cii\ 158\micron\ emission is widely distributed in this region \citep{Stutzki88,Matsuhara89}.  \citet{Stutzki88} investigated the clumpiness of the edge-on \hii\ region/PDR interface using the spatial distribution of \cii\ 158\micron\ and CO rotational line emission.  The PDR structures and morphology were also investigated by \citet{Ando02} and \citet{Povich07}.

\subsection{Data reduction}

\begin{figure*}
\centering
\resizebox{0.45\linewidth}{!}{\includegraphics{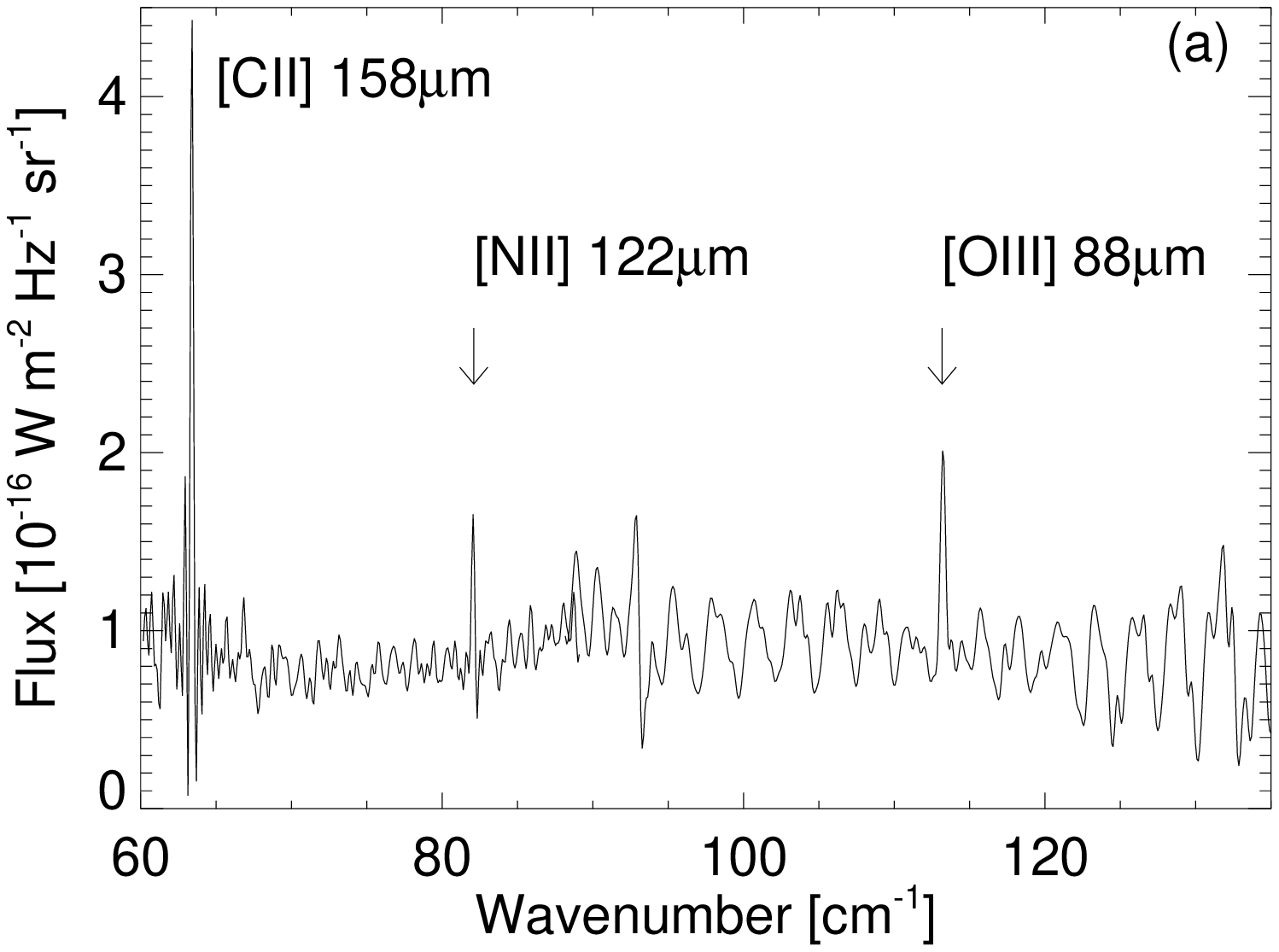}}
\resizebox{0.45\linewidth}{!}{\includegraphics{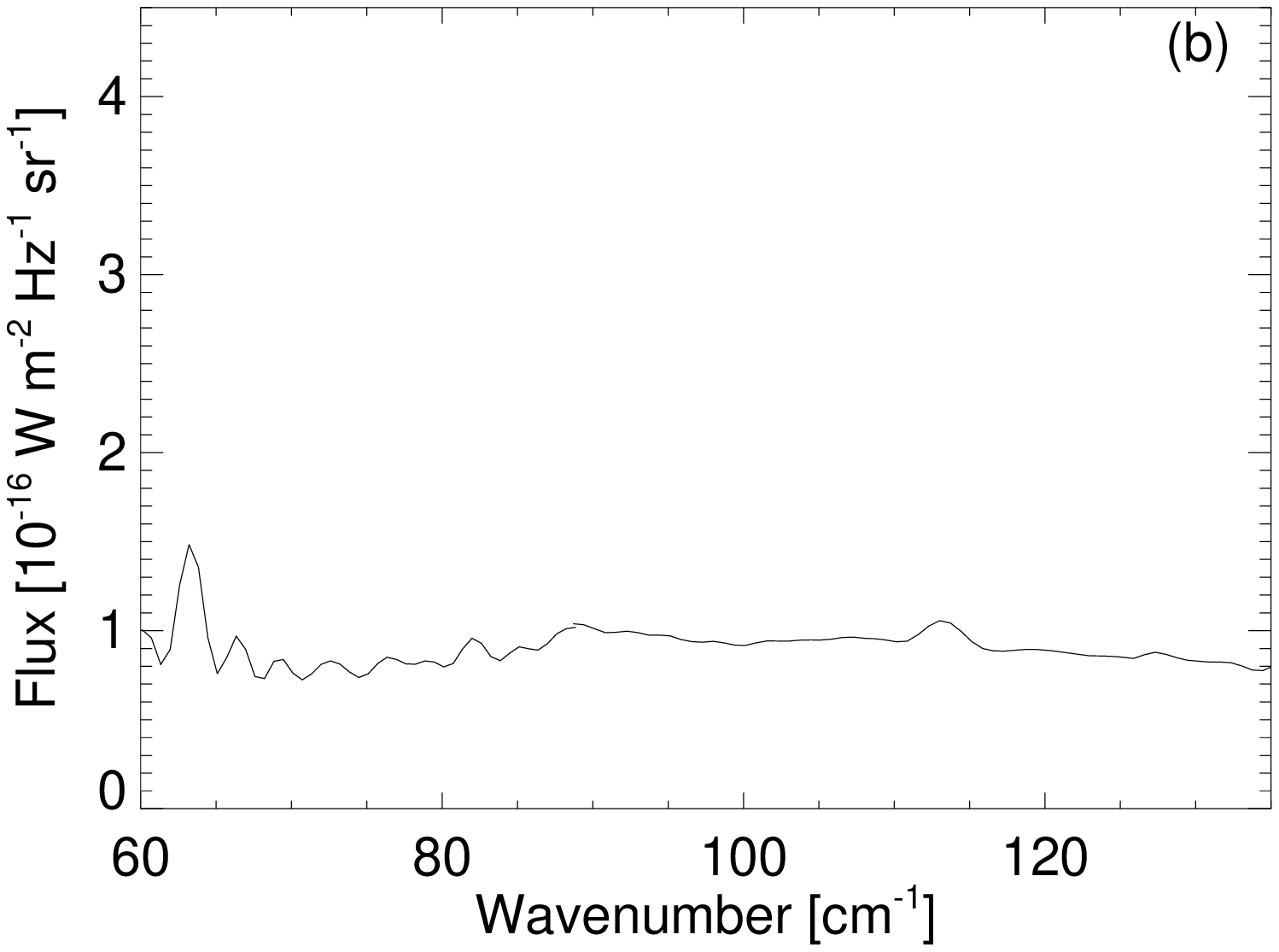}}
\caption{(a) An example of the obtained full-resolution spectra in G333.6-0.2 region.  Three emission lines are labeled.  For the spectrum of 60--89\pcm, obtained by LW detector, we apply the defringing process with fringe parameters of around \nii\ 122\micron\ line and do not apply the apodization.  For the spectrum of 88-135\pcm, obtained by SW detector, we apply the defringing process with fringe parameters around \oiii\ 88\micron\ line and apply the `Hanning' apodization (see Appendix \ref{app:linefit}). (b) The lower resolution version of (a) obtained by the Fourier transformation of the shorter range of the interferogram (see text).}\label{ex_spectrum}
\end{figure*}

We performed the data reduction of the time series data of the FIS-FTS using the official pipeline.  Details of the processes and calibration are described in Murakami et al. (in prep.).  The time series data are provided as integrated ramp signals.  The interferogram is derived from the differentiation of the time series data after correcting for the non-linearity of the ramp curve.  We remove glitches in the interferogram and apply the discrete Fourier transform with the self-determined zero optical path difference to obtain the spectra.  After the flux calibration, we have a spectrum of each forward and backward scan of the driving mirror.  We then take an average of 7 scans of the individual scanning directions.

\oiii\ 88\micron\ is detected by the SW detector and \nii\ 122\micron\ and \cii\ 158\micron\ are detected by the LW detector.  The spectra obtained by both detectors are affected by fringe patterns \citep[][Murakami et al. in prep.]{Kawada08}.  The defringing process and the fitting function are optimized for each line.  Details of the line-fitting process are described in Appendix \ref{app:linefit}.  Figure~\ref{ex_spectrum}a shows an example of a spectrum (an average of the forward and backward scans).  Since the parameters of the defringing process are optimized for fringes around emission lines, the residual of the fringe pattern becomes large at wavenumbers far from the emission lines.  After obtaining the line intensities for each pixel within the detectors, we combine the results from 2 or 3 pointing observations of the same region (table~\ref{obs_summary}) and compile maps with a regular grid of half the size of the original pixel scale.  Details of constructing line maps and estimating the uncertainties are described in Appendix \ref{app:const_map}.  The final line maps are shown in Figs.~\ref{linemap_G3}, \ref{linemap_G333}, \ref{linemap_NGC3603}, and \ref{linemap_M17} for individual regions, which are discussed in the next section.

The maps of the continuum emission at specific wavenumbers are created by a similar procedure to that of the line maps.  Instead of the defringed full-resolution spectra, we use low-resolution spectra, which are obtained by the Fourier transformation of a narrower range of the optical path ($\pm 0.35$~cm for SW and $\pm 0.4$~cm for LW) to avoid the effect of fringes (Fig.~\ref{ex_spectrum}b).  The obtained continuum maps are also shown in Figs.~\ref{linemap_G3}, \ref{linemap_G333}, \ref{linemap_NGC3603}, and \ref{linemap_M17}.

\section{Results and discussion}

In the following, we discuss the structure and the nature of the ISM in each star-forming region from the line and continuum emission maps obtained by the present observations as well as radio and MIR continuum data.  For the radio continuum map, we use NRAO VLA Sky Survey (NVSS) at 1.4~GHz \citep{Condon98} if available, otherwise the 843~MHz continuum by the Sydney University Molonglo Sky Survey \citep[SUMSS;][]{Bock99} is employed.  The resolutions of these surveys are $45^{\prime\prime}$ and $43^{\prime\prime}$, respectively, which are similar to those of the FIS-FTS.  For the MIR continuum maps, we use 9\micron\ and 18\micron\ data obtained by the All-Sky Survey of the Infrared Camera \citep[IRC;][]{Onaka07,Ishihara06,Ishihara07,Ishihara08} onboard AKARI.  When either of the IRC bands data is saturated, we use the MSX A-band and E-band instead.  The resolution of the IRC All-Sky Survey and MSX is $9.4^{\prime\prime}$ and $18.3^{\prime\prime}$ \citep{Egan99}, respectively.

\subsection{G3.270-0.101}

\begin{figure*}
\centering
\resizebox{0.45\linewidth}{!}{\includegraphics{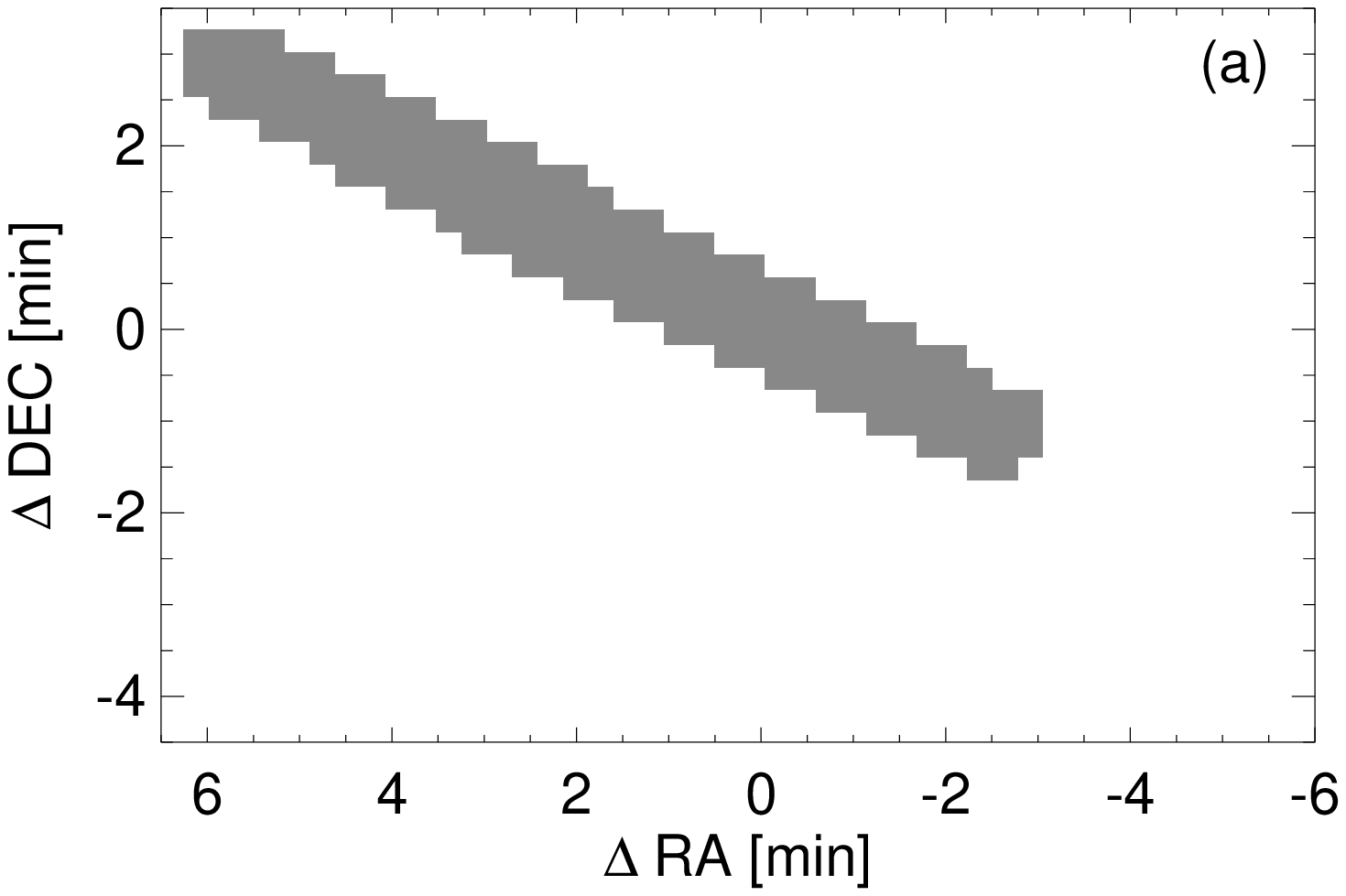}}
\resizebox{0.45\linewidth}{!}{\includegraphics{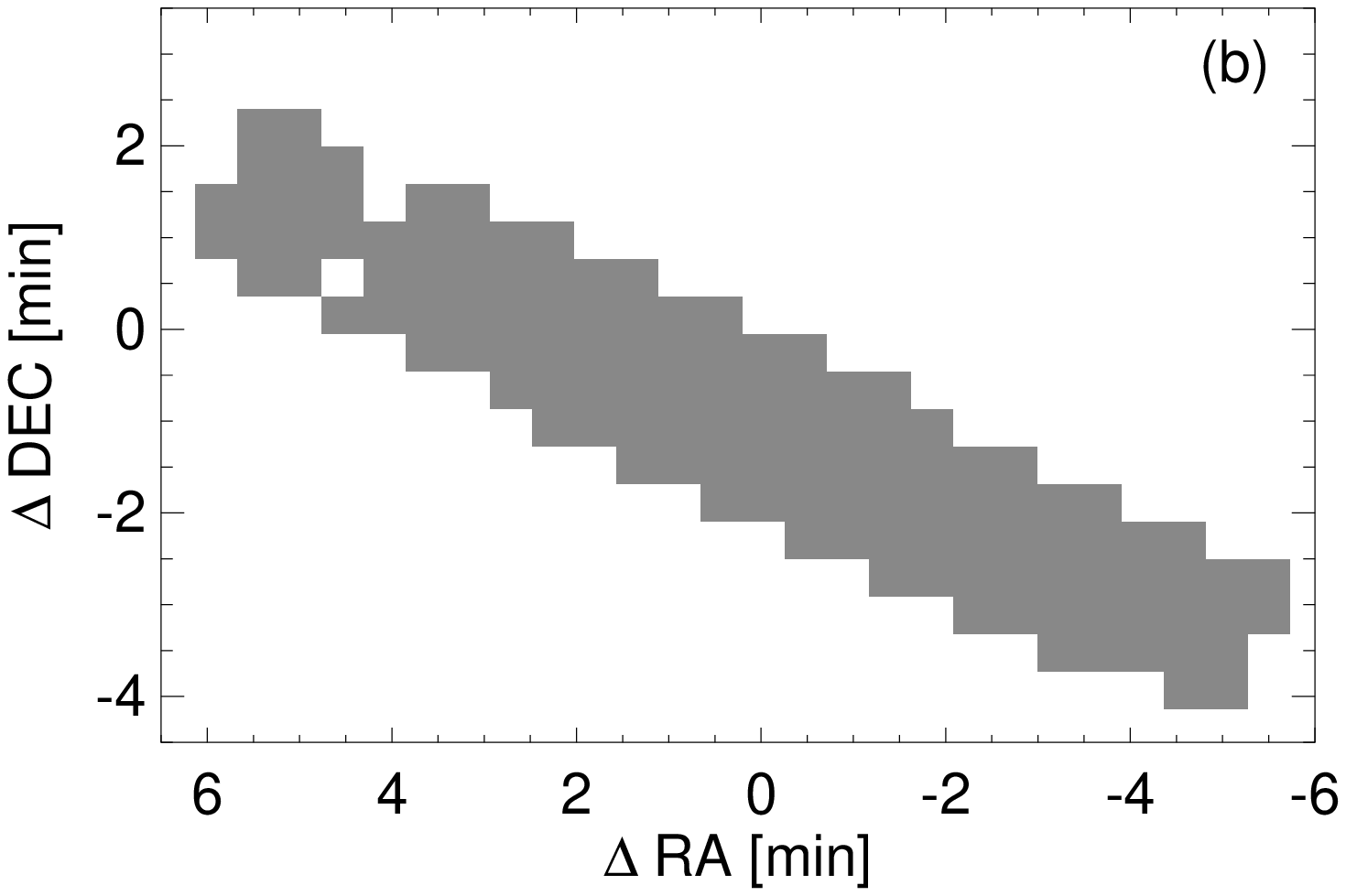}}
\resizebox{0.45\linewidth}{!}{\includegraphics{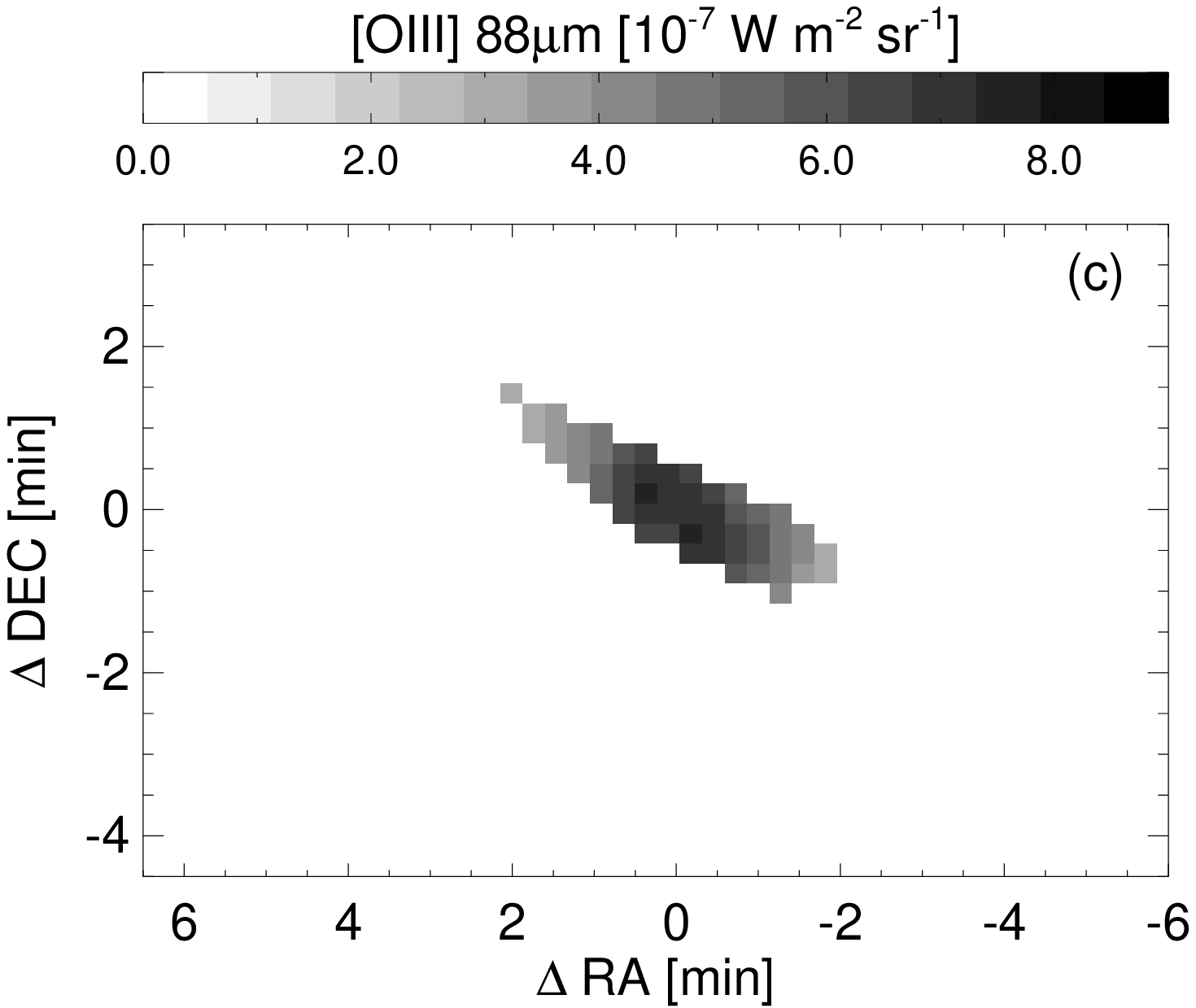}}
\resizebox{0.45\linewidth}{!}{\includegraphics{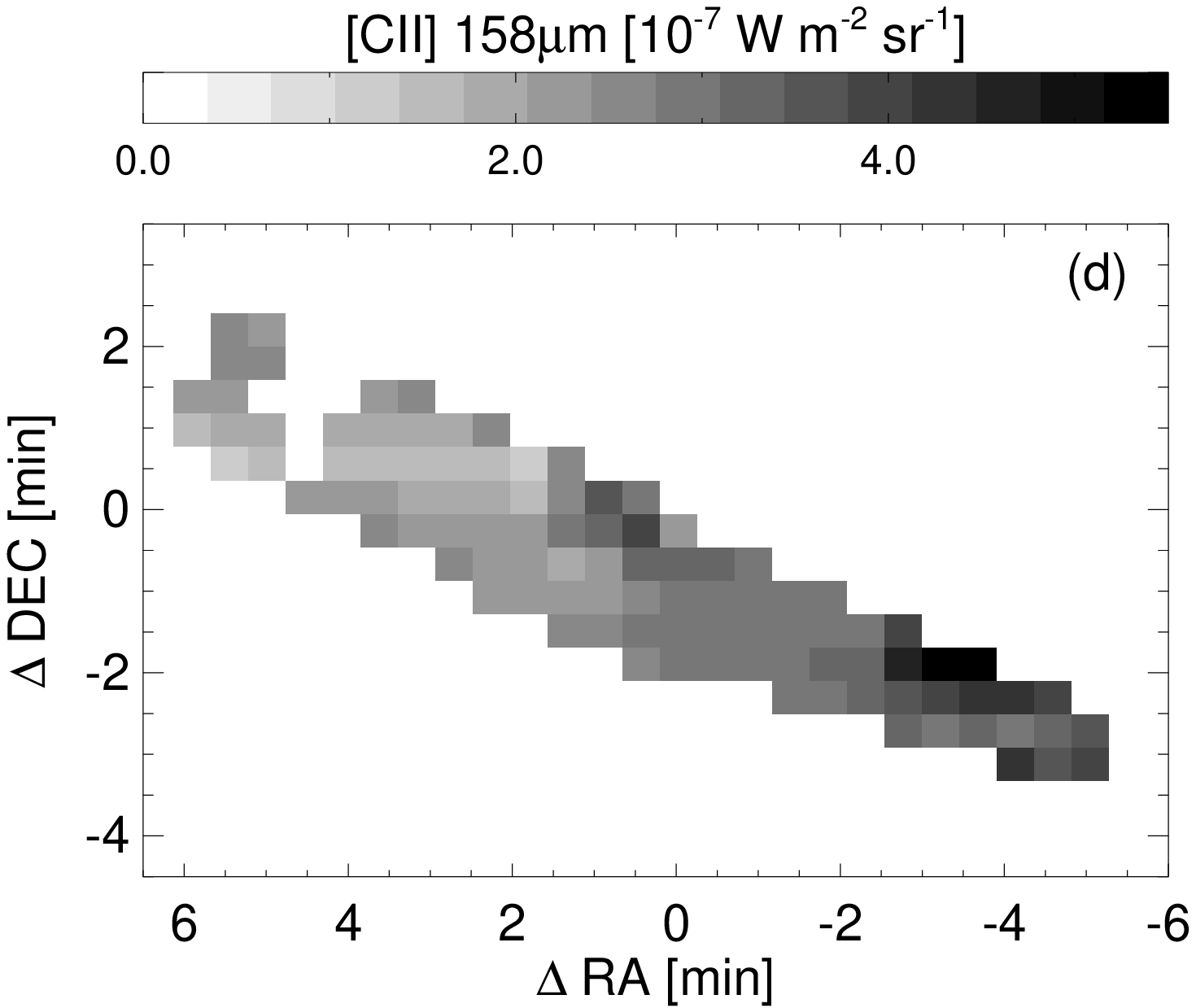}}
\resizebox{0.45\linewidth}{!}{\includegraphics{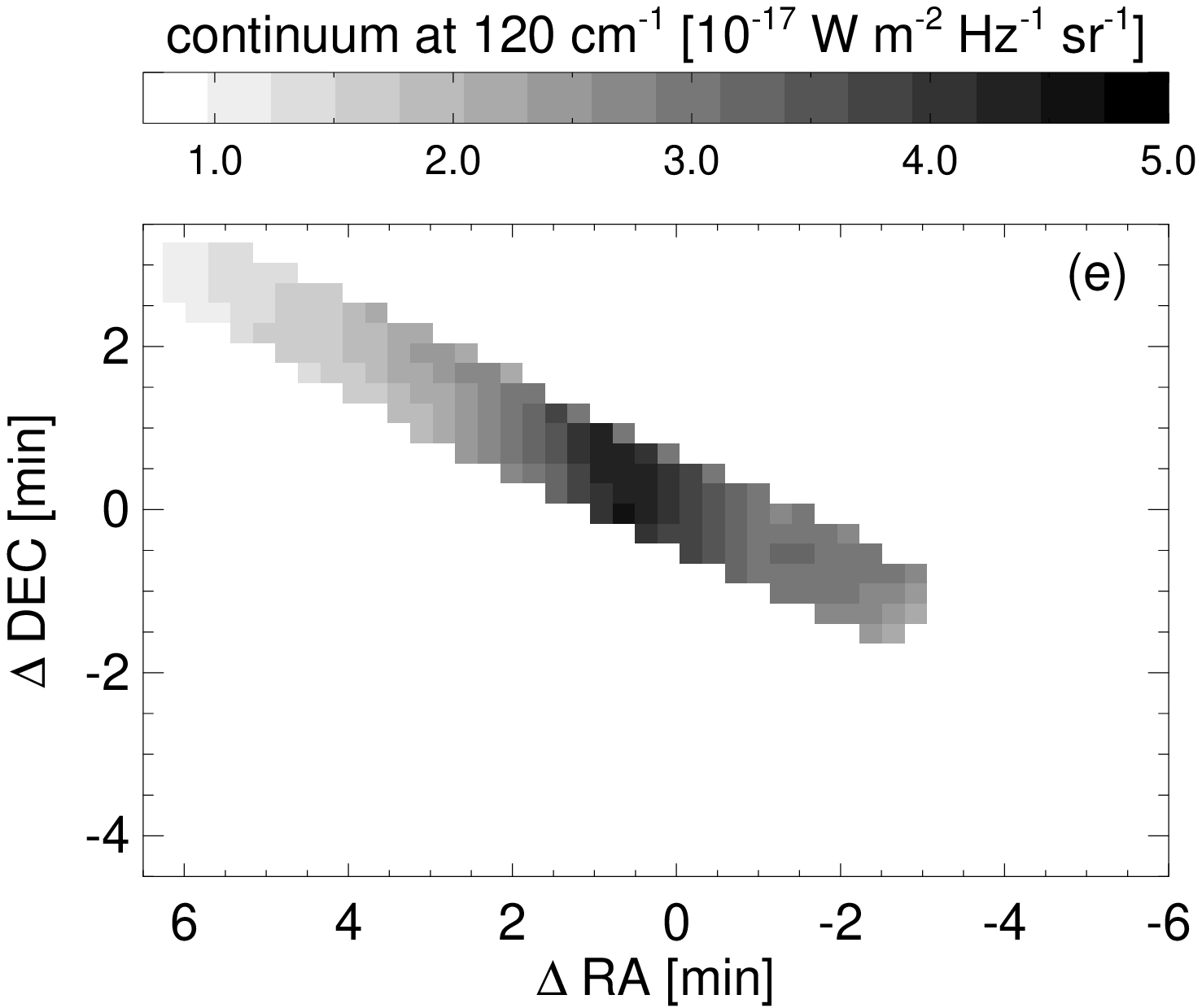}}
\resizebox{0.45\linewidth}{!}{\includegraphics{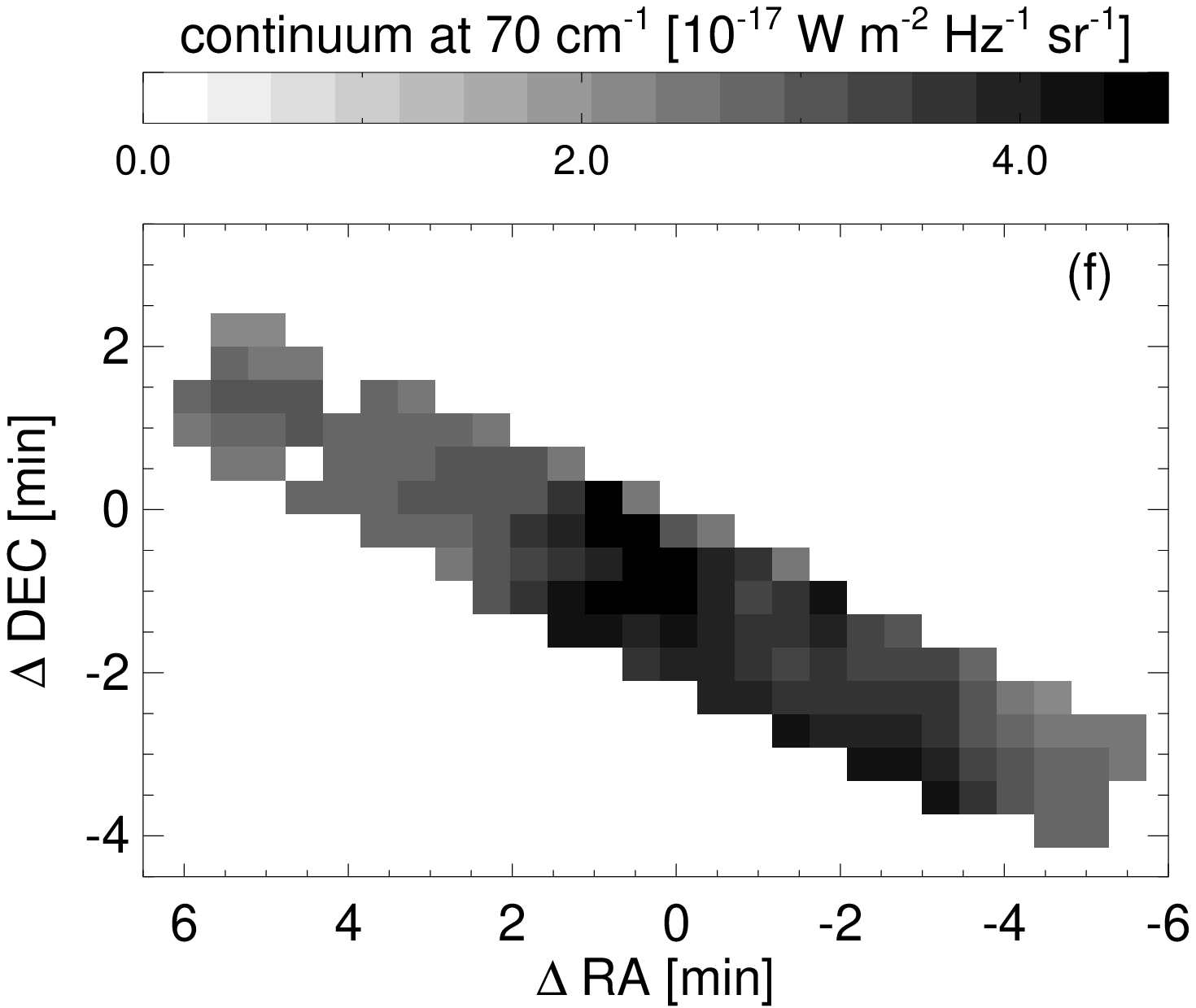}}
\resizebox{0.45\linewidth}{!}{\includegraphics{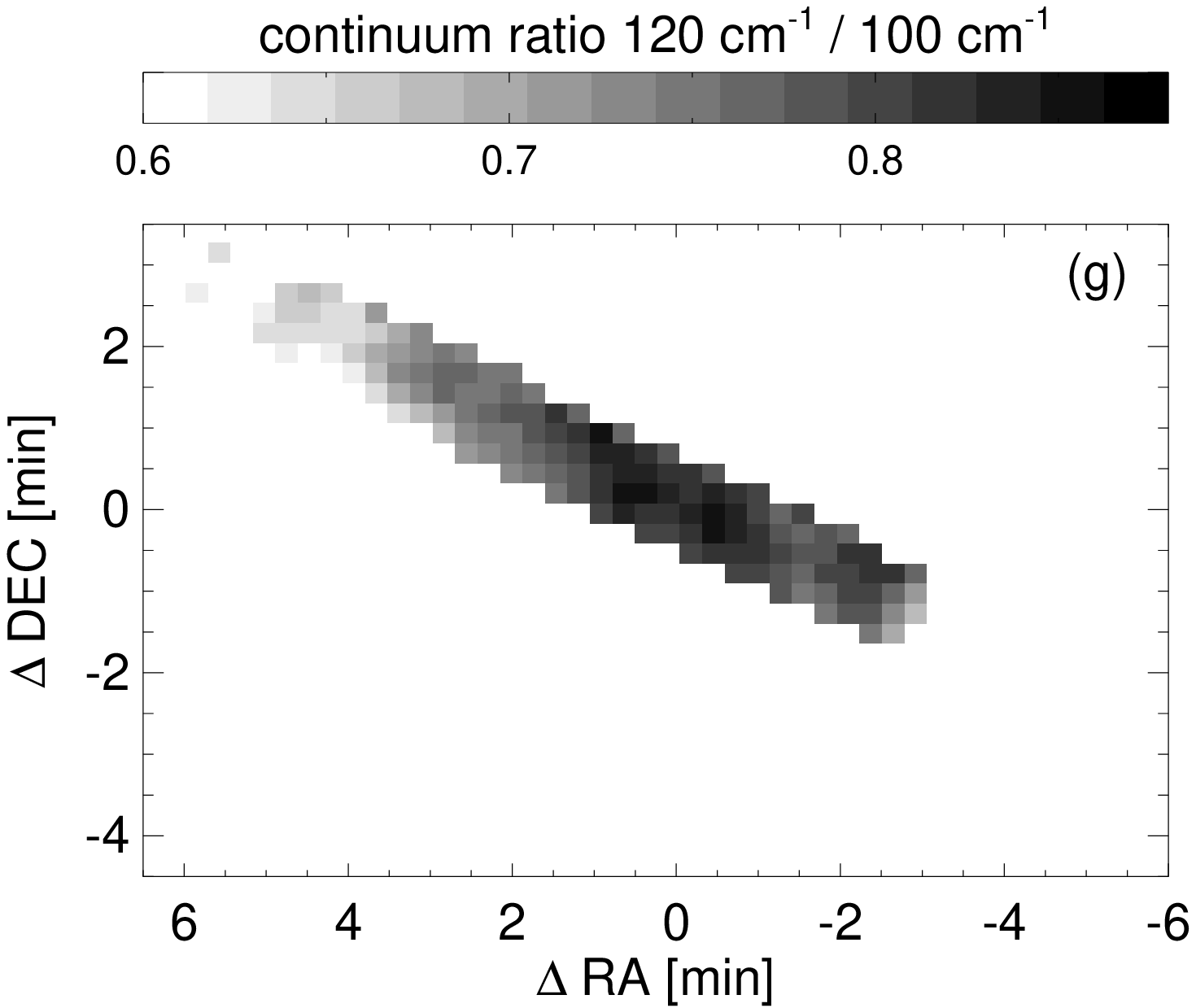}}
\caption{(a) The observed (gray) area of SW and (b) that of LW, (c) \oiii\ 88\micron, and (d) \cii\ 158\micron\ intensity map and continuum maps at (e) 120\pcm\ and (f) 70\pcm\ and (g) the continuum ratio of 120\pcm/100\pcm\ in G3.270-0.101.  The origin of the coordinates is RA = $17^\mathrm{h}53^\mathrm{m}35^\mathrm{s}.8$ and Dec = $-26^{\circ}10^{\prime}59^{\prime\prime}$ (J2000)\citep{Lockman89}.}\label{linemap_G3}
\end{figure*}

\begin{figure*}
\centering
\resizebox{0.45\linewidth}{!}{\includegraphics{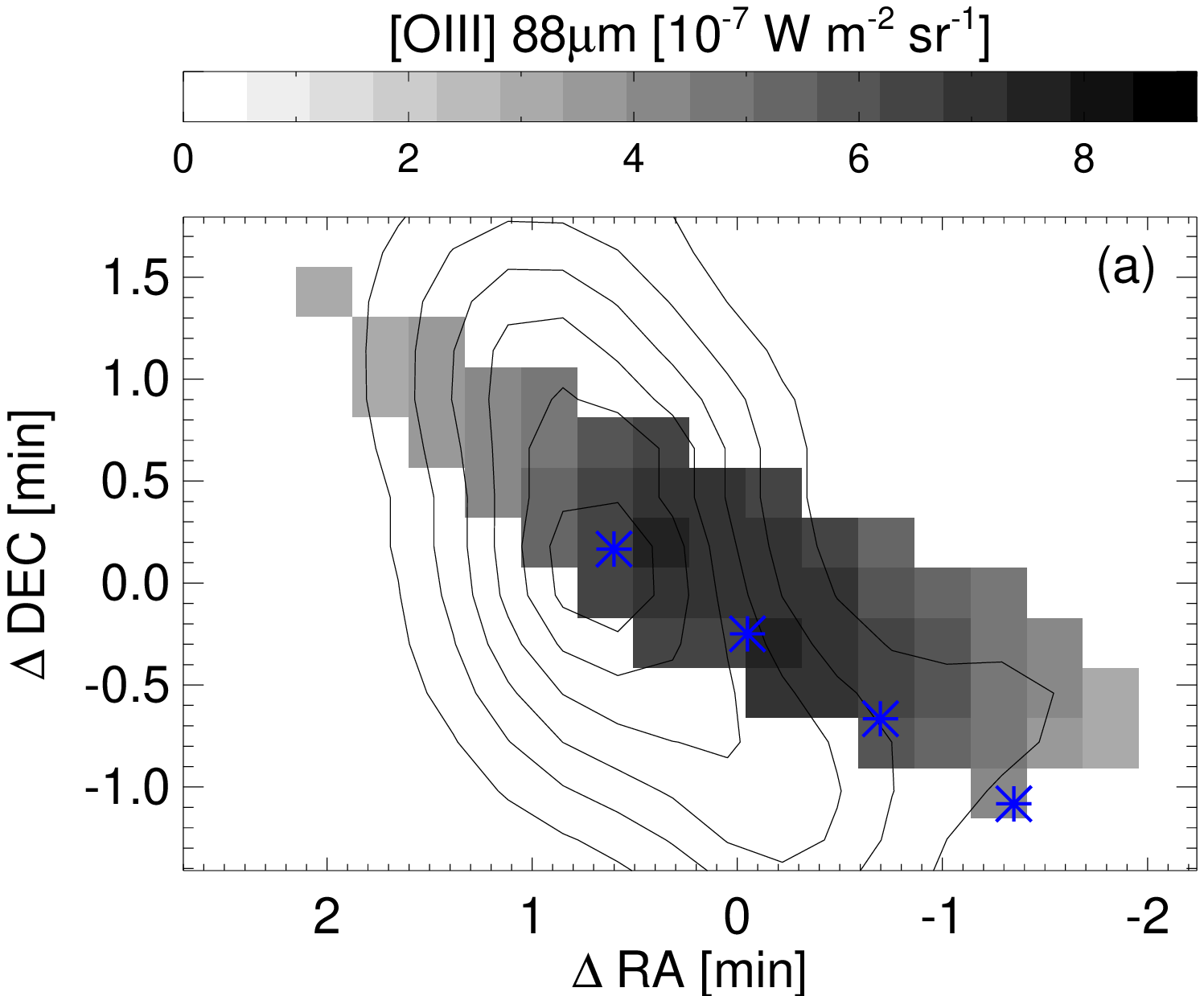}}
\resizebox{0.45\linewidth}{!}{\includegraphics{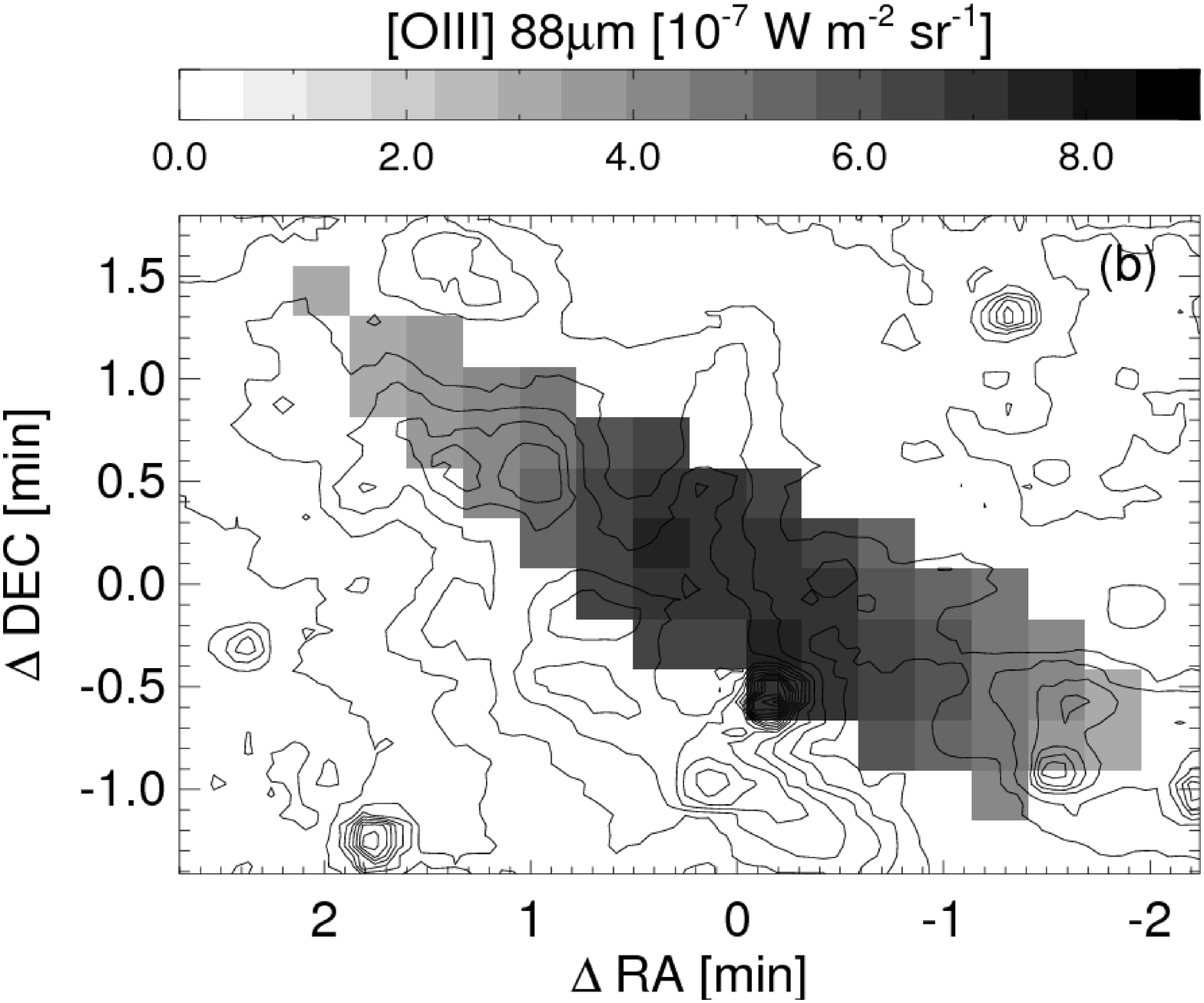}}
\resizebox{0.45\linewidth}{!}{\includegraphics{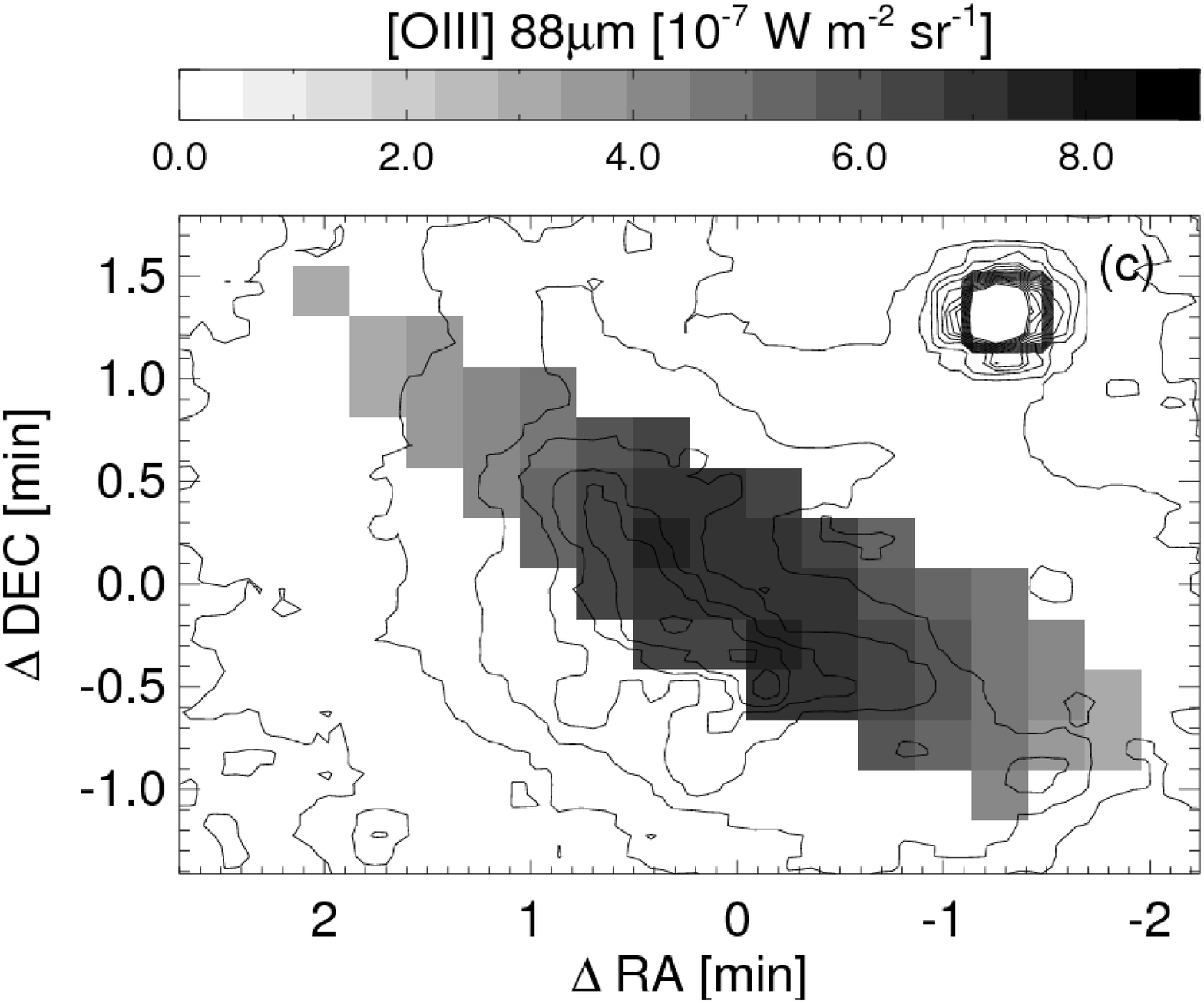}}
\resizebox{0.45\linewidth}{!}{\includegraphics{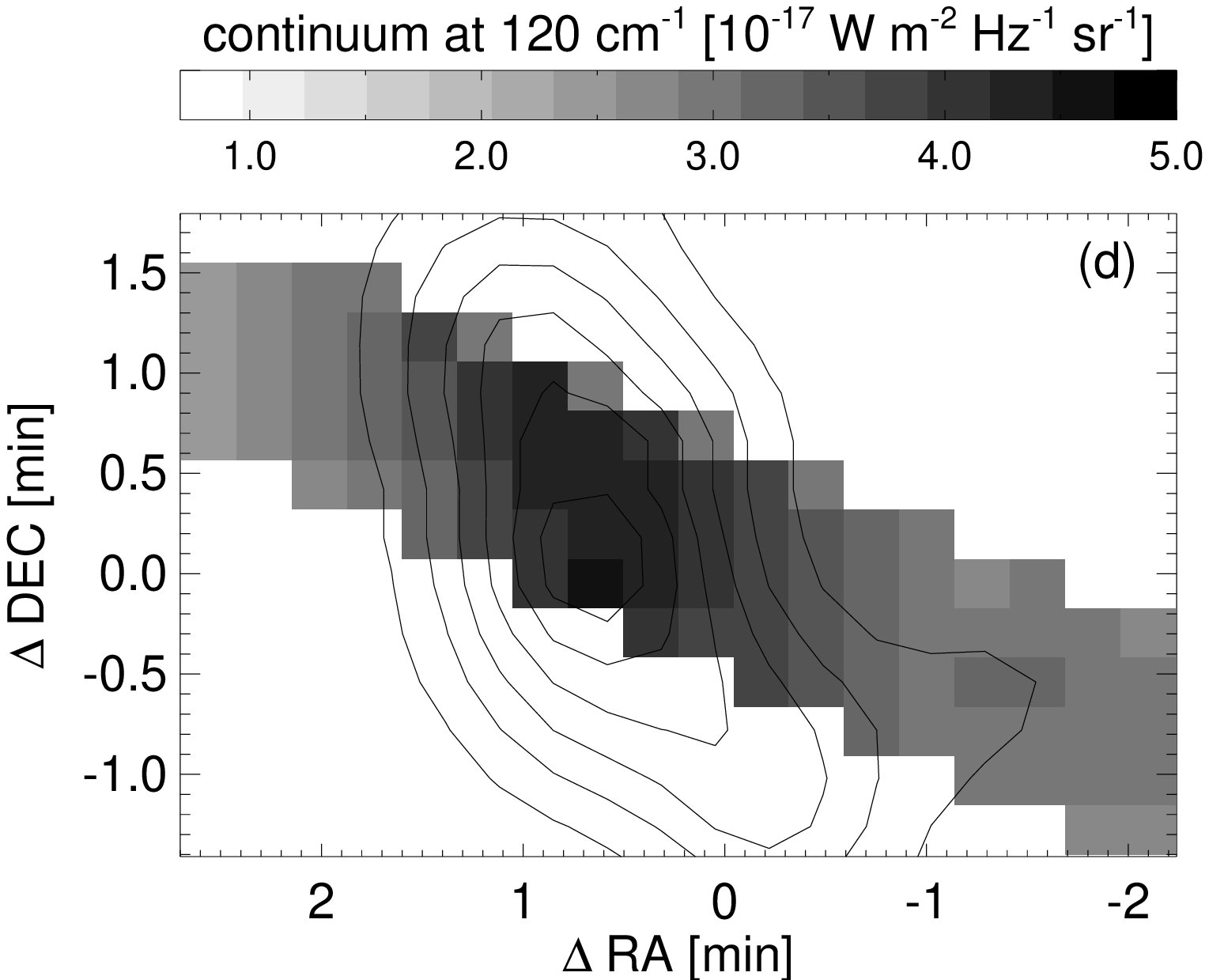}}
\resizebox{0.45\linewidth}{!}{\includegraphics{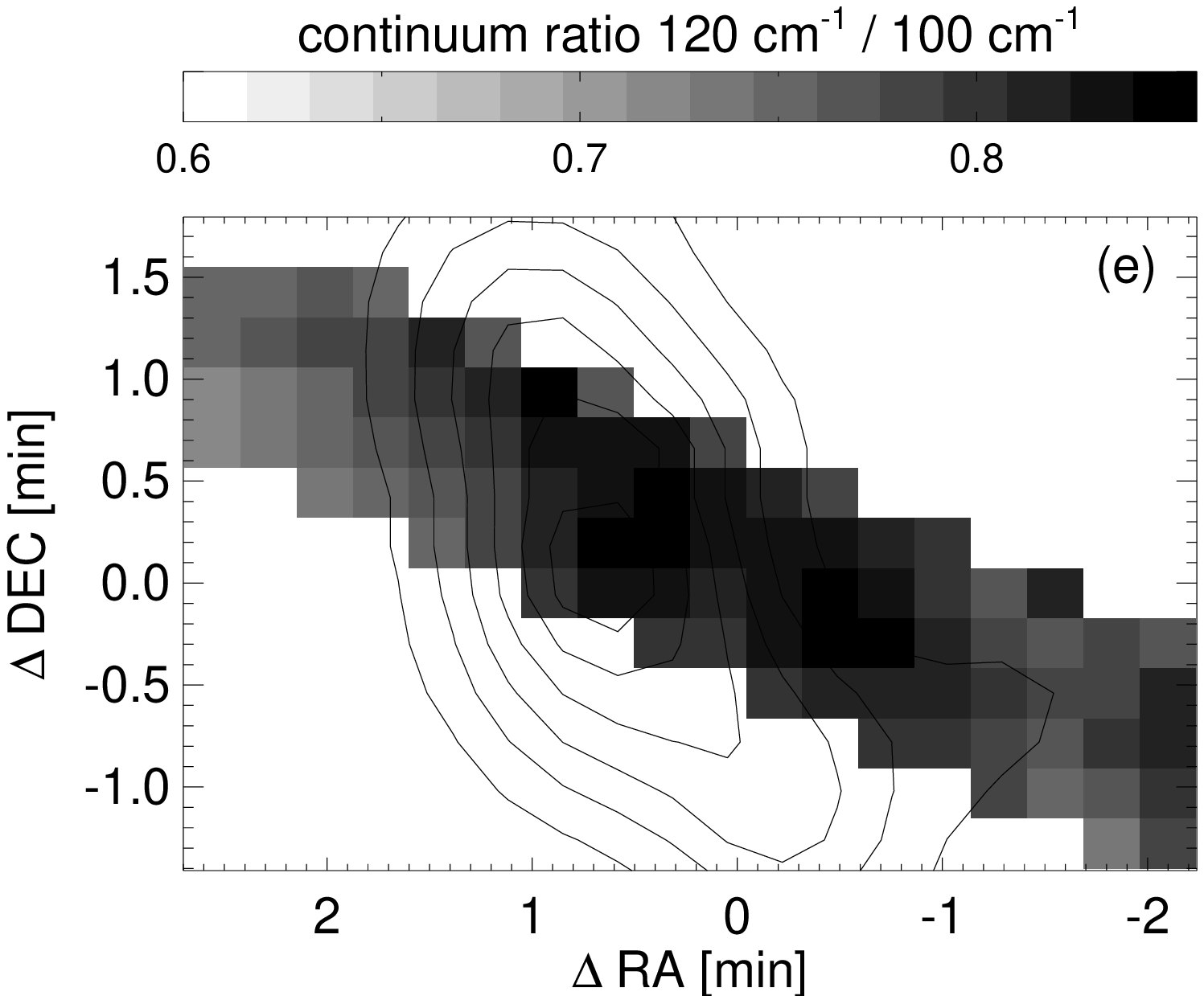}}
\caption{The gray scale shows the \oiii\ 88\micron\ intensity ((a)--(c)), the continuum emission of 120\pcm\ (d), and the continuum ratio of 120\pcm/100\pcm\ (e) for G3.270-0.101.  In (a) the NVSS 1.4~GHz intensity, in (b) the IRC 9\micron\ intensity, and in (c) IRC 18 \micron\ intensity are overlaid in contours, respectively.  In (d) and (e) the contours indicate the NVSS 1.4~GHz intensity.  The contours of the NVSS are drawn from 0.02 Jy\ beam$^{-1}$ in a 0.02 Jy\ beam$^{-1}$ interval, and those of the IRC 9\micron\ and 18\micron\ maps are on a linear scale of arbitrary units.  Asterisks in (a) show the positions observed by the {\it Spitzer}/IRS \citep{Okada08}.  The origin of the coordinates is the same as Fig.~\ref{linemap_G3}.}\label{G3_OIII_w_contours}
\end{figure*}

\begin{figure}
\centering
\resizebox{\linewidth}{!}{\includegraphics{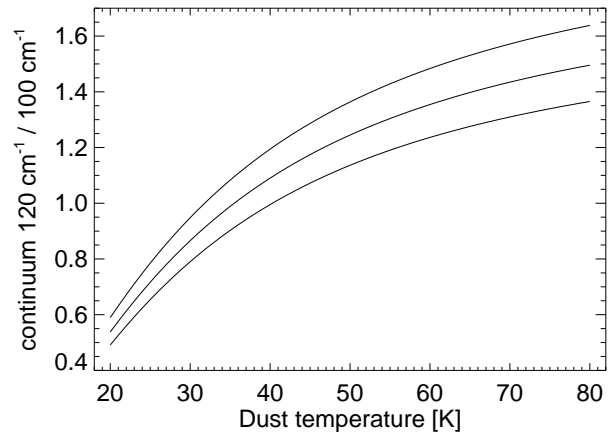}}
\caption{The calculated ratio of 120\pcm/100\pcm\ versus the dust temperature with the dust emissivity index of $-1$, $-1.5$, and $-2$ from the lower to upper curves.}\label{calc_120div100}
\end{figure}

\begin{figure}
\centering
\resizebox{\linewidth}{!}{\includegraphics{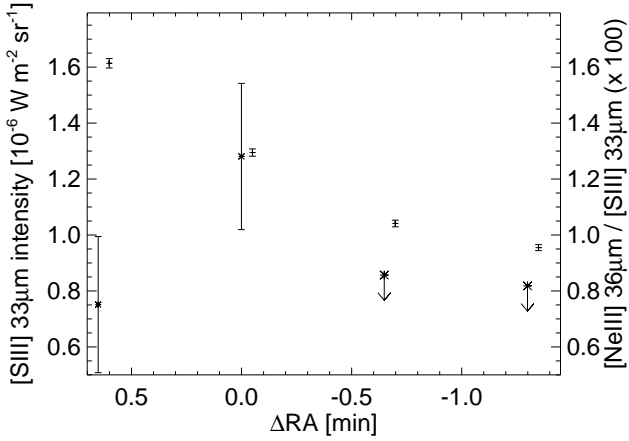}}
\caption{The \suiii\ 33\micron\ intensity (plus points) and the intensity ratio of \neiii\ 36\micron/\suiii\ 33\micron\ (asterisk points) in G3.270-0.101 observed by the {\it Spitzer}/IRS.  The latter is shifted slightly to the left for clarity.  These four positions are shown in Fig.~\ref{G3_OIII_w_contours}a.}\label{G3_SSTline}
\end{figure}

\begin{figure}
\centering
\resizebox{\linewidth}{!}{\includegraphics{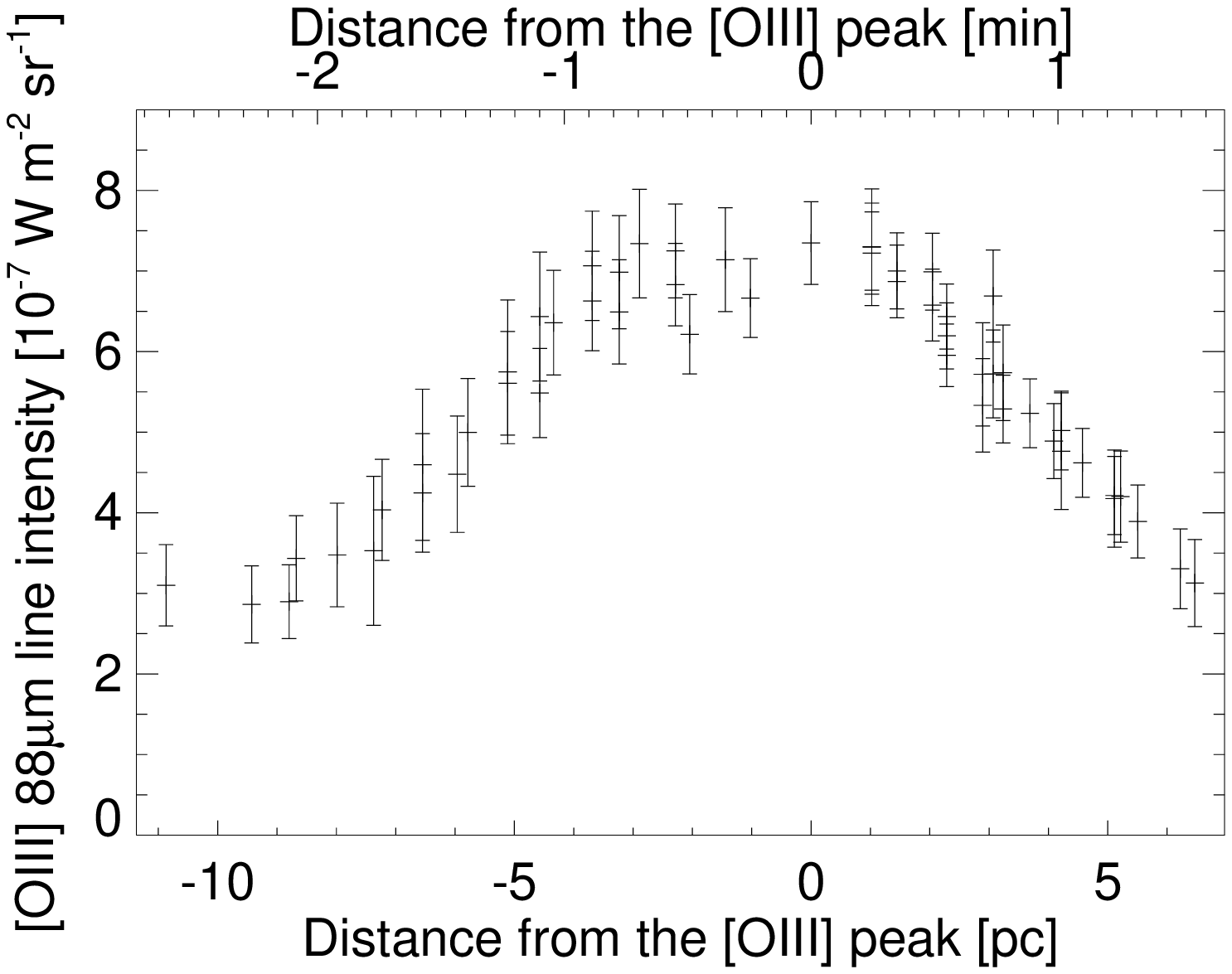}}
\caption{The spatial distribution of the \oiii\ 88\micron\ emission in G3.270-0.101.  All the data of grids in Fig.~\ref{linemap_G3}c are plotted, with the origin being the center of the brightest grids.}\label{distribution_G3}
\end{figure}

We present the highest spatial resolution maps \oiii\ 88\micron\ and \cii\ 158\micron\ line emission (Figs.~\ref{linemap_G3}c,d) and continuum (Figs.~\ref{linemap_G3}e,f) obtained for this region to date.  Figures~\ref{G3_OIII_w_contours}a--e show the correlation of both \oiii\ 88\micron\ and the FIR continuum emissions with the radio continuum from the NVSS at 1.4~GHz, and 9\micron\ and 18\micron\ emission detected by the IRC all-sky survey.  The peak of the radio continuum is located at the peak of the FIR continuum emission (Fig.~\ref{G3_OIII_w_contours}d), but it has an offset from that of the \oiii\ 88\micron\ (Fig.~\ref{G3_OIII_w_contours}a).  This offset is larger than the uncertainty of the position, which is smaller than the grid size in the figures.  Since it takes 35.12~eV to form O$^{2+}$, the \oiii\ 88\micron\ line emission probably peaks close to the ionization source.  However, the free-free continuum is only sensitive to the emission measure, not the ionization structure of the \hii\ region.  Therefore, it is likely that the ionization source is located at the \oiii\ peak, and that the radio continuum (free-free) peak is just a local peak in the emission measure, not necessarily the location of an ionization source.  The \oiii\ 88\micron\ emission also shows a peak close to the radio peak, which indicates that some O$^{2+}$ ions still exist, and the large emission measure compensates for the decrease in the number of O$^{2+}$ ions, although we cannot exclude another ionization source existing at the radio peak.  This picture is supported by several other observational results.  First, the 18\micron\ continuum emission, which is understood to trace the hot dust distribution \citep{Ishihara07}, has a broad peak including the radio continuum peak, but also extends toward the south-west (Fig.~\ref{G3_OIII_w_contours}c).  Second, Fig.~\ref{G3_OIII_w_contours}e shows the continuum ratio of 120\pcm/100\pcm, which is a good tracer of a dust temperature in the range 20--50~K (Fig.~\ref{calc_120div100}).  For example, for a ratio of 0.8, the dust temperature ($T_\mathrm{d}$) is 30~K, and for a ratio of 1.0, $T_\mathrm{d}\sim 40$~K with the emissivity index of $-1$.  The high temperature (high ratio) region is as widely distributed as the \oiii\ and the radio continuum emission.  Observations of MIR emission lines using the {\it Spitzer}/IRS have also shown a spatial distribution that supports this hypothesis.  Figure~\ref{G3_SSTline} shows the intensity of the \suiii\ 33\micron\ line and the intensity ratio of \neiii\ 36\micron/\suiii\ 33\micron\ at 4 observed positions indicated by asterisks in Fig.~\ref{G3_OIII_w_contours}a.  The line intensities are averages of the dithering pairs \citep{Okada08}.  The ionization potential of S$^{2+}$ and Ne$^{2+}$ is 23.33~eV and 40.96~eV, respectively.  Thus, the intensity ratio of \neiii\ 36\micron/\suiii\ 33\micron\ indicates the ionization degree, if a constant elemental abundance within the region is assumed, whereas the \suiii\ 33\micron\ intensity depends on the electron density and/or the gas column density as the radio continuum emission does under optically thin conditions.  Figure~\ref{G3_SSTline} shows that the peak of \suiii\ 33\micron\ is located at the peak of the radio continuum, and the peak of the ratio \neiii\ 36\micron/\suiii\ 33\micron\ is similar to that of \oiii\ 88\micron.  

In summary, the peak of the radio continuum is likely to be illuminated externally by the exciting source, and the position of the exciting source is probably at the peak of the \oiii\ 88\micron\ line intensity map.

In Fig.~\ref{distribution_G3}, the spatial distribution of the \oiii\ 88\micron\ emission is shown, according to which the excitation energy is investigated, assuming a distance is 14.3~kpc \citep{Conti04}.  The Lyman photon flux of $3\times 10^{50}$\,s$^{-1}$ \citep{Conti04} is indicative of 23 O7-type stars or their equivalent.  Assuming an excitation source of 23 O7-type stars, an electron density of $31$\pcm\ \citep{Conti04}, an oxygen abundance of $4.57\times 10^{-4}$ \citep{Asplund05}, and the photoionization cross-section and the recombination coefficient in \citet{Osterbrock89}, we derive the O$^{++}$ Str\"{o}mgren radius to be 4.6~pc, which is compatible with the observed distribution of \oiii\ 88\micron\ emission (Fig.~\ref{distribution_G3}).  As a check, we also compiled a model using a Salpeter initial mass function that matches the total Lyman continuum flux, and found that the O$^{++}$ Str\"{o}mgren sphere radius would then be about 4.7~pc, very close to our value derived above.

\subsection{G333.6-0.2}

\begin{figure*}
\centering
\resizebox{0.45\linewidth}{!}{\includegraphics{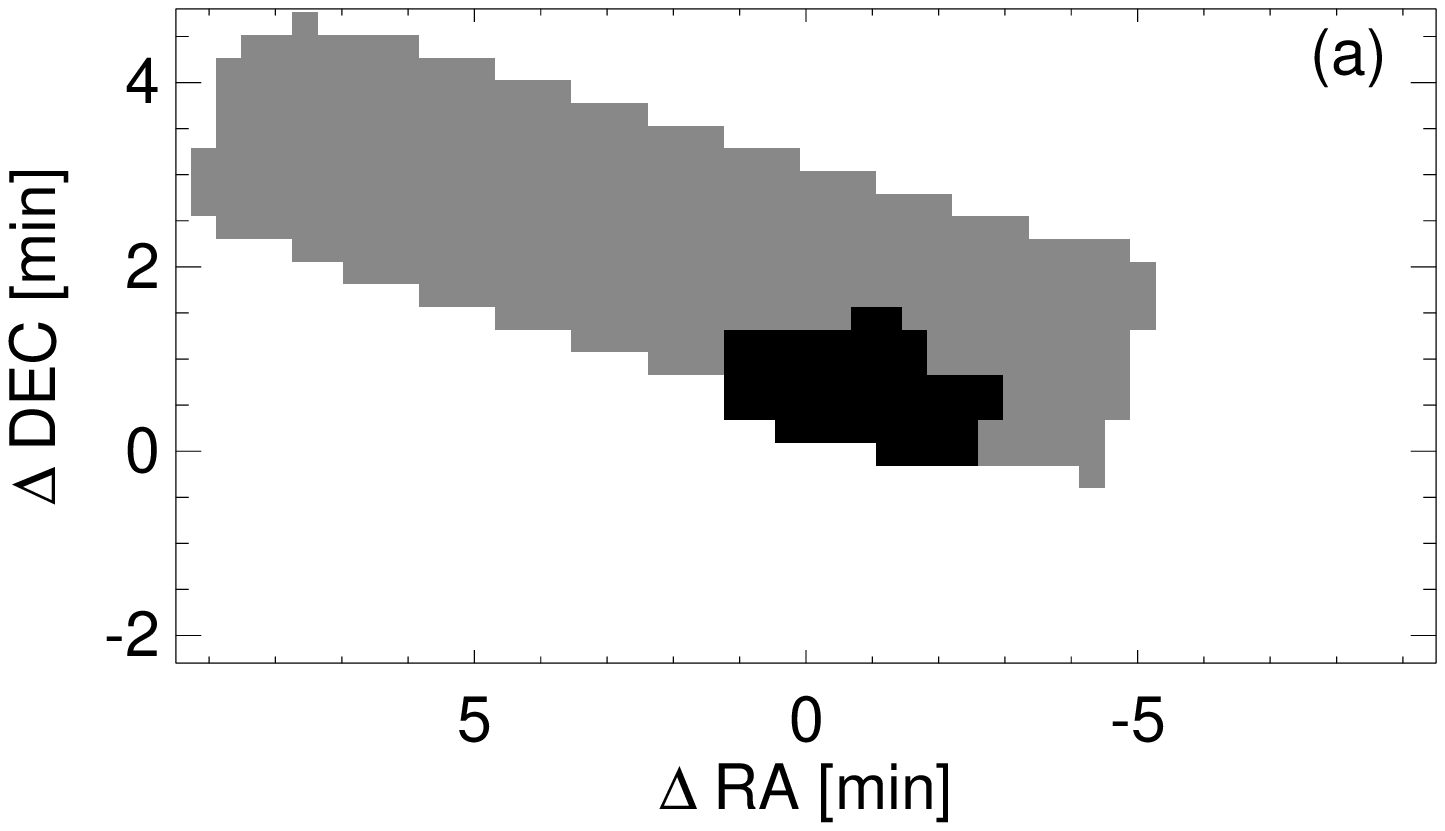}}
\resizebox{0.45\linewidth}{!}{\includegraphics{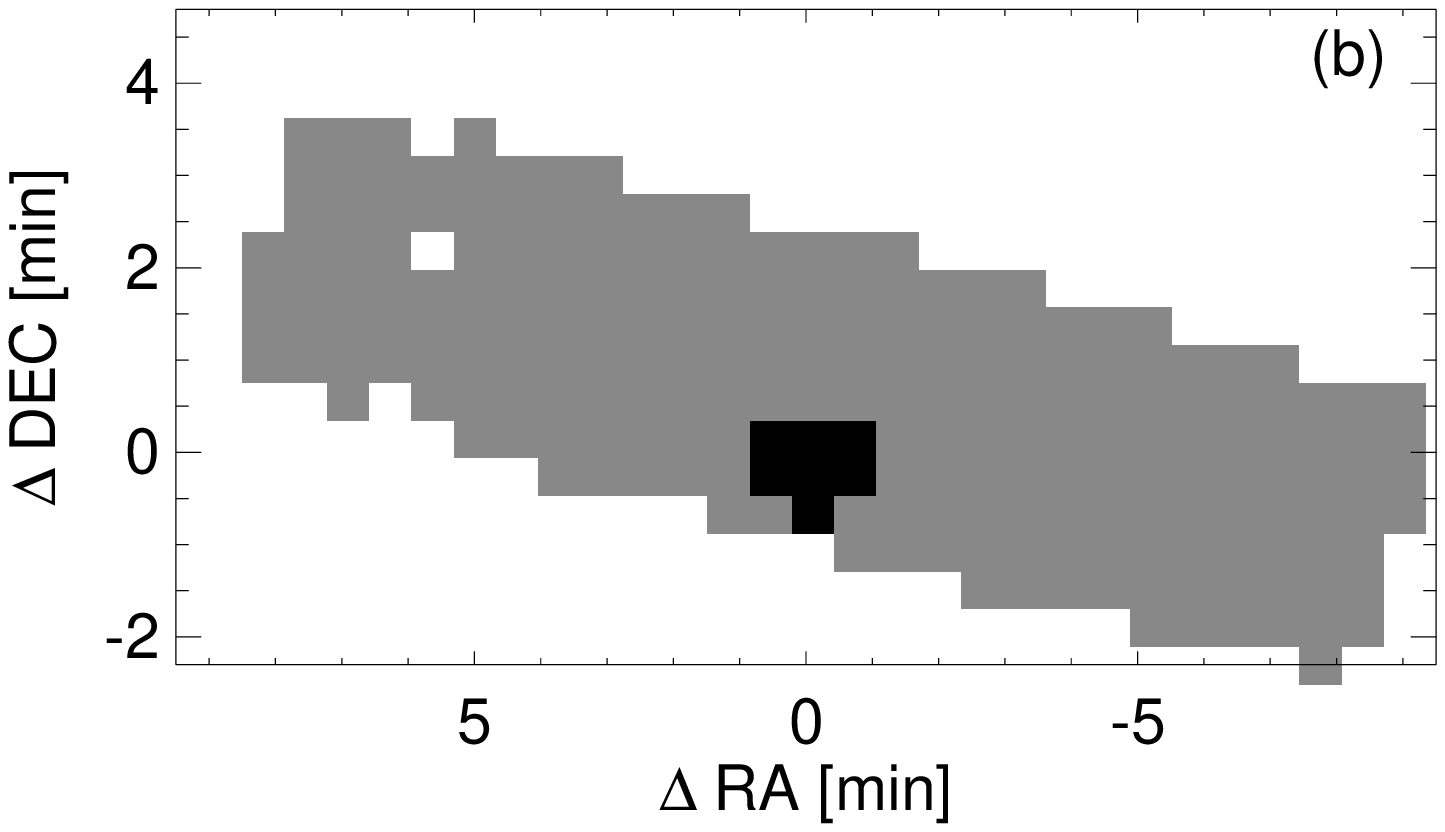}}
\resizebox{0.45\linewidth}{!}{\includegraphics{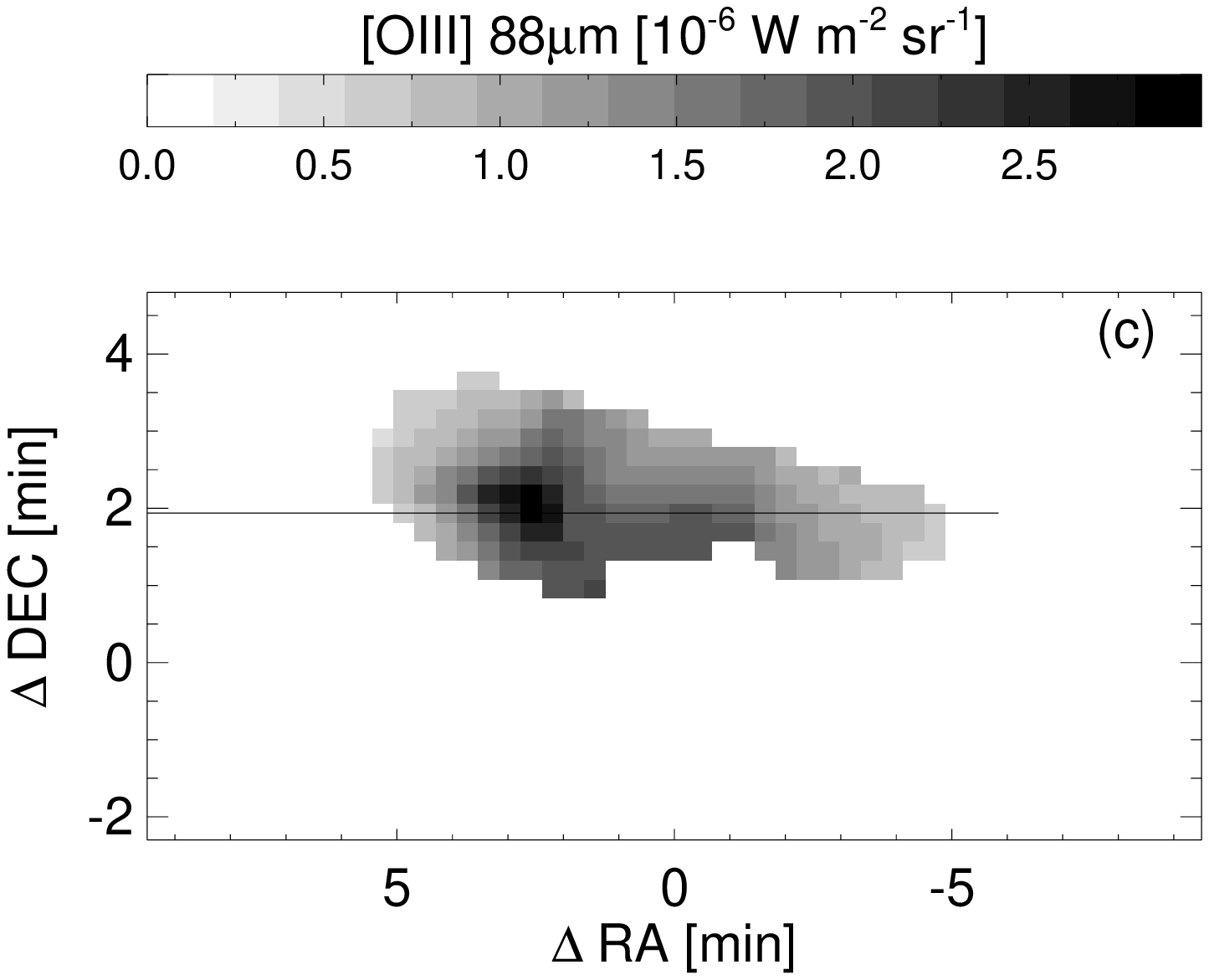}}
\resizebox{0.45\linewidth}{!}{\includegraphics{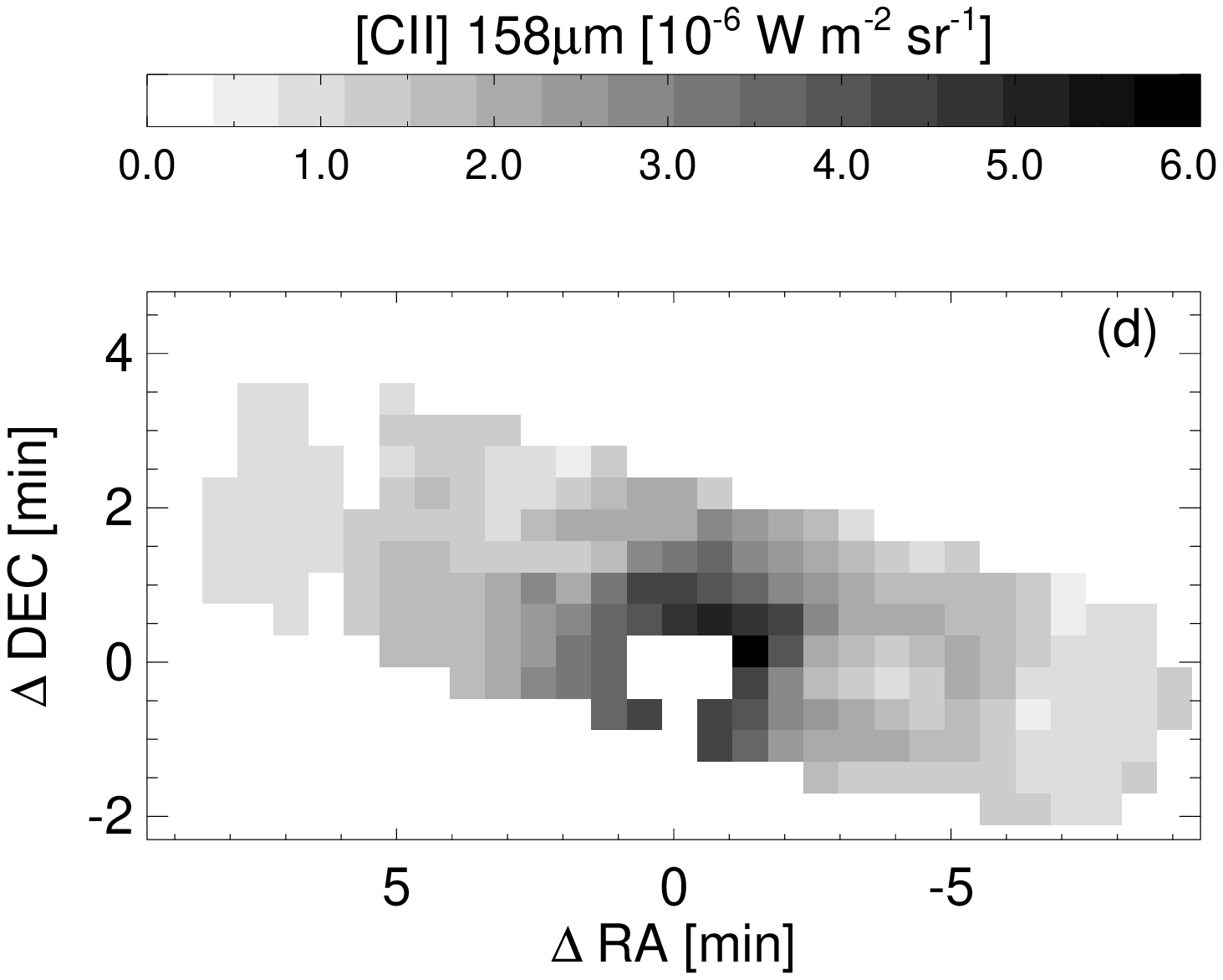}}
\resizebox{0.45\linewidth}{!}{\includegraphics{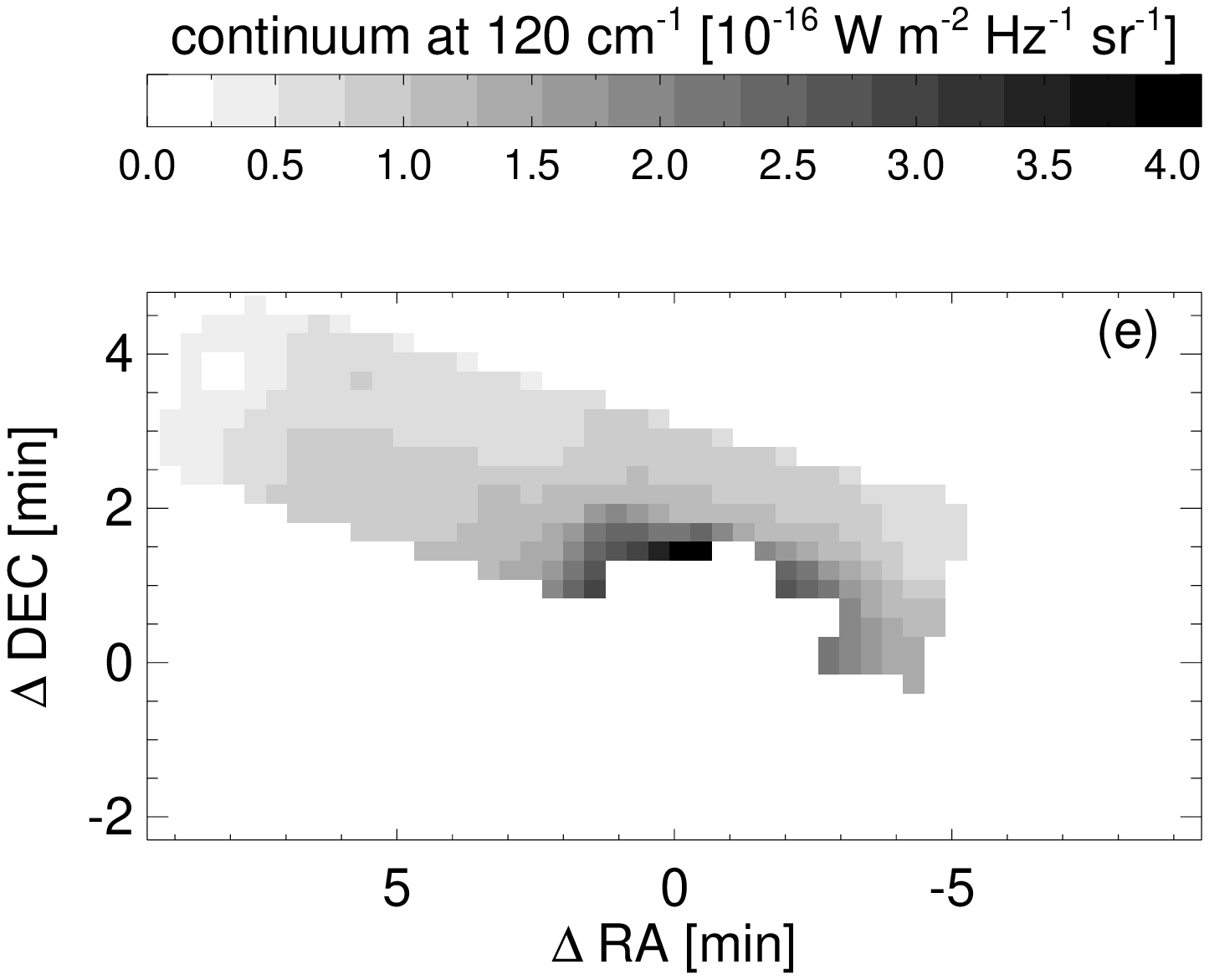}}
\resizebox{0.45\linewidth}{!}{\includegraphics{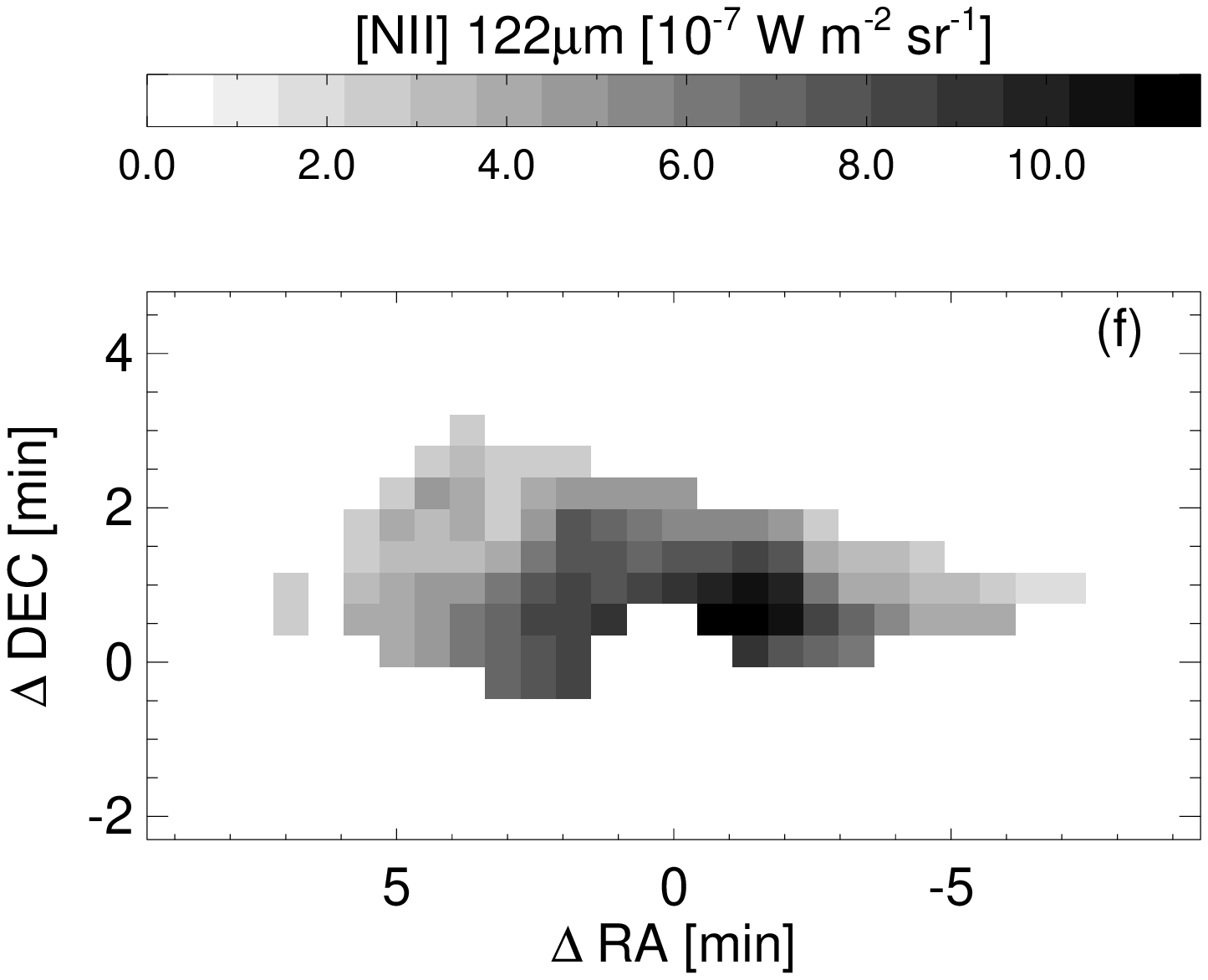}}
\resizebox{0.45\linewidth}{!}{\includegraphics{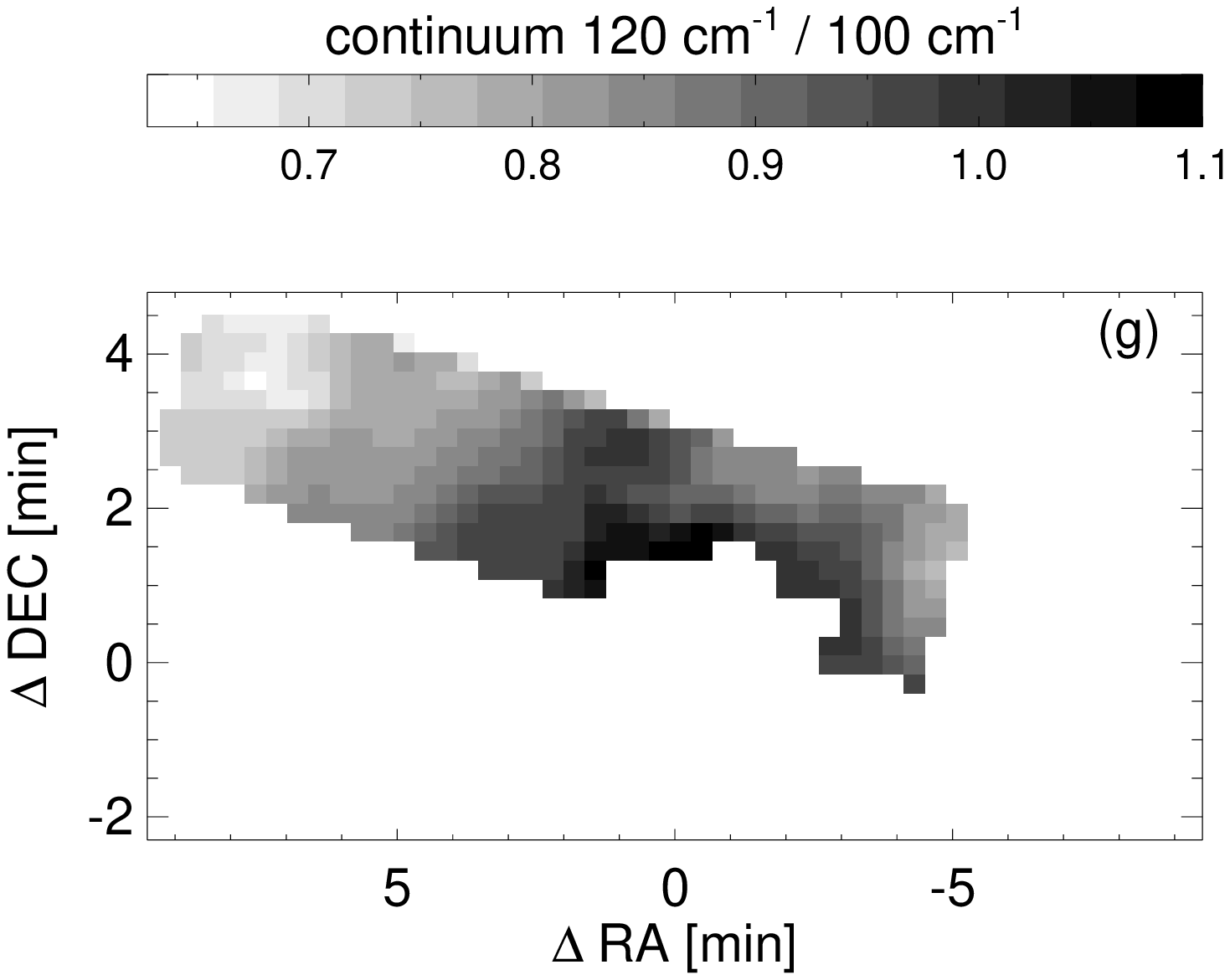}}
\resizebox{0.45\linewidth}{!}{\includegraphics{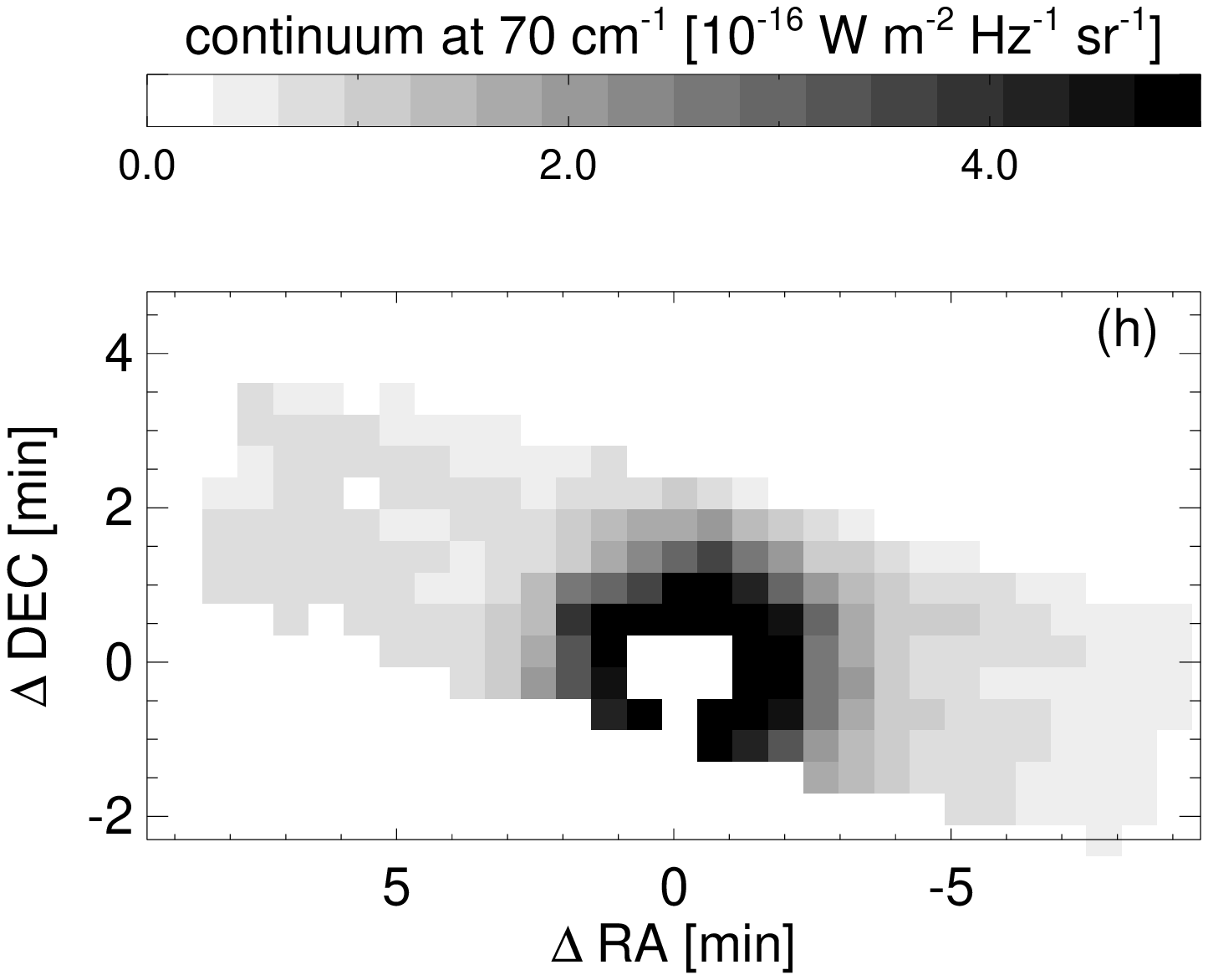}}
\caption{(a) The observed (gray) and saturated (black) area of SW and (b) that of LW, (c) \oiii\ 88\micron, (d) \cii\ 158\micron, and (f) \nii\ 122\micron\ intensity map and continuum maps at (e) 120\pcm\ and (h) 70\pcm\ and (g) the continuum ratio of 120\pcm/100\pcm\ in G333.6-0.2.  The origin of the coordinates is the radio main peak [RA = $16^\mathrm{h}22^\mathrm{m}09^\mathrm{s}.5$ and Dec = $-50^{\circ}05^{\prime}58^{\prime\prime}$ (J2000)] \citep{Fujiyoshi06}.  The solid line in (c) is a cut, along which the distribution of \oiii\ emission is shown in Fig.~\ref{G333_OIII_cut}.}\label{linemap_G333}
\end{figure*}

\begin{figure*}
\centering
\resizebox{0.45\linewidth}{!}{\includegraphics{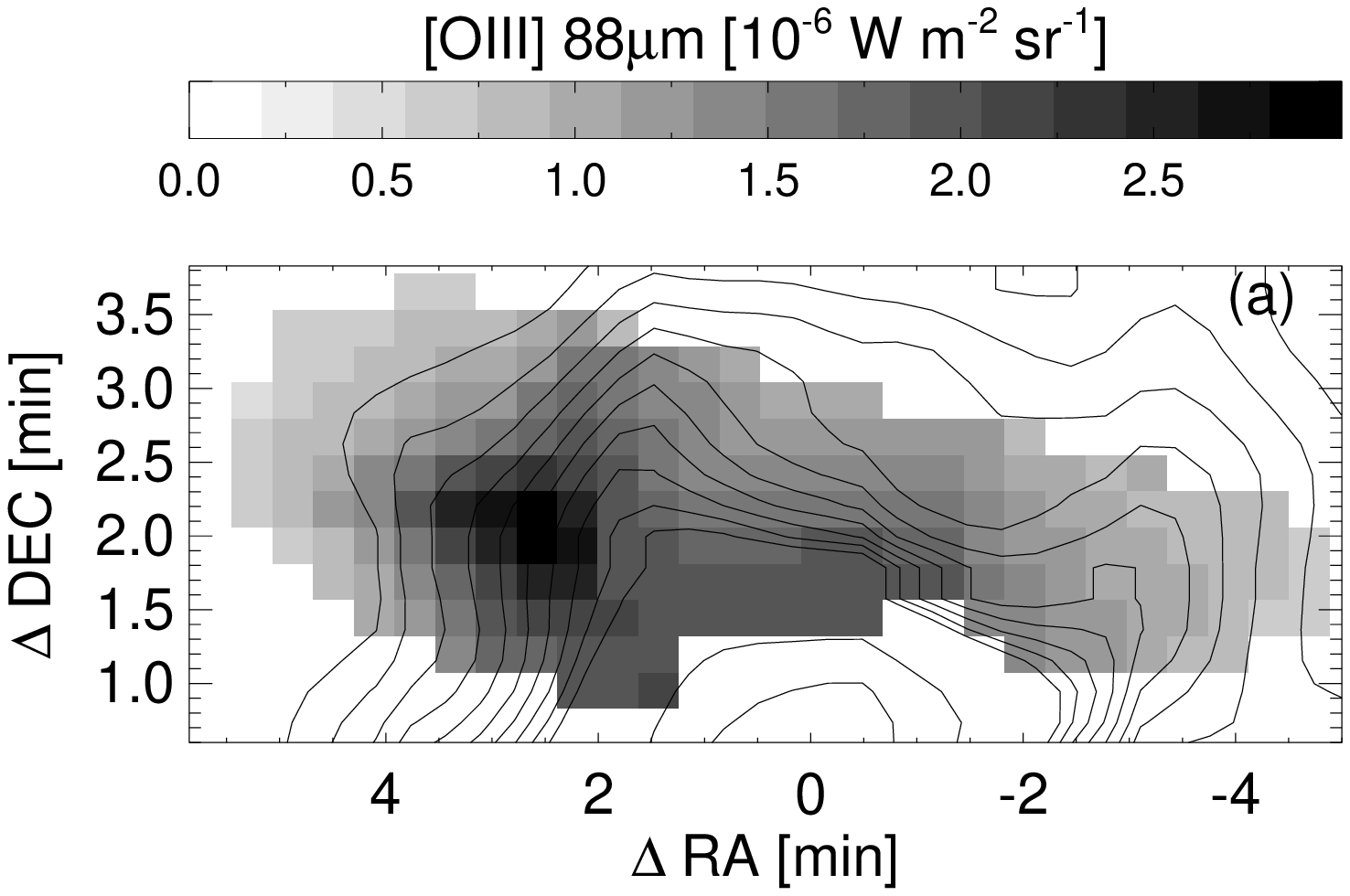}}
\resizebox{0.45\linewidth}{!}{\includegraphics{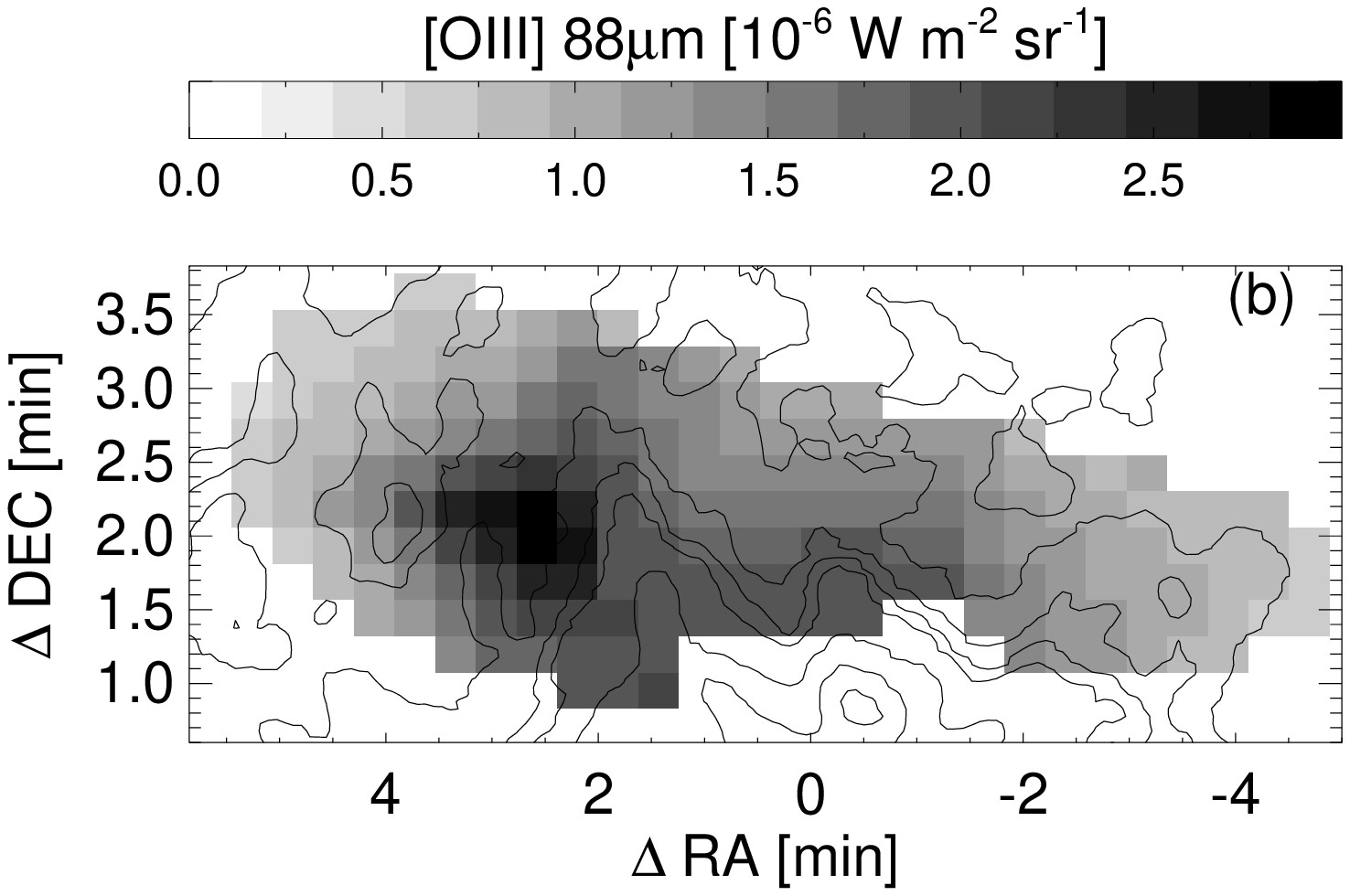}}
\resizebox{0.45\linewidth}{!}{\includegraphics{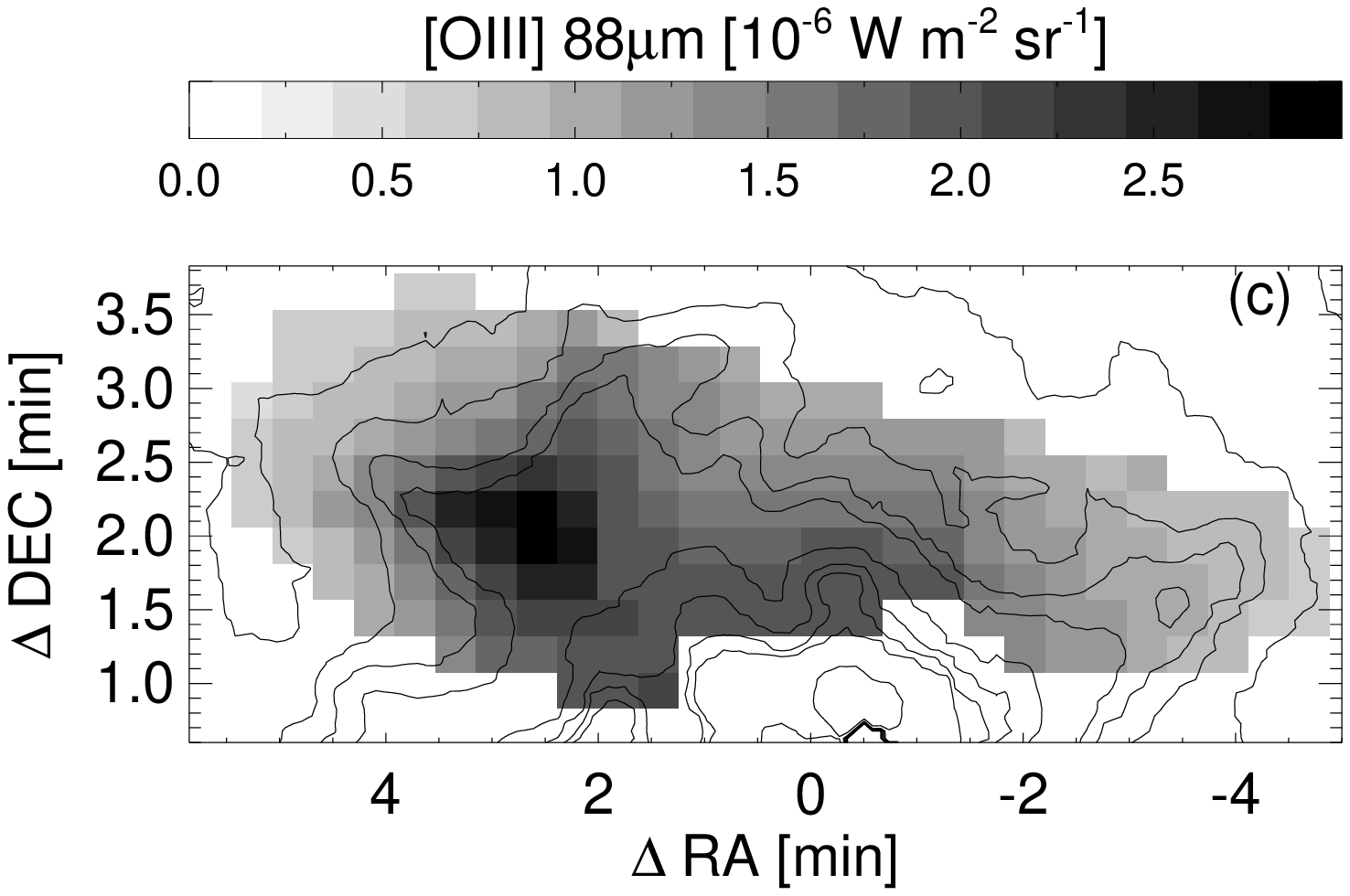}}
\resizebox{0.45\linewidth}{!}{\includegraphics{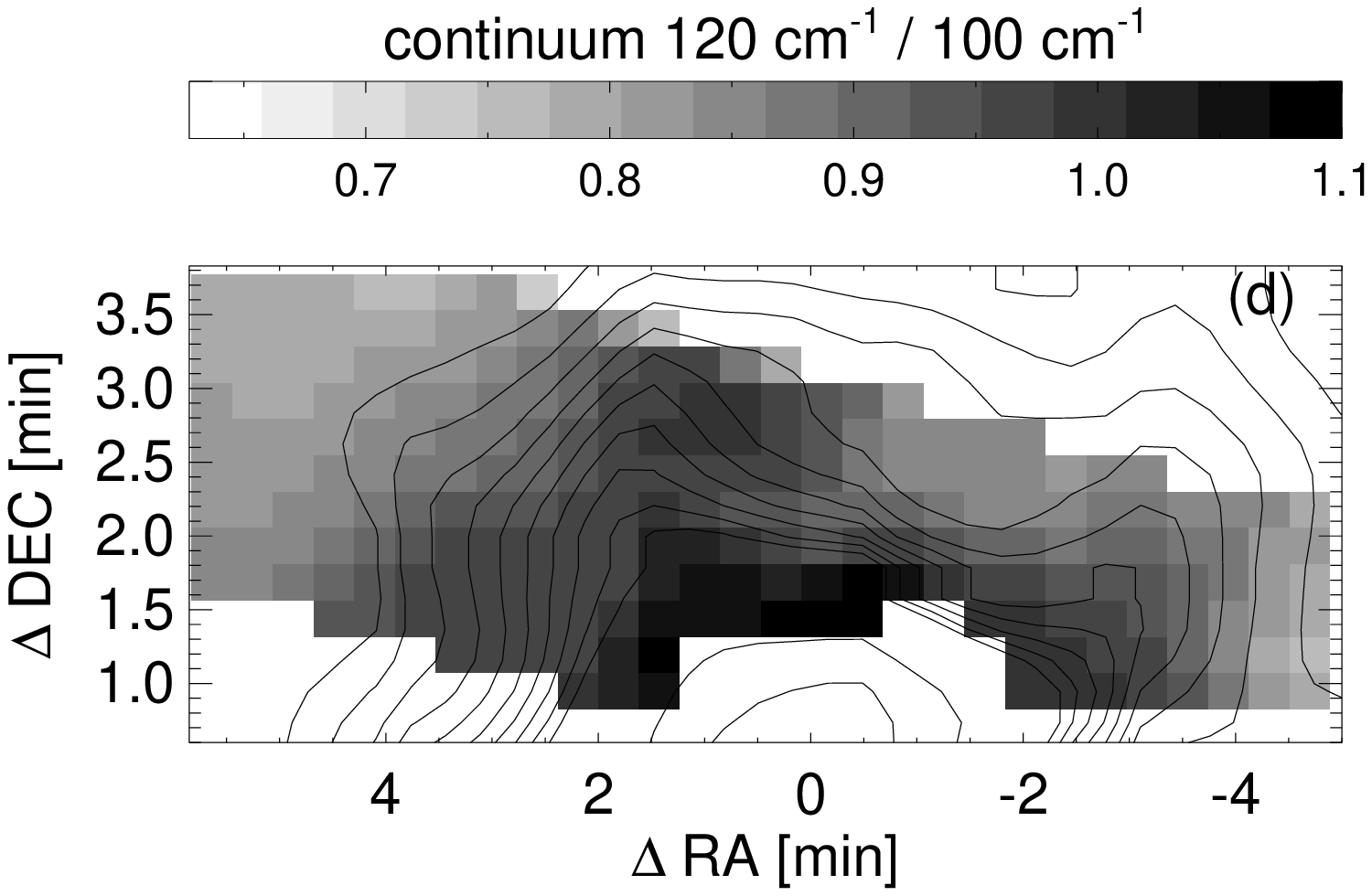}}
\caption{The gray scale shows the \oiii\ 88\micron\ intensity ((a)--(c)) and continuum ratio of 120\pcm/100\pcm\ (d) for G333.6-0.2. In (a), the NVSS 1.4GHz intensity, in (b), the IRC 9\micron\ intensity, and in (c), the IRC 18\micron\ intensity are overlaid in contours, respectively.  In (d), the contours indicate the NVSS 1.4GHz intensity.  The contours of SUMSS are drawn from 0.1 Jy\ beam$^{-1}$ to 1 Jy\ beam$^{-1}$ in a 0.1 Jy\ beam$^{-1}$ interval and 2 Jy\ beam$^{-1}$ and 3 Jy\ beam$^{-1}$.  The contours of the IRC 9\micron\ and 18\micron\ maps are on a linear scale with arbitrary units.  The origin of the coordinates is the same as Fig.~\ref{linemap_G333}.}\label{G333_OIII_w_contours}
\end{figure*}

\begin{figure}
\centering
\resizebox{\linewidth}{!}{\includegraphics{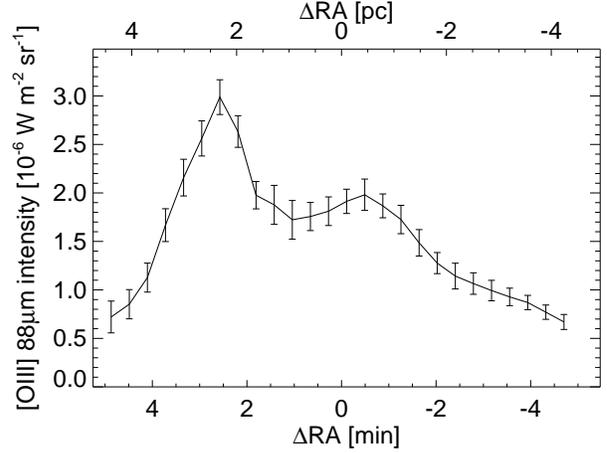}}
\caption{The distribution of the \oiii\ 88\micron\ emission line along the solid line in Fig.~\ref{linemap_G333}c.}\label{G333_OIII_cut}
\end{figure}

\begin{figure}
\centering
\resizebox{\linewidth}{!}{\includegraphics{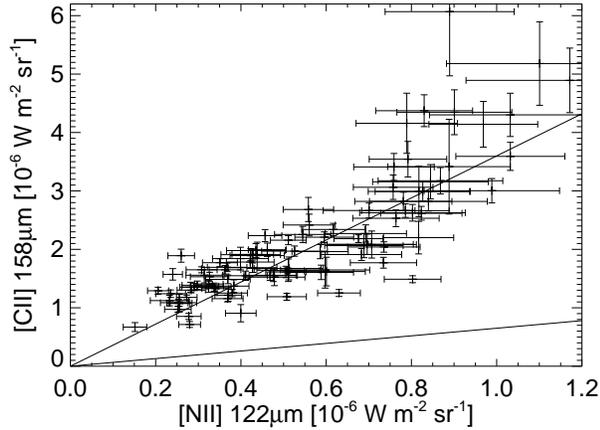}}
\caption{The observed \cii\ 158\micron\ emission versus the \nii\ 122\micron\ emission in G333.6-0.2.  The lines show the calculated intensity ratio assuming all emission arises from ionized gas, for two electron densities, 16\cc\ (upper) and 977\cc\ (lower), the electron temperature of $10^4$~K, and the solar abundance by \citet{Asplund05}.}\label{G333_NIICII}
\end{figure}

\begin{figure*}
\centering
\resizebox{0.45\linewidth}{!}{\includegraphics{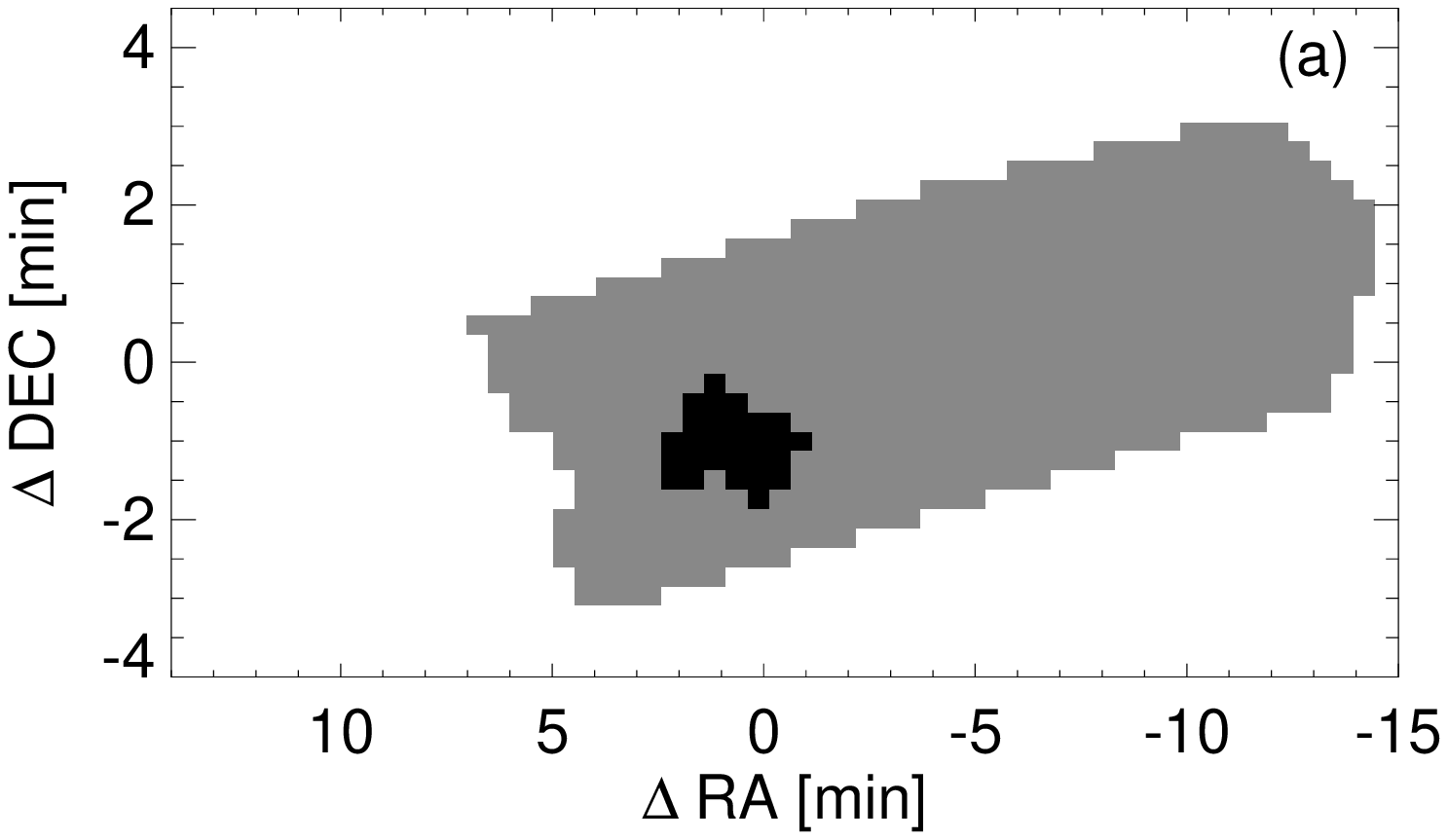}}
\resizebox{0.45\linewidth}{!}{\includegraphics{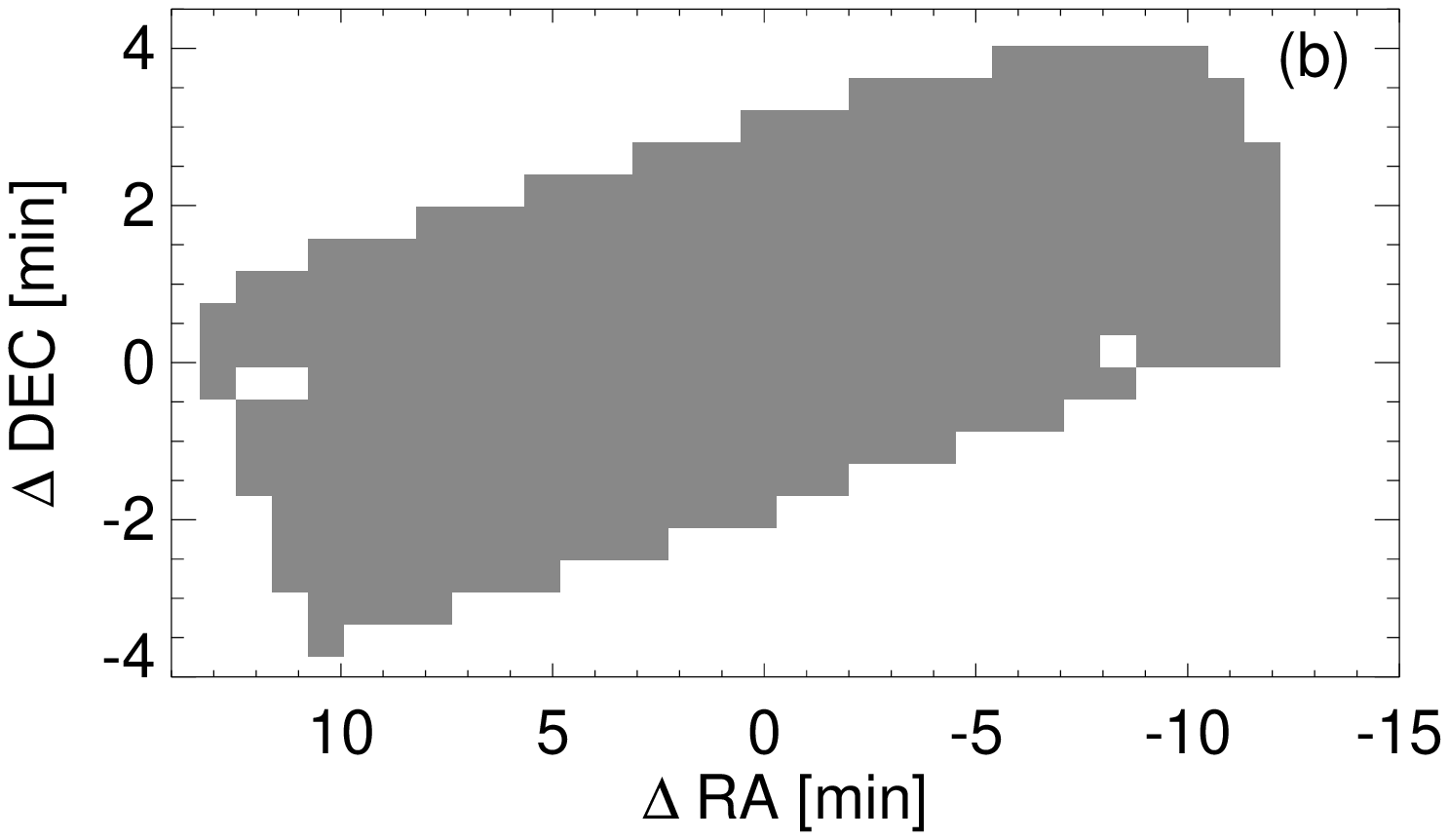}}
\resizebox{0.45\linewidth}{!}{\includegraphics{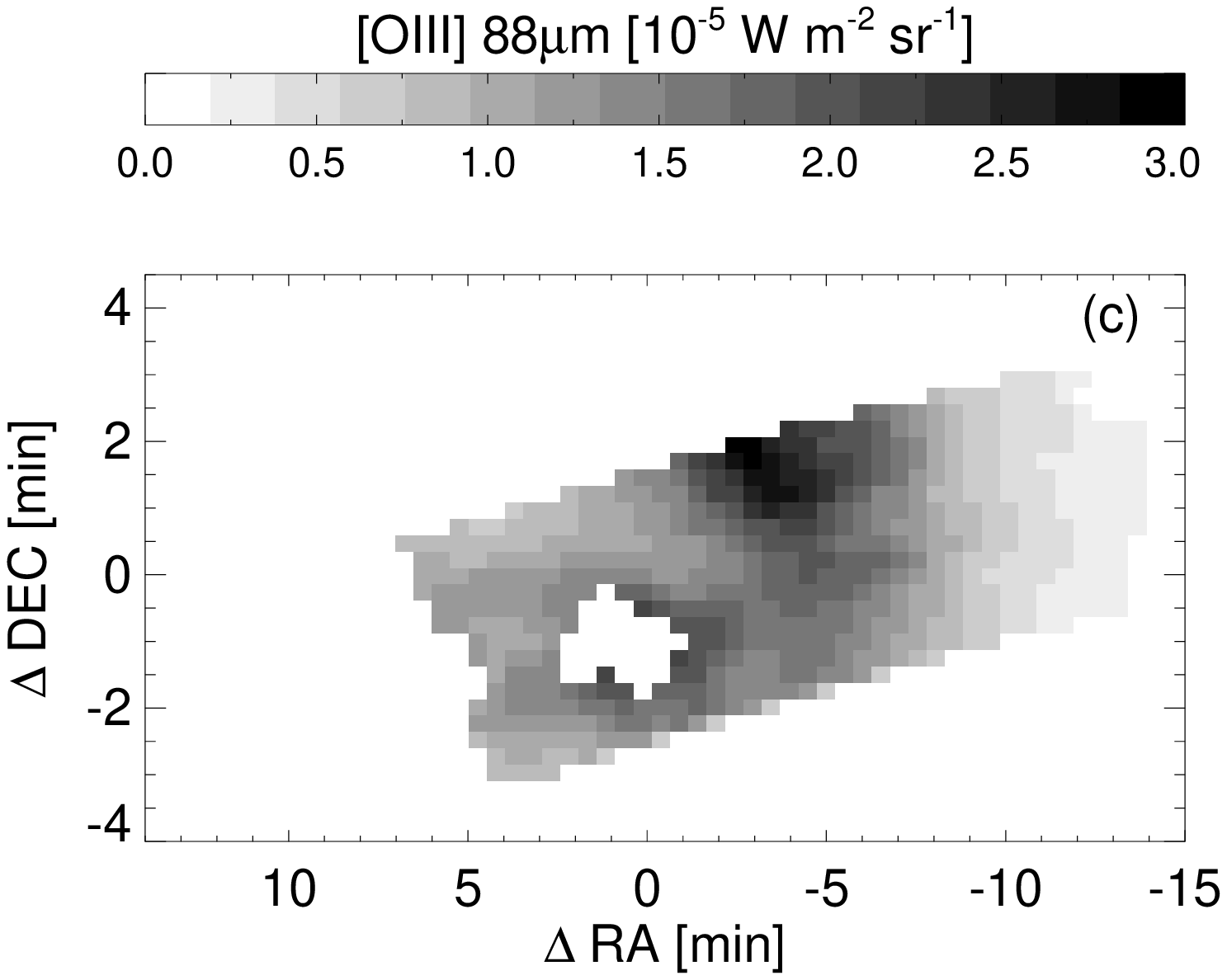}}
\resizebox{0.45\linewidth}{!}{\includegraphics{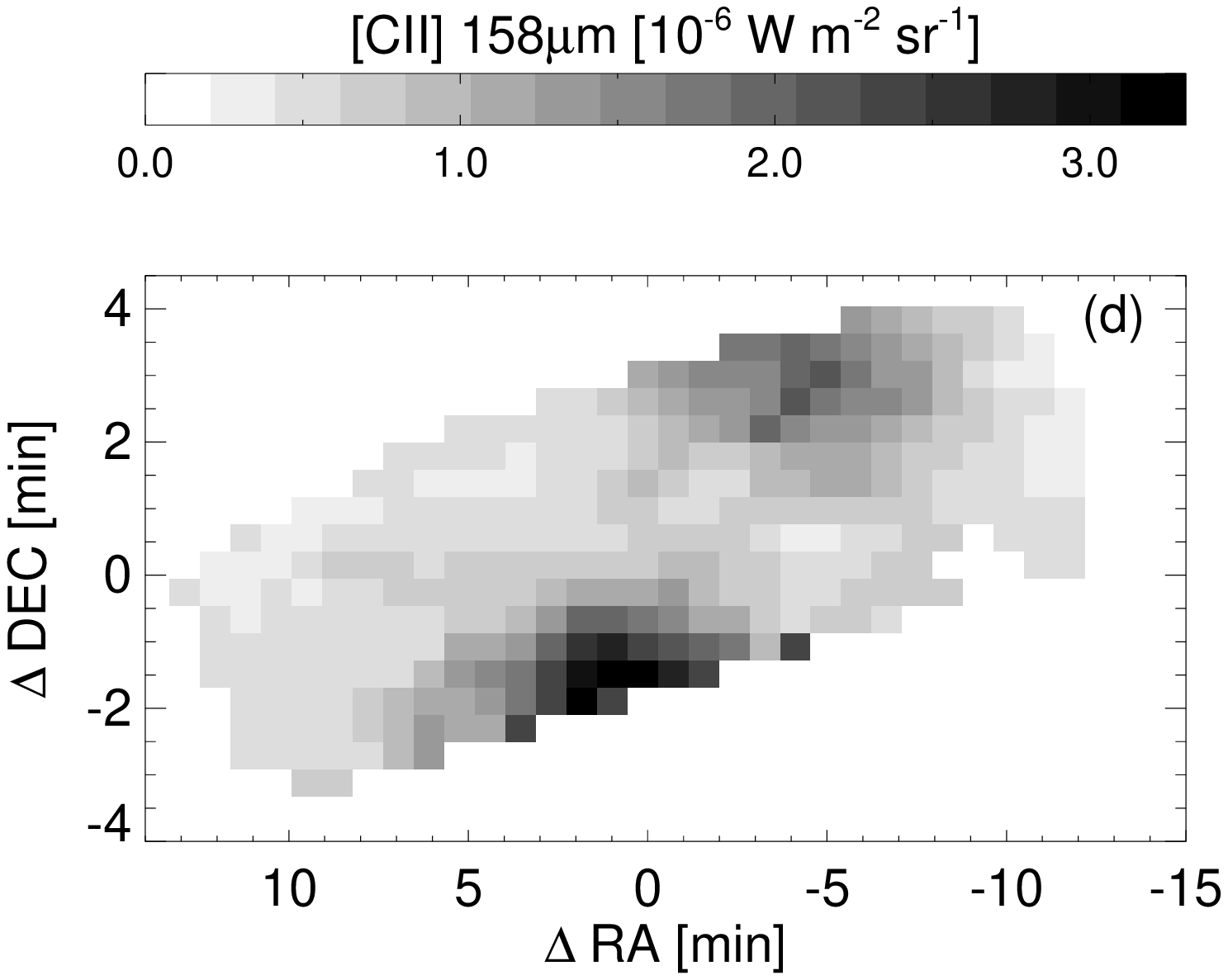}}
\resizebox{0.45\linewidth}{!}{\includegraphics{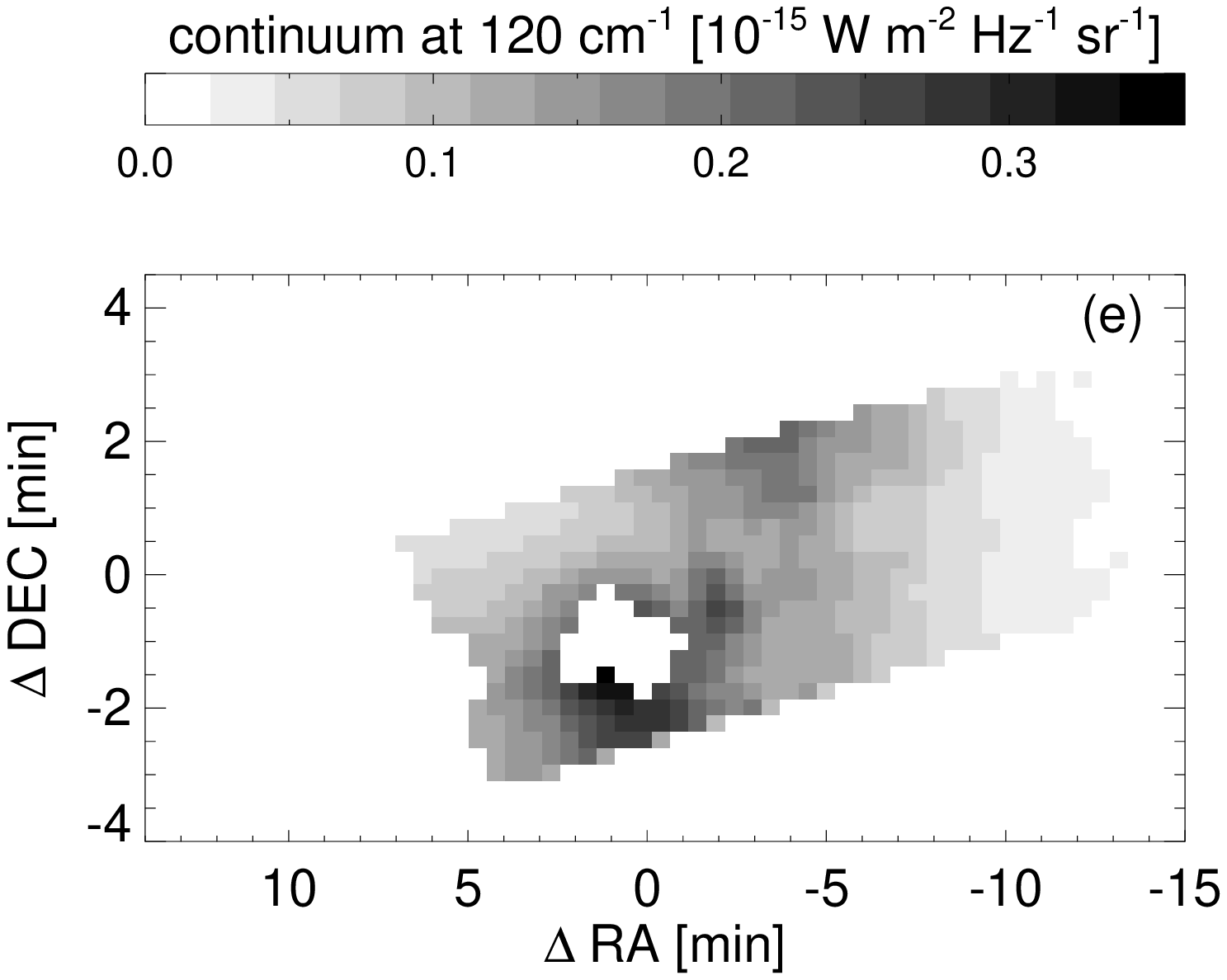}}
\resizebox{0.45\linewidth}{!}{\includegraphics{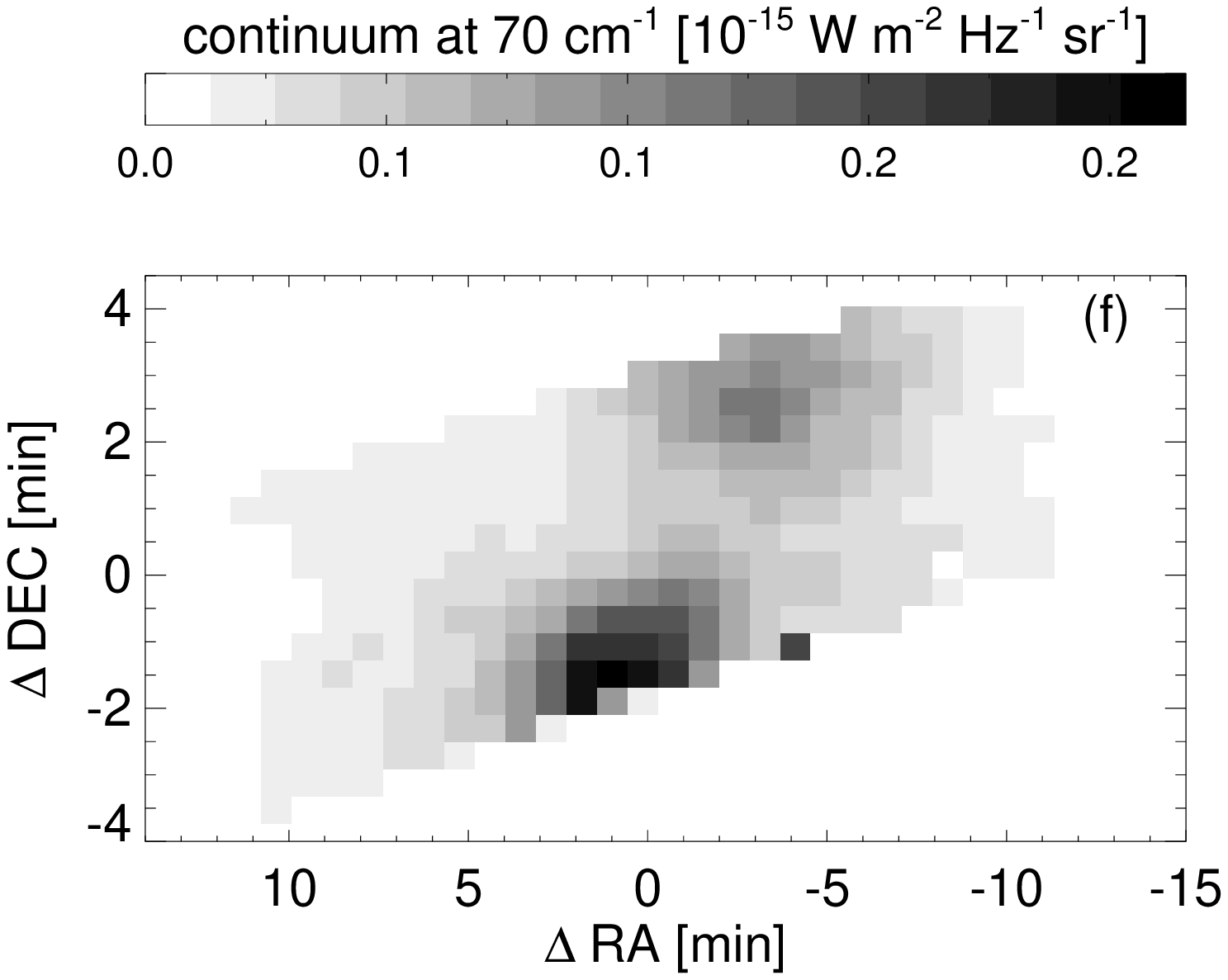}}
\caption{(a) The observed (gray) and saturated (black) area of SW and (b) that of LW, (c) \oiii\ 88\micron, and (d) \cii\ 158\micron\ intensity map and continuum maps at (e) 120\pcm\ and (f) 70\pcm\  in NGC~3603.  The origin of the coordinates is HD97950 [RA = $11^\mathrm{h}15^\mathrm{m}07^\mathrm{s}.3$ and Dec = $-61^{\circ}15^{\prime}38^{\prime\prime}.52$ (J2000)].}\label{linemap_NGC3603}
\end{figure*}

The exciting cluster of G333.6-0.2 is located around the origin of Fig.~\ref{linemap_G333}, where both the SW and LW detectors are saturated.  Figure~\ref{G333_OIII_w_contours} shows the correlation of both \oiii\ 88\micron\ and the continuum emission ratio of 120\pcm/100\pcm\ with the 843~MHz radio continuum measured by SUMSS and the 9\micron\ and 18\micron\ emission detected by the IRC all-sky survey.  Globally, the line emission or continuum exhibit a large gradient in the observed area and become intense towards the exciting cluster.  The \oiii\ 88\micron\ emission has a local peak around ($\Delta$RA, $\Delta$Dec) $\sim$ (2$^\prime$.5, 2$^\prime$) (Fig.~\ref{linemap_G333}c).  Figure~\ref{G333_OIII_cut} shows the intensity distribution along the line shown in Fig.~\ref{linemap_G333}c, which confirms that this local peak is real even taking into account the uncertainty in the line intensity (Appendix \ref{app:const_map}).  Is this local \oiii\ peak indicative of an excitation source?

Although the radio continuum does not show a clear feature corresponding to the \oiii\ local peak (Fig.~\ref{G333_OIII_w_contours}a), the 9\micron\ emission appears as a wall-like structure that sandwiches the eastern and western side of the \oiii\ 88\micron\ peak with a hole at the local \oiii\ peak (Fig.~\ref{G333_OIII_w_contours}b).  Since the 9\micron\ emission is dominated by PAH emission, this hole in the 9\micron\ emission is understood to represent a depletion of the PAHs near the excitation sources, and an enhancement at the PDR/\hii\ region interface.  The 18\micron\ emission has a corresponding feature at the \oiii\ local peak (Fig.~\ref{G333_OIII_w_contours}c), which is consistent with heating sources that produce hot dust.  On the other hand, \cii\ 158\micron\ (Fig.~\ref{linemap_G333}d), \nii\ 122\micron\ (Fig.~\ref{linemap_G333}f), the radio continuum (Fig.~\ref{G333_OIII_w_contours}a), and the continuum emission ratio of 120\pcm/100\pcm\ (Figs.~\ref{linemap_G333}g, \ref{G333_OIII_w_contours}d) do not exhibit a clear peak around the \oiii\ local peak.  If the \oiii\ peak corresponds to the excitation source as in G3.270-0.101, the dust temperature should be high and thus the ratio of 120\pcm/100\pcm\ emission should also peak there.  On the other hand, \nii\ and \cii\ emission should not show a peak at the local \oiii\ peak since most N and C ions should be in higher ionization states.  The radio continuum is sensitive to the density and/or column density (i.e, the emission measure) of the ionized gas, and therefore may not show a peak at the excitation source.  However, the flux ratio of 120\pcm/100\pcm\ is sensitive to the temperature in a narrow range, and thus the absence of the peak in the ratio at the \oiii\ peak may be understood by this region having a much higher dust temperature than G3.270-0.101.  The 120\pcm/100\pcm\ range around the \oiii\ peak is higher than that in G3.270-0.101 (Figs.~\ref{linemap_G3}g and \ref{linemap_G333}g).  Unfortunately, other continuum flux ratios available in our data will not improve our understanding because of the poor S/N in other spectral ranges and misalignment of SW and LW detector arrays.  Hence, we surmise but do not prove that the \oiii\ emission indicates the position of the exciting source(s).  The distribution of the 18\micron\ emission supports this hypothesis.

Assuming that there is an excitation source at the \oiii\ local peak, we estimate the O$^{++}$-Str\"{o}mgren radius in the same manner as for G3.270-0.101.  With an electron density of $200$\pcm\ \citep{Simpson04}, it is estimated to be 0.9~pc, 0.45~pc, and 0.2~pc for an O3, O5, and O7 star, respectively, which correspond to 62\arcsec, 30\arcsec, and 14\arcsec, respectively for a distance of 3.1~kpc.  The width of the \oiii\ 88\micron\ emission emitting area in Fig.~\ref{G333_OIII_cut} is comparable with the FWHM of the SW in spectroscopic mode, which is $39^{\prime\prime}$--$44^{\prime\prime}$, thus the observed width is an upper limit.  Therefore, an O3 star is too early to be the excitation source at the local peak of \oiii\ 88\micron.  On the other hand, the total 88\micron\ luminosity that each O$^{++}$-Str\"{o}mgren sphere provides is $2\times 10^{-13}$, $6\times 10^{-14}$, and $1\times 10^{-14}$\,W\,m$^2$, respectively.  The observed 88\micron\ excess around the \oiii\ local peak is $\sim 8\times 10^{-14}$\,W\,m$^2$.  Therefore, with taking account of the uncertainty in the absolute flux calibration of the FIS-FTS (50\%; Murakami et al. in prep.), a single O5 type star can consistently account for the radius of the Str\"{o}mgren sphere and the total flux of the excess 88\micron\ emission.

Figure~\ref{G333_NIICII} shows the correlation between the observed \cii\ 158\micron\ and \nii\ 122\micron\ intensities.  They are basically well correlated with each other.  We calculate the intensity ratio in the ionized gas for an electron density of 16--977\cc, which is derived from \nii\ 122\micron/205\micron\ at the central position and two northern positions \citep{Colgan93,Simpson04}, the electron temperature of $10^4$~K, and the solar abundance for C and N by \citet{Asplund05}.  The results are overplotted in Fig.~\ref{G333_NIICII}.  Since the \cii\ 158\micron/\nii\ 122\micron\ ratio largely depends on the electron density, 16--91\% of the observed \cii\ intensity comes from the ionized gas for the range of $n_e=16$--$977$\cc.  The contribution of the ionized gas to the \cii\ 158\micron\ emission has been investigated in some star-forming regions: 27\% or 20$\pm$10\% in the Carina nebula \citep{Oberst06,Mizutani04}, 20--70\% in S171 \citep{Okada03}, and 30--80\% in the $\sigma$ Sco region \citep{Okada06}.  Model calculations indicate that it is at least 10\% and up to 50--60\% according to \citet{Abel06}, and $\sim 10$\% for \citet{Kaufman06}.  The present result is compatible with these previous studies, although it does not provide a strong constraint.

\subsection{NGC~3603}

\begin{figure*}
\centering
\resizebox{0.45\linewidth}{!}{\includegraphics{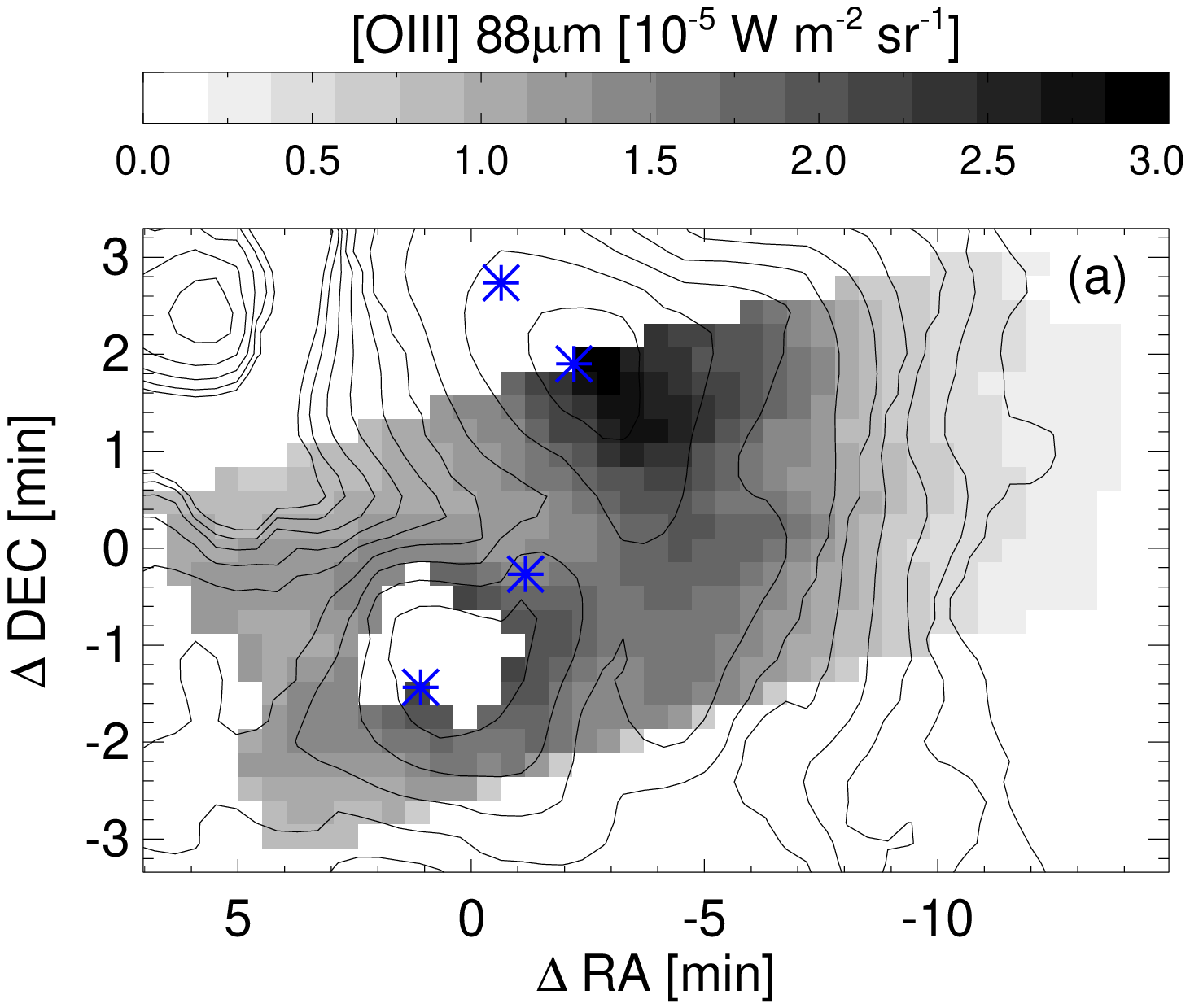}}
\resizebox{0.45\linewidth}{!}{\includegraphics{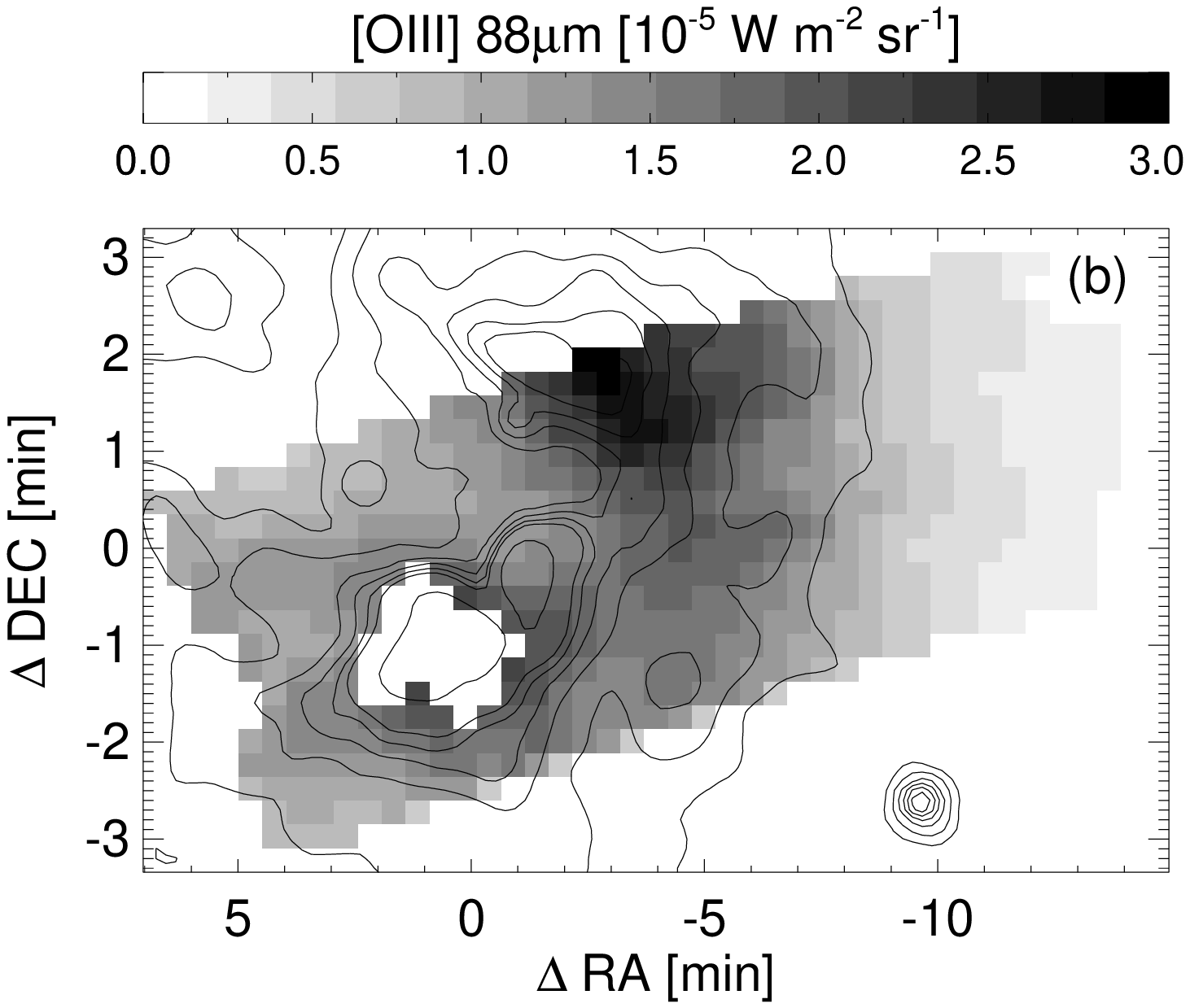}}
\resizebox{0.45\linewidth}{!}{\includegraphics{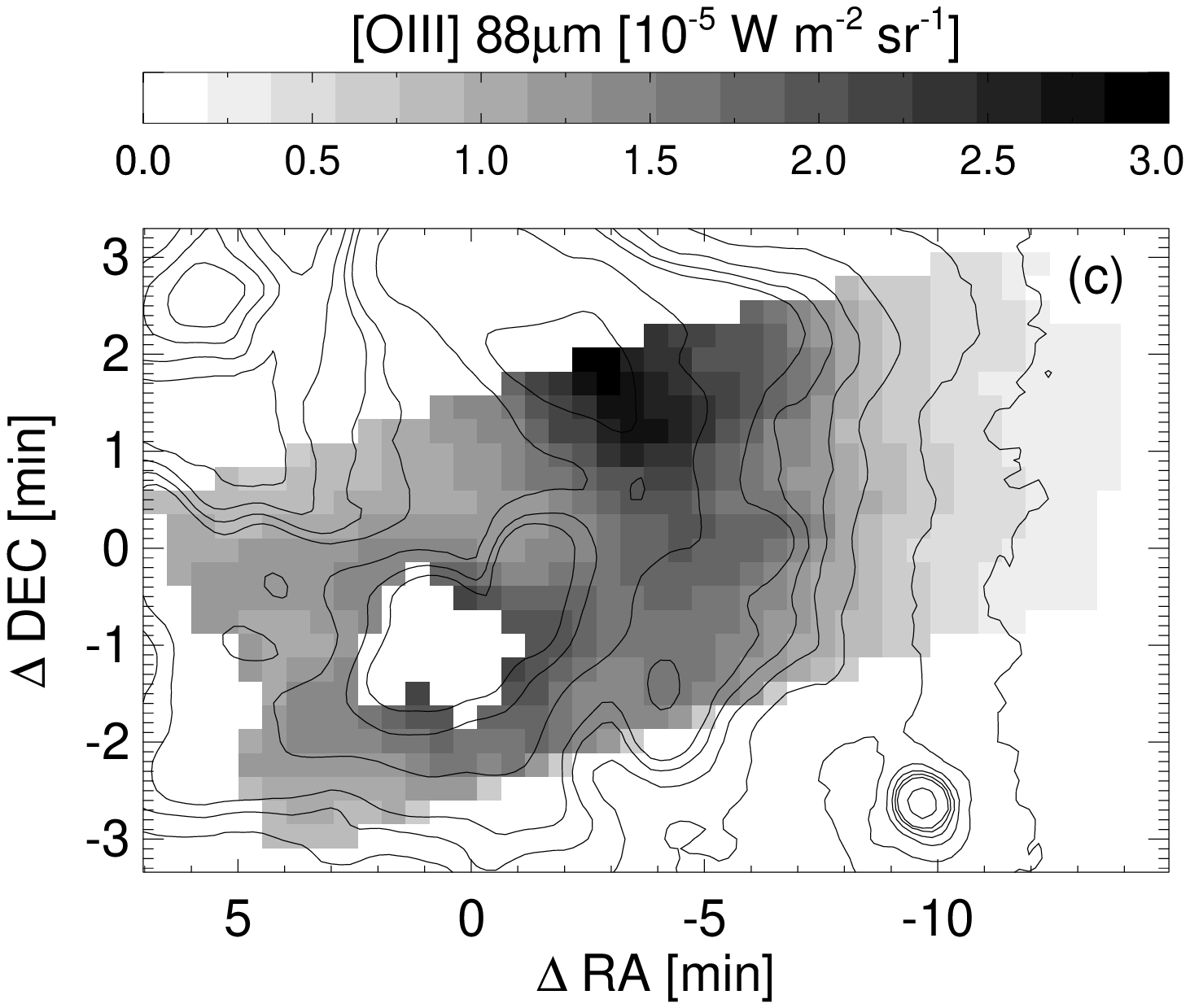}}
\caption{The \oiii\ 88\micron\ intensity map (gray scale) with contours of (a) SUMSS at 843~MHz, (b) MSX band-A at 8\micron, and (c) band-E at 21\micron\ in NGC~3603.  The contours of the SUMSS are drawn from $10^{-1}$ Jy\ beam$^{-1}$ to $10^{0.6}$ Jy\ beam$^{-1}$ in a 0.2 interval of index.  The contours of MSX band-A is $5\times 10^{-6}$, $1\times 10^{-5}$, $2\times 10^{-5}$, $3\times 10^{-5}$, $4\times 10^{-5}$, $5\times 10^{-5}$, and $1\times 10^{-4}$\ W\,m$^2$\,sr$^{-1}$, and those of MSX band-E is the same with the additional two contours of $2\times 10^{-4}$ and $3\times 10^{-4}$\ W\,m$^2$\,sr$^{-1}$.  The asterisks in (a) show the molecular clumps reported by \citet{Nurnberger02} as MM7, MM6, MM1, and MM2 from north to south.  The origin of the coordinates is the same as in Fig.~\ref{linemap_NGC3603}.}\label{NGC3603_OIII_w_contours}
\end{figure*}

\begin{figure*}
\centering
\resizebox{0.45\linewidth}{!}{\includegraphics{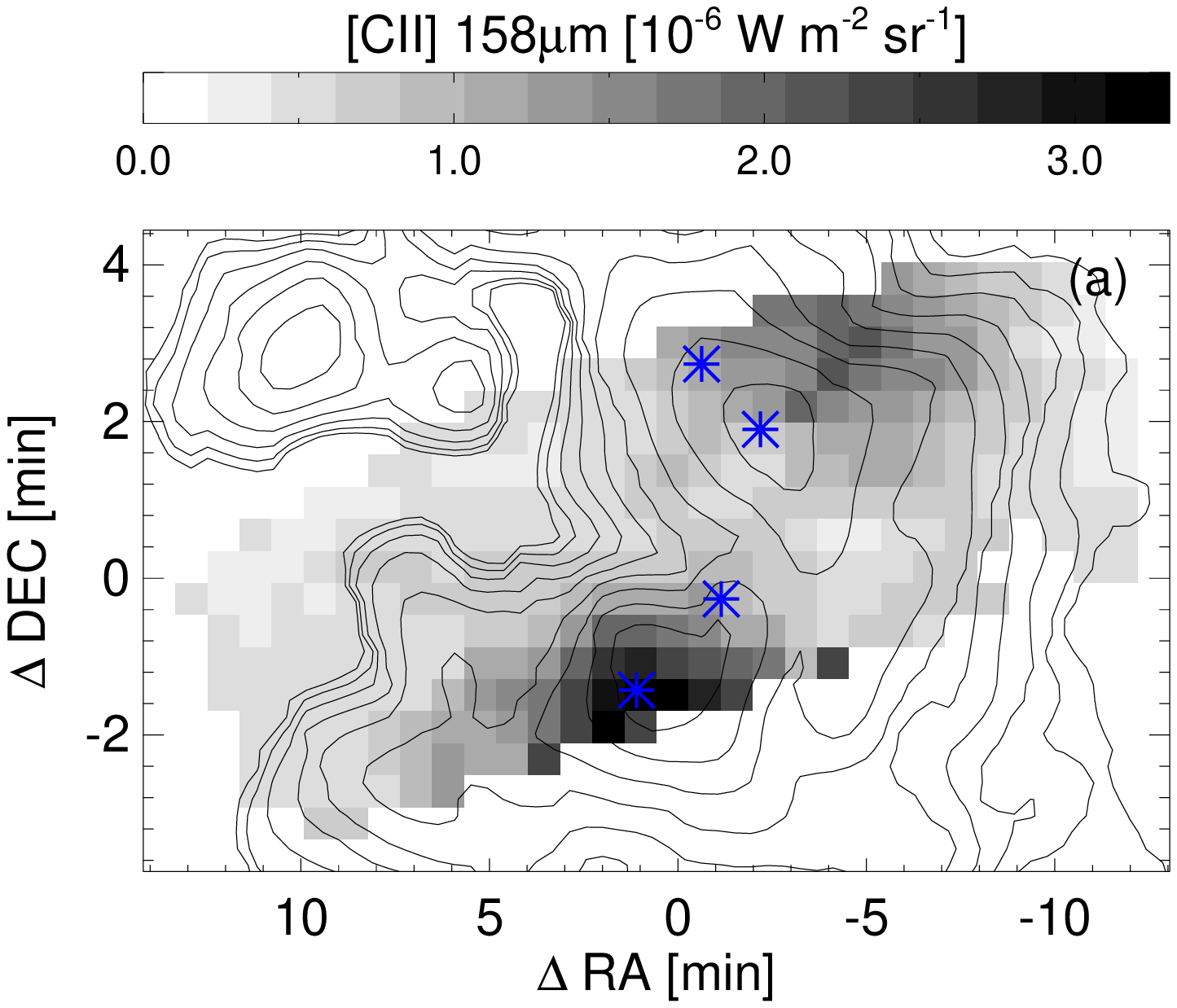}}
\resizebox{0.45\linewidth}{!}{\includegraphics{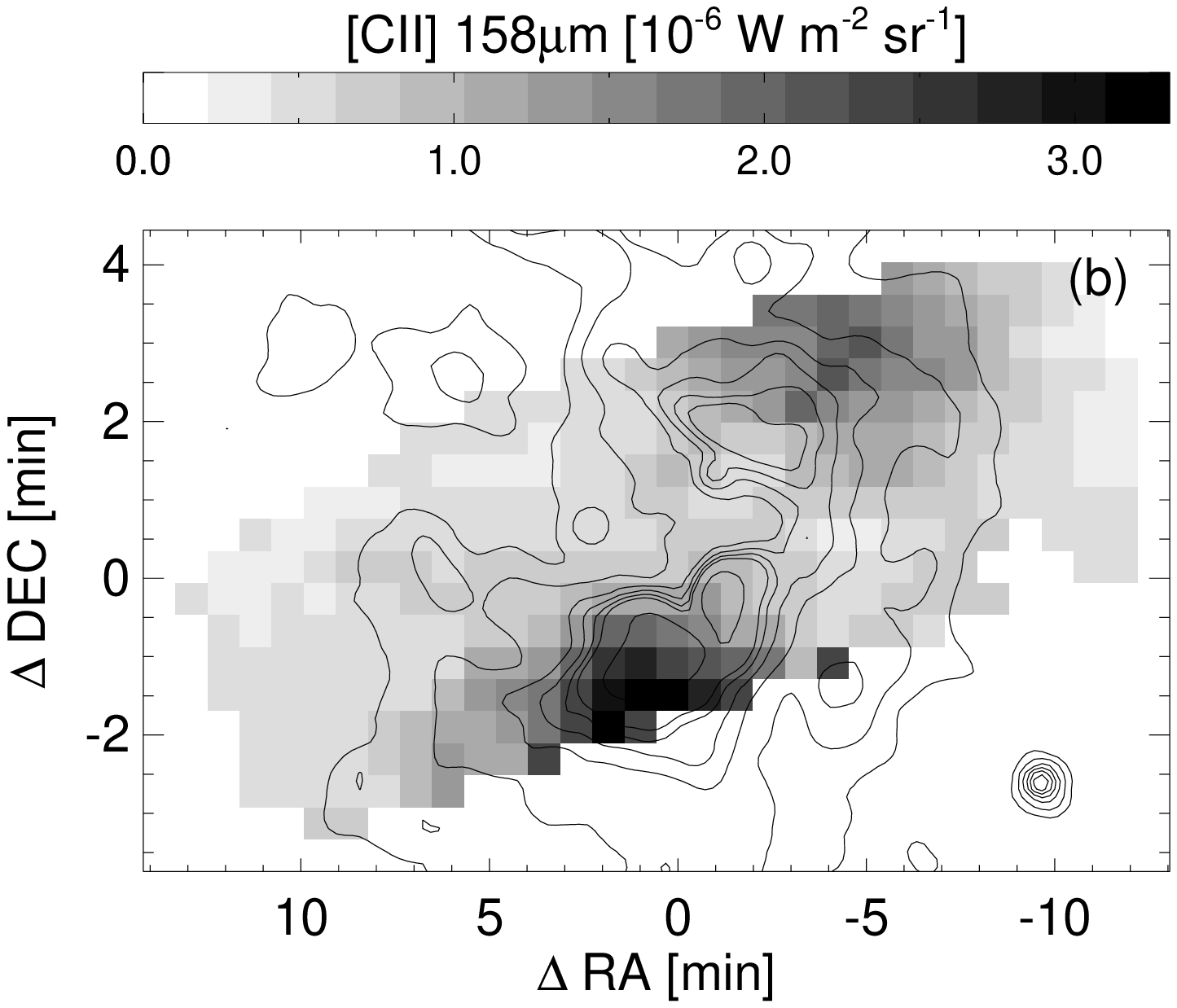}}
\resizebox{0.45\linewidth}{!}{\includegraphics{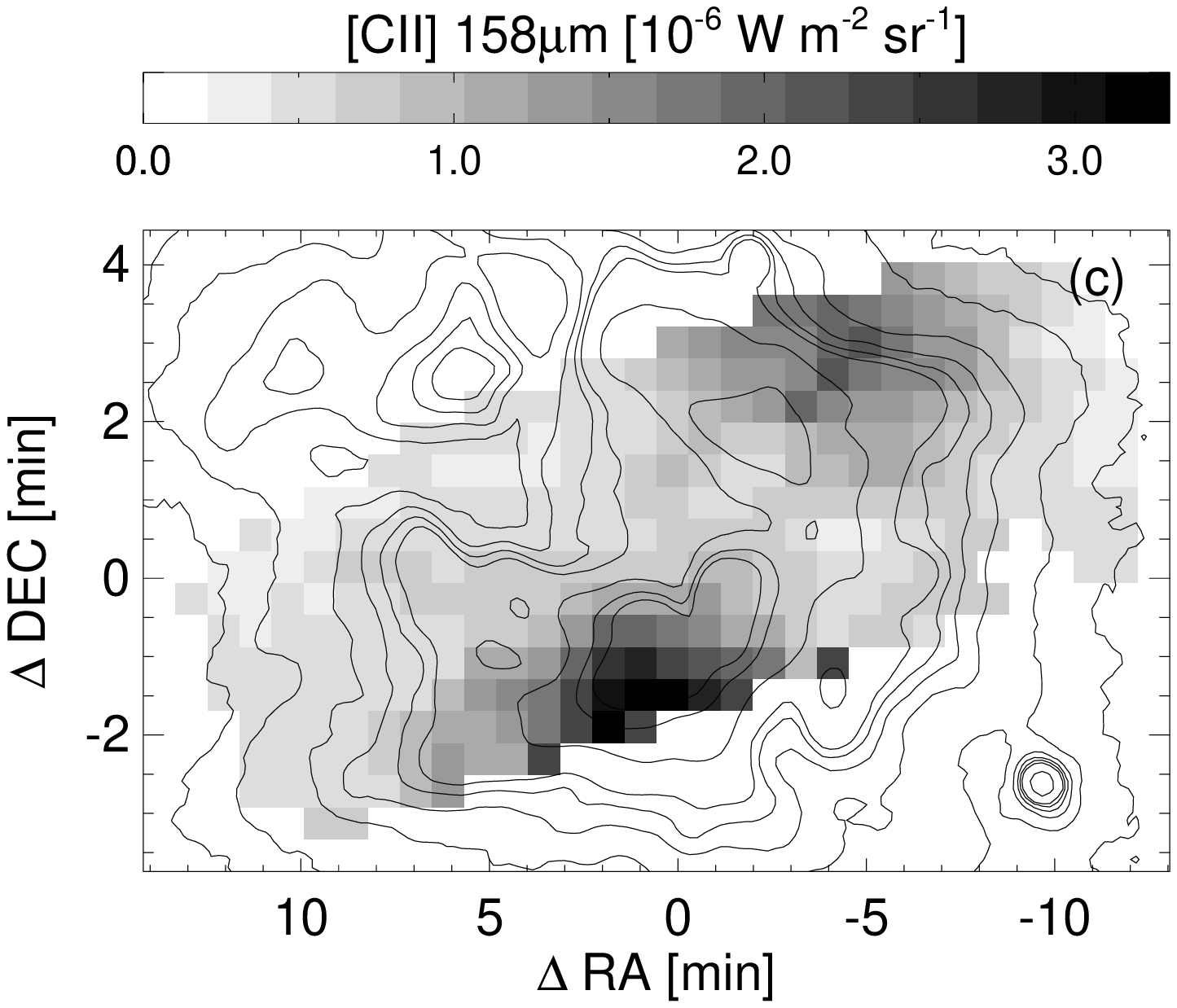}}
\caption{The \cii\ 158\micron\ intensity map (gray scale) with contours of (a) SUMSS at 843~MHz, (b) MSX band-A at 8\micron, and (c) at band-E at 21\micron\ in NGC~3603.  The contours are the same as those in Fig.~\ref{NGC3603_OIII_w_contours}.  The asterisks in (a) are the same as in Fig.~\ref{NGC3603_OIII_w_contours}.  The origin of the coordinates is the same as Fig.~\ref{linemap_NGC3603}.}\label{NGC3603_CII_w_contours}
\end{figure*}

Figures~\ref{linemap_NGC3603}, \ref{NGC3603_OIII_w_contours}, and \ref{NGC3603_CII_w_contours} show the distribution of \oiii\ 88\micron\ and \cii\ 158\micron\ line emission and radio, MIR, and FIR continuum emission in NGC~3603.  These figures indicate that the overall spatial distribution of these tracers is similar to each other; these line and continuum emissions have two peaks toward the south of the central cluster (the origin in figures) and at around ($\Delta$RA, $\Delta$Dec) $\sim$ ($-3^\prime$, $2^\prime$), although some pixels of the SW detectors are saturated around the former positions.  The positions of the molecular clump in \citet{Nurnberger02} indicated by the CS observations are shown in Figs.~\ref{NGC3603_OIII_w_contours}a and \ref{NGC3603_CII_w_contours}a.  Their positions are close to the peaks of the radio, MIR, and FIR continuum emission, indicating the coexistence of the molecular clumps and the ionized gas, and the warm dust.

The central cluster is located at the origin of Fig~\ref{linemap_NGC3603}.  Previous observations suggest that a low density bubble of about 50\arcsec\ or less in radius is formed around it facing to two dense cores of south-east and south-west \citep{Balick80,Clayton86,Clayton90}.  Because of the spatial resolution of the present observations, the bubble detection cannot be confirmed, but we speculate that Figs.~\ref{linemap_NGC3603}e and \ref{NGC3603_OIII_w_contours} show that the radio, MIR, and FIR continuum emission correspond to the structures of these dense cores.  Based on the same discussion we made for G3.270-0.101, \oiii\ 88\micron\ should have a peak at around the excitation sources, but a peak is clearly not seen because of the saturation of the pixels (Fig.~\ref{linemap_NGC3603}c).

On the other hand, there is another peak at around ($\Delta$RA, $\Delta$Dec) $\sim$ ($-3^\prime$, $2^\prime$) in the \oiii\ map as well as the radio, MIR, and FIR continuum emissions, which corresponds to MM6 and MM7 in \citet{Nurnberger02}.  It indicates that the excitation source is within the molecular cloud.  Here only the \cii\ 158\micron\ shows a slightly different distribution, which extends to the western side (Fig.~\ref{NGC3603_CII_w_contours}) and is indicative of an extended PDR.

\subsection{M17}

\begin{figure*}
\centering
\resizebox{0.45\linewidth}{!}{\includegraphics{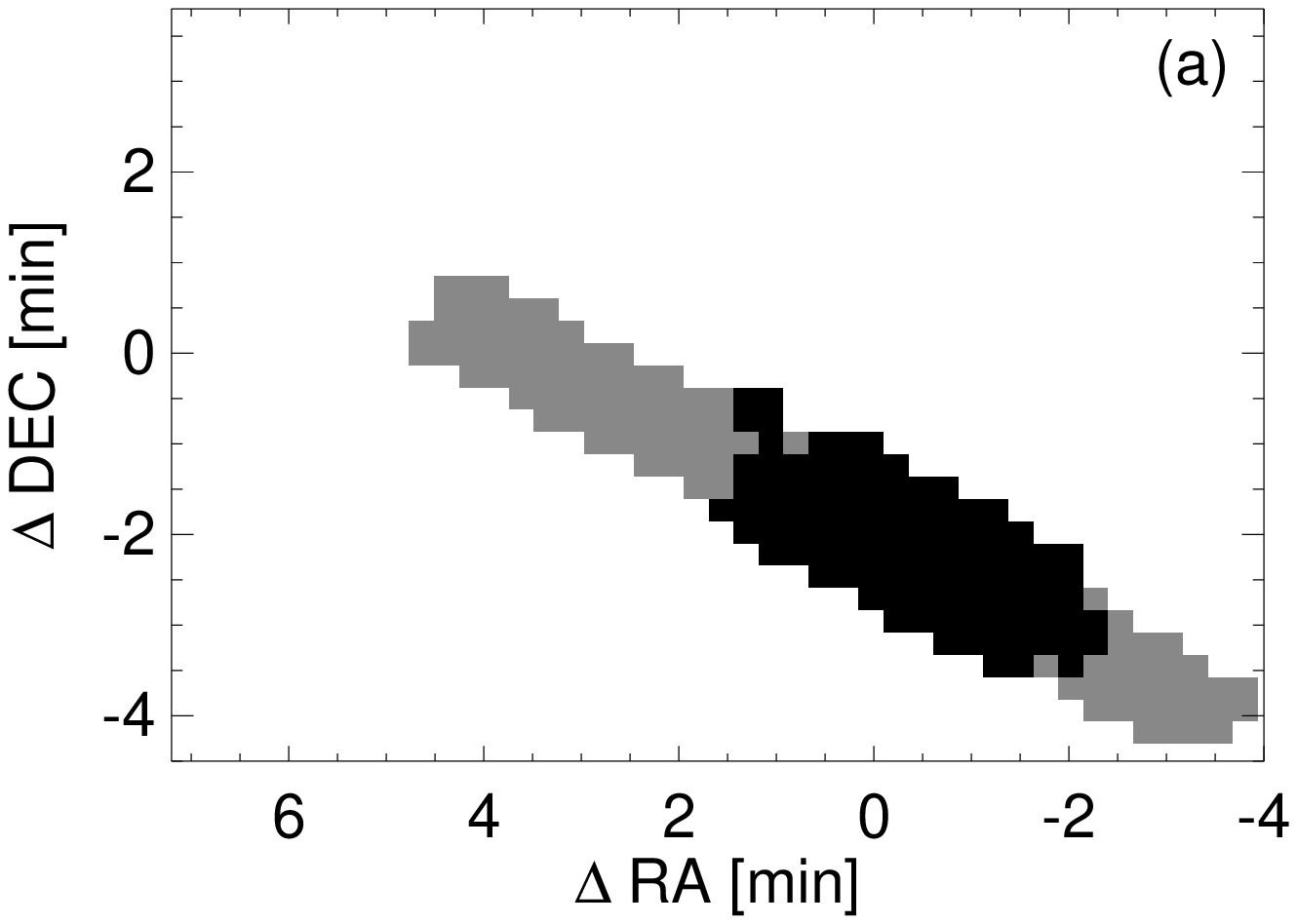}}
\resizebox{0.45\linewidth}{!}{\includegraphics{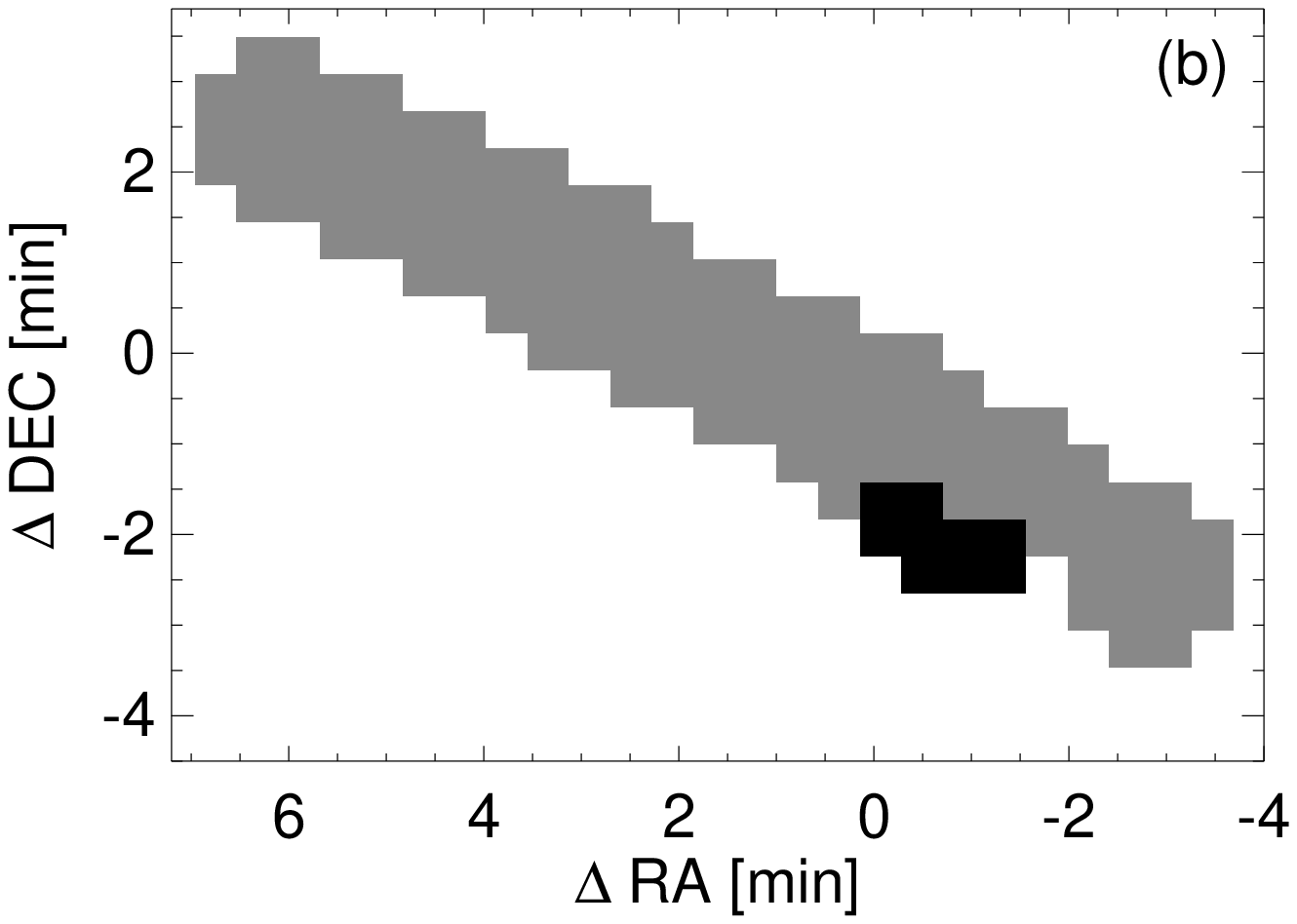}}
\resizebox{0.45\linewidth}{!}{\includegraphics{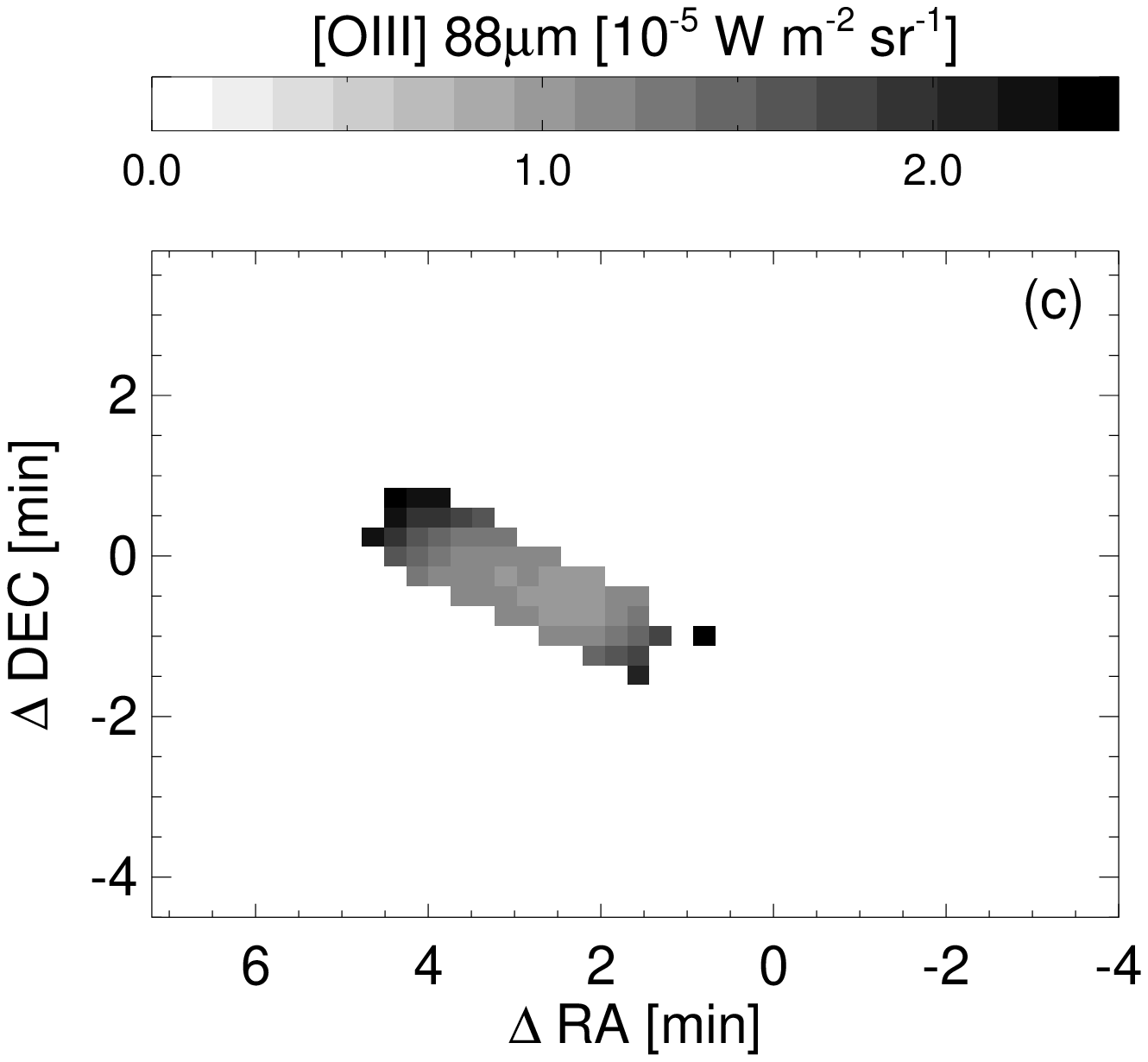}}
\resizebox{0.45\linewidth}{!}{\includegraphics{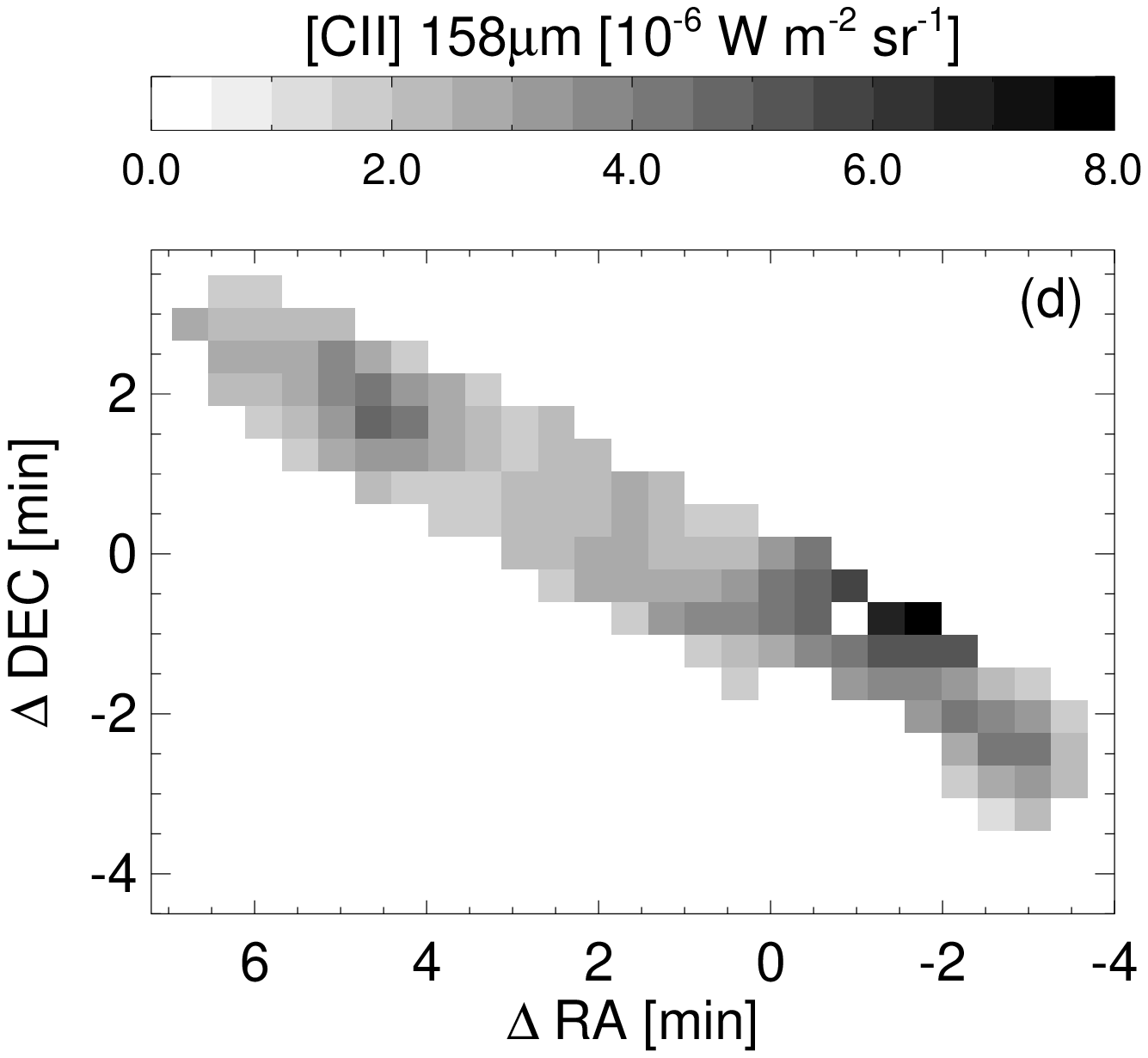}}
\resizebox{0.45\linewidth}{!}{\includegraphics{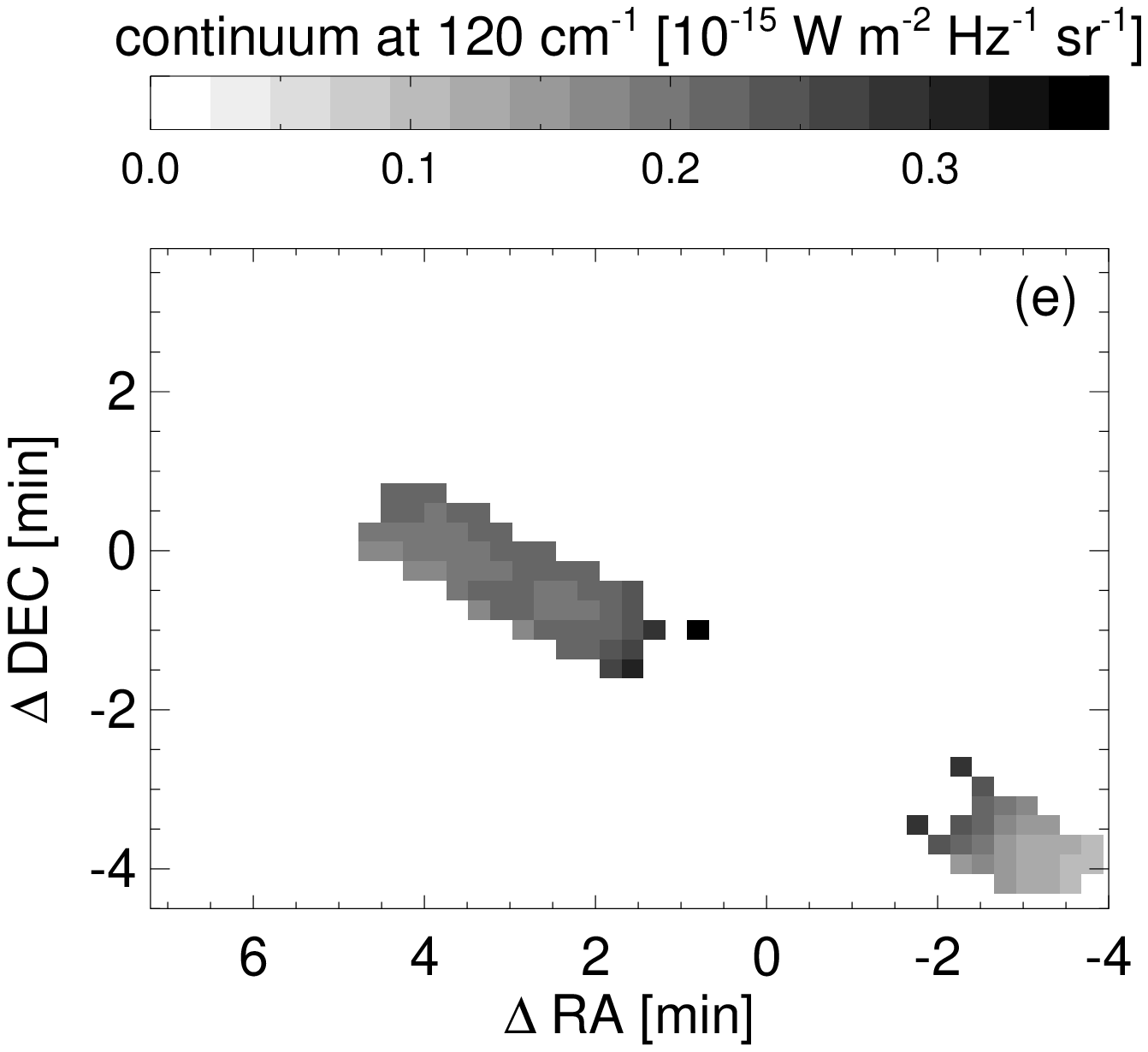}}
\resizebox{0.45\linewidth}{!}{\includegraphics{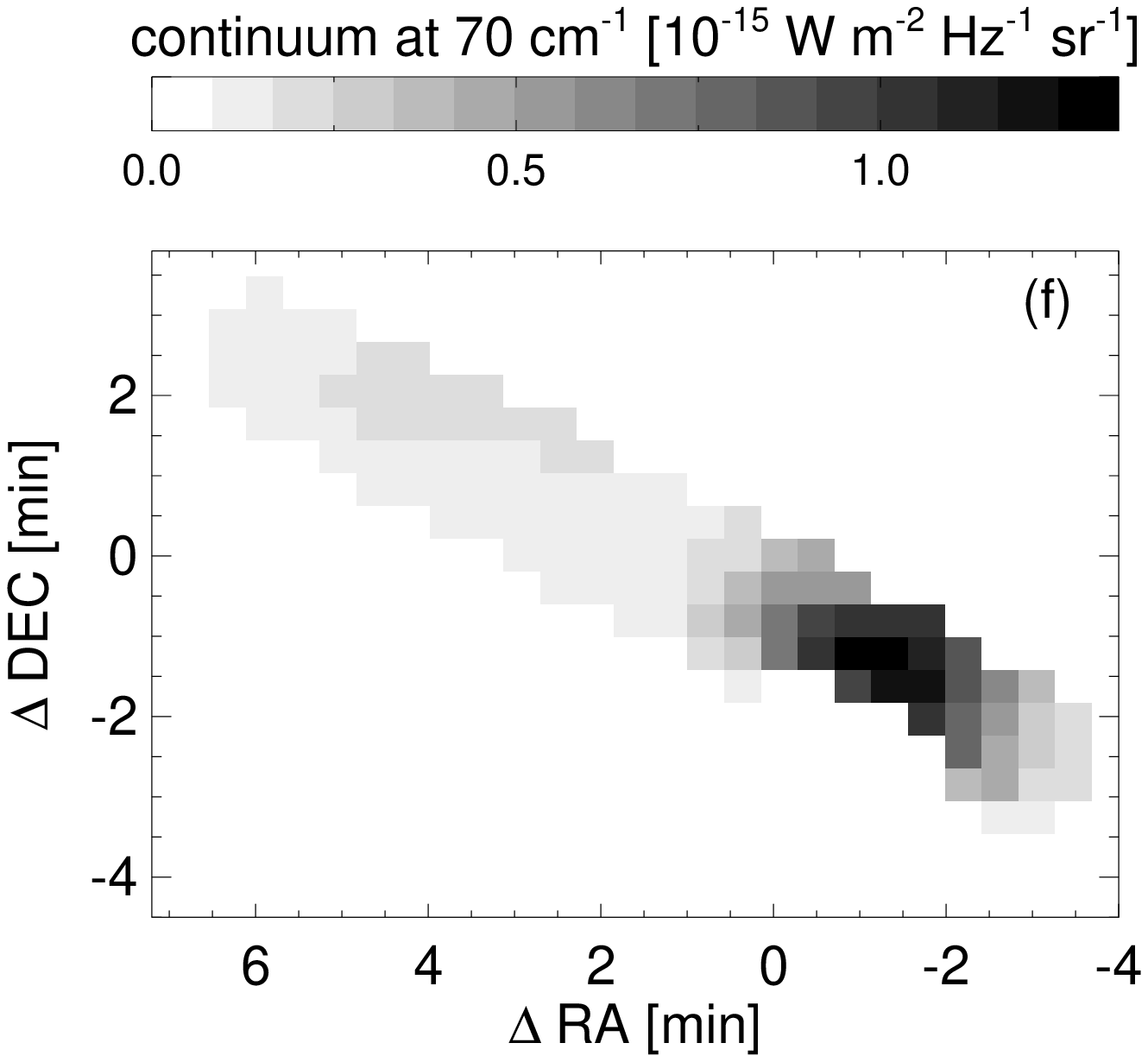}}
\caption{(a) The observed (gray) and saturated (black) area of SW and (b) that of LW, (c) \oiii\ 88\micron, and (d) \cii\ 158\micron\ intensity map and continuum maps at (e) 120\pcm\ and (f) 70\pcm\  in M17.  The origin of the coordinates is NGC6618 [RA = $18^\mathrm{h}20^\mathrm{m}26^\mathrm{s}$ and Dec = $-16^{\circ}10^{\prime}36^{\prime\prime}$ (J2000)].}\label{linemap_M17}
\end{figure*}

\begin{figure*}
\centering
\resizebox{0.45\linewidth}{!}{\includegraphics{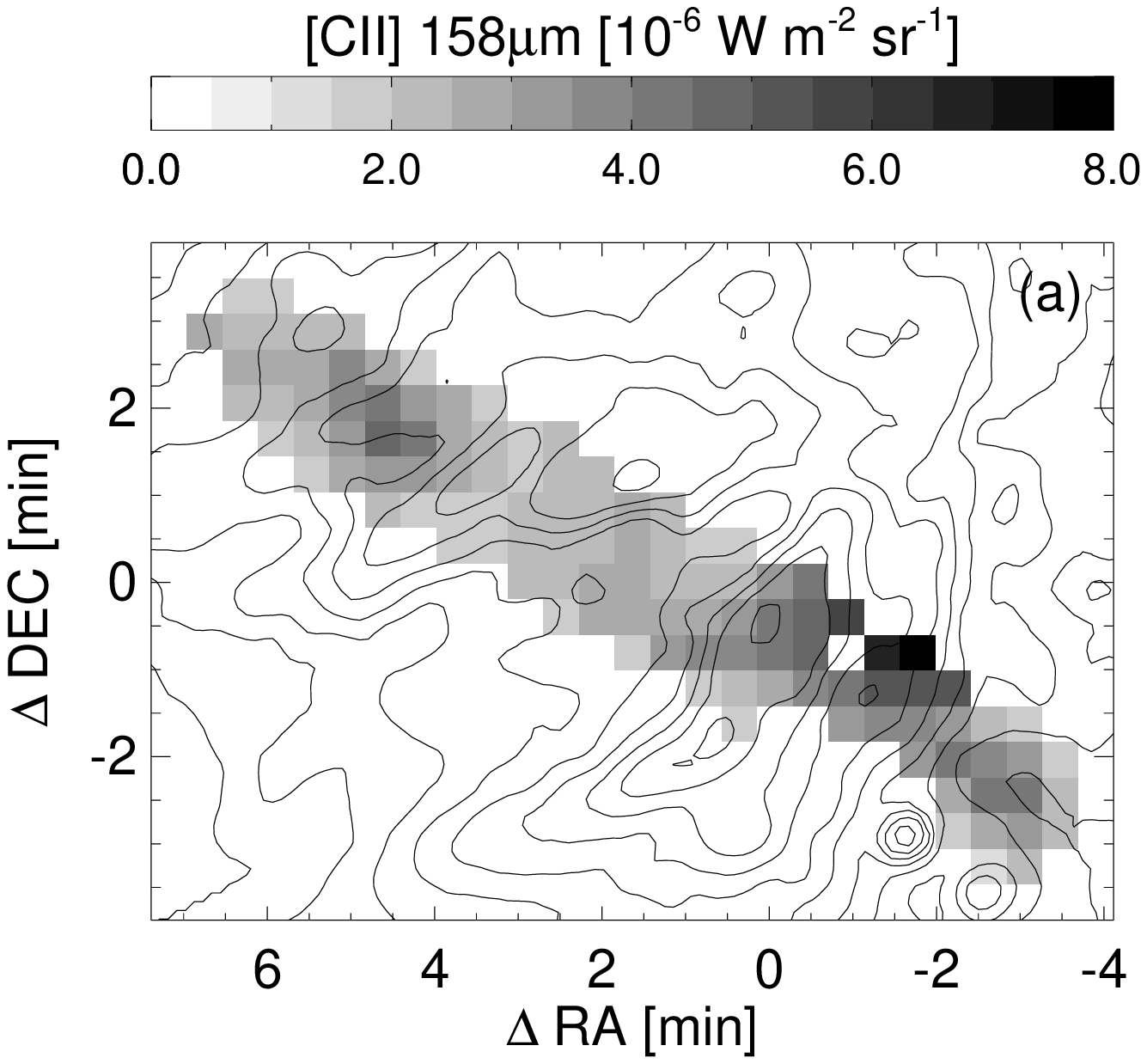}}
\resizebox{0.45\linewidth}{!}{\includegraphics{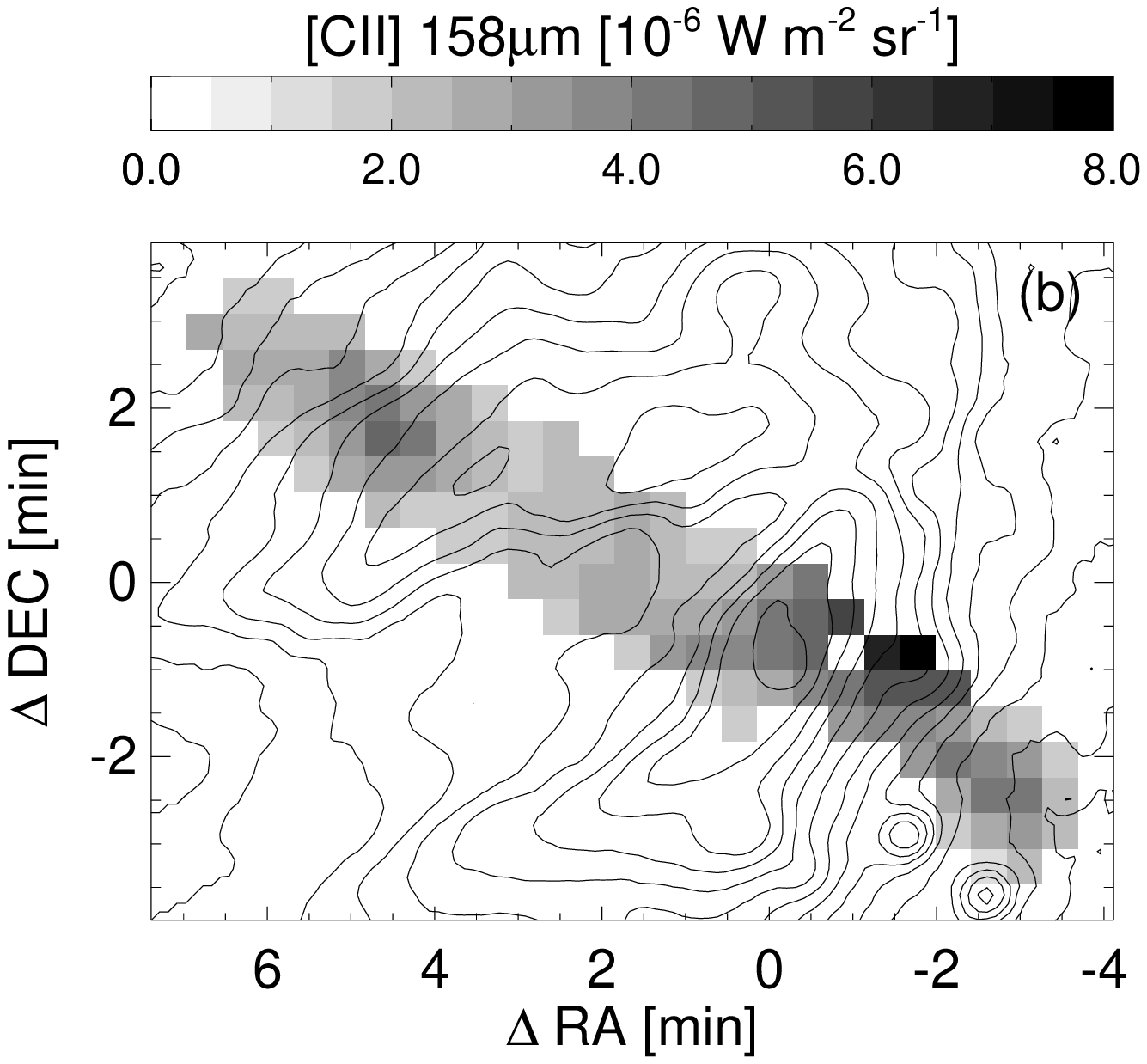}}
\caption{The \cii\ 158\micron\ intensity map (gray scale) with contours of (a) MSX band-A at 8\micron\ and (b) band-E at 21\micron\ in M17.  The contours of MSX band-A are drawn from $10^{-5.4}$ to $10^{-3.6}$\ W\,m$^2$\,sr$^{-1}$ in a $0.2$ interval in logscale, and those of MSX band-E is the same but to $10^{-3.0}$\ W\,m$^2$\,sr$^{-1}$.  The origin of the coordinates is the same as in Fig.~\ref{linemap_M17}.}\label{M17_CII_w_contours}
\end{figure*}

Observations were made across the southern bar region and part of the northern bar region in M17.  The derived line and continuum maps are shown in Fig.~\ref{linemap_M17}, and the MIR continuum emission by MSX is shown in Fig.~\ref{M17_CII_w_contours}.  The NVSS radio continuum can be obtained only for part of the observed region of the FIS-FTS and not shown in the figures.  Unfortunately, nearly half of the pixels of the SW detector are saturated in the southern bar region (Fig.~\ref{linemap_M17}a).  Toward the east of the southern bar corresponds to an \hii\ region and the \oiii\ 88\micron\ emission is detected, whereas the area west of southern bar contains molecular clouds and lacks any \oiii\ 88\micron\ detection.  The contrast between the east and the west side of the southern bar is also clearly seen in the continuum ratio of 120\pcm/100\pcm\ (not shown in the figure), which is $\sim 1.2$ in the east side and below 1 in the west side, indicative of hotter dust on the east side (56~K and 40~K, respectively, with the emissivity index of $-1$ from Fig.~\ref{calc_120div100}).  At the eastern edge of the observed region of the \oiii\ emission, an increase in the line intensity can be seen, which corresponds to the northern bar and is consistent with previous observations \citep{Emery83}.  The \cii\ 158\micron\ line emission is widespread in this region and peaks at the northern bar and the southern bar (Fig.~\ref{linemap_M17}d), in a good agreement with previous observations \citep{Stutzki88,Matsuhara89}.  \citet{Ando02} pointed out that a double ridge of the northern bar is distinct in the MSX band-A and the 3.3\micron\ UIR band emission corresponds to the outer ridge, again consistent with band-A tracing PAHs, hence PDR interfaces. Figure~\ref{M17_CII_w_contours}a shows that the \cii\ 158\micron\ emission is also strong at the outer ridge, which indicates the presence of the PDR region.

The FIR continuum emission shows a strong peak at the southern bar, but there is no significant peak at the northern bar (Figs.~\ref{linemap_M17}e,f).  This indicates that the northern bar has a much lower column density than the southern bar \citep{Elmegreen76,Stutzki88,Ando02,Wilson03}

\section{Summary}
We have presented the results of imaging spectroscopic observations of 4 giant Galactic star-forming regions with FIS-FTS onboard AKARI.  We obtained \oiii\ 88\micron\ and \cii\ 158\micron\ line intensity maps of all the regions.  In G3.270-0.101, we found that the \oiii\ 88\micron\ emission peaks at the excitation source, whereas the radio continuum emission shows a peak at a region 0.8\arcmin (3~pc) away.  In G333.6-0.2, we found a local \oiii\ 88\micron\ emission peak 3\arcmin (3~pc) away from the central cluster, which is likely to be an embedded O5 star.  In NGC~3603, the distribution of the MIR, FIR, and radio continuum emission as well as the \oiii\ 88\micron\ emission are similar to each other.  For all regions, \cii\ 158\micron\ emission is widely distributed, as suggested by previous observations of star-forming regions.  We have pointed out that the \oiii\ 88\micron\ emission shows a stronger correlation with the 18\micron\ continuum emission than 9\micron\ emission, which indicates that the 18\micron\ emission originates in hot dust grains close to early-type stars, in contrast to the 9\micron\ emission, which traces PAH emission, and PAHs are destroyed in the relatively hard UV environments in which O$^{2+}$ forms.  This work indicates that the \oiii\ 88\micron\ emission traces the excitation sources more closely than the radio continuum emission, especially when there is a large density and/or column density gradient.  The FIR spectroscopy provides a powerful tool for understanding the nature of star-forming regions including the position of the excitation sources.

\begin{acknowledgements}
This work is based on observations with AKARI, a JAXA project with the participation of ESA.  We thank all the members of the AKARI project for their continuous help and support.  FIS was developed in collaboration with Nagoya University, ISAS, University of Tokyo, National Institute of Information and Communications Technology (NICT), the National Astronomical Observatory of Japan (NAOJ), and other research institutes.  We thank S. Makiuti for his help in data analysis. 
\end{acknowledgements}

\begin{appendix}
\section{Line-fitting procedure}\label{app:linefit}
Full-resolution spectra of both the SW and LW detectors are affected by strong channel fringes, which can be modeled by an Airy function \citep[][Murakami et al. in prep.]{Kawada08}.  The defringing process is included in the official pipeline, in which the spectrum is divided by an Airy function with the parameter being optimized for each emission line (Murakami et al. in prep.).  We apply them to the present study.  Since the \oiii\ emission is strong and the line width is sufficiently smaller than the fringe period, we apply the `Hanning' apodization to avoid strong tails around the emission line.  After defringing, we derive the line intensities measured by Gaussian fits, assuming a linear baseline.  For the \nii\ and \cii\ emission, which are detected by the LW detector, we do not apply the apodization to improve the detectability since the line width is comparable to the fringe period.  The fringe pattern around the \cii\ emission is more irregular than that around \nii, and we perform a defringing for the \nii\ emission but not for \cii\ emission.  The amplitude of the fringe pattern around \cii\ line is less than $\sim 30$~\%.  We derive the \nii\ and \cii\ line intensities by fitting with a sinc function, assuming a linear baseline after defringing and without defringing, respectively.  We then take the average of the line intensities of the forward and backward scans.  The uncertainty in the line intensities is estimated from the uncertainties in both the baseline and the differences between the intensity of the forward and backward scans.

\end{appendix}
\begin{appendix}

\section{Constructing line maps}\label{app:const_map}

To create a map of a regular grid from distorted detector array images, and to combine images of several pointings toward the same region, we use a Delaunay triangulation algorithm using an IDL routine, interpolating the values from nearby points.  The coordinates of each pixel of the detector array are estimated from the attitude of the satellite with the alignment of the detector arrays, including the distortion in the optics.

The uncertainty in the line intensity of each pointing observation is estimated from the uncertainties in both the baseline and the differences between the intensity of the forward and backward scans as described in Appendix \ref{app:linefit}.  We estimate the uncertainty in the constructed maps by simulations.  We assume a Gaussian distribution with the width of the uncertainty in the line intensity as the probability function for each pixel value.  For each trial, we randomly set the value at each pixel according to this probability function, and create a map of a regular grid.  After a thousand trials, we have a certain distribution for the value at each grid of the constructed maps.  We fit the distribution with a Gaussian profile for each grid and consider the width of the obtained Gaussian to present the uncertainty.

\end{appendix}

\end{document}